\documentclass{jfm}
\usepackage{graphicx}
\usepackage{xcolor}
\usepackage{epstopdf, epsfig}
\newcommand{\appropto}{\mathrel{\vcenter{
  \offinterlineskip\halign{\hfil$##$\cr
    \propto\cr\noalign{\kern2pt}\sim\cr\noalign{\kern-2pt}}}}}

\shorttitle{Nonlinear limiting dynamics of a shrinking interface}
\shortauthor{M. Zhao, S. Li, J, Lowengrub, and Z. Niroobakhsh}

\title{Nonlinear limiting dynamics of a shrinking interface in a Hele-Shaw cell}

\author{Meng Zhao\aff{1},
   Zahra Niroobakhsh\aff{2},
 John Lowengrub\aff{1,3}
  \corresp{\email{lowengrb@math.uci.edu}}
 \and Shuwang Li\aff{4}
  \corresp{\email{sli@math.iit.edu}}
 }

\affiliation{\aff{1}Department of Mathematics, University of California at Irvine,
Irvine, CA 92617, USA
\aff{2} Department of Civil and Mechanical Engineering, University of Missouri-Kansas City, Kansas City, Missouri 64110, USA
\aff{3} Department of Biomedical Engineering; Center for Complex Biological Systems, University of California at Irvine,
Irvine, CA 92617, USA
\aff{4}Department of Applied Mathematics, Illinois institute of technology, Chicago, IL 60616, USA

}

\begin{document}

\maketitle

\begin{abstract}
The flow in a Hele-Shaw cell with a time-increasing gap poses a unique shrinking interface problem. When the upper plate of the cell is lifted perpendicularly at a prescribed speed, the exterior less viscous fluid penetrates the interior more viscous fluid, which generates complex, time-dependent interfacial patterns through the Saffman-Taylor instability. The pattern formation process sensitively depends on the lifting speed and is still not fully understood. For some lifting speeds, such as linear or exponential speed, the instability is transient and the interface eventually shrinks as a circle. However, linear stability analysis suggests there exist shape invariant shrinking patterns if the gap $b(t)$ is increased more rapidly: $\displaystyle b(t)=\left(1-\frac{7}{2}\tau \mathcal{C} t\right)^{-{2}/{7}}$, where $\tau$ is the surface tension and $\mathcal{C}$ is a function of the interface perturbation mode $k$. Here, we use a spectrally accurate boundary integral method together with an efficient time adaptive rescaling scheme, which for the first time makes it possible to explore the nonlinear limiting dynamical behavior of a vanishing interface. When the gap is increased at a constant rate, our numerical results quantitatively agree with experimental observations (Nase {\it et al., Phys. Fluids}, vol. 23, 2011, pp. 123101). When we use the shape invariant gap $b(t)$, our nonlinear results reveal the existence of $k$-fold dominant, one-dimensional, web-like networks, where the fractal dimension is reduced to almost one at late times. We conclude by constructing a morphology diagram for pattern selection that relates the dominant mode $k$ of the vanishing interface and the control parameter $\mathcal{C}$.

\end{abstract}

\begin{keywords}
Fingering instability, Pattern formation, Boundary integral method. 
\end{keywords}

\section{Introduction}

 Saffman-Taylor instabilities \citep{PG} occur when a less viscous fluid is injected into a more viscous fluid confined in a fixed narrow gap between the two parallel plates (Hele-Shaw cell). During injection, the inner less viscous fluid displaces the outer viscous fluid and the interface separating the two fluids exhibits fingering patterns \citep{Langer89,Cummins}. Through repeated tip-splitting events, new fingers develop and the proliferation of fingers leads to dense branching morphologies as the system is driven out of equilibrium \citep{Chuoke,McLean,Park,Jacob86,OH,ShuwangJCP}. Viscous fingering is considered a paradigm for a variety of pattern forming phenomena such as bacterial colony growth and snowflake formation as the physical mechanisms and mathematical structure are similar \citep{Langer80,Langer89,Jacob-1}.
 
One variant of the conventional Hele-Shaw set-up that provides another way to produce viscous fingering patterns is the so-called lifting plate problem \citep{MS,CCJ,AA,Sinha08,Sinha09,EJ10,JDA}. In the lifting plate problem, the top plate in a Hele-Shaw cell is lifted perpendicularly at a prescribed speed and the bottom plate remains at rest. This setup has been used to study adhesion-related problems such as debonding \cite{Francis,Poivet,DAD,MD} and the associated probe tack test \citep{Zosel,Lakrout}.

 In the lifting plate problem, the gap $b(t)$ between the two plates is increasing in time but uniform in space. As the plate is lifted, an inner viscous fluid rushes inward between the two plates and increases in the $z$-direction to preserve volume.  An outer less viscous fluid (usually air) invades the more viscous fluid and generates fingering patterns. The patterns are visually similar to those in the classical radial Hele-Shaw problem, but the driving physics is different in the sense that the flow in this problem is extensional (e.g., free-surface instabilities seen in \citep{McKinley}).  Viscous fingering patterns can also be observed using a Hele-Shaw cell where only one edge of the plate is lifted, which makes the gap width a function of time and space \citep{SEOF,EDJM}.



To characterize pattern formation in the lifting plate problem, the number of fingers has been studied using experiments and theory using Darcy's law as an approximation. Theoretical studies  \citep{MD,ADM,Sinha08,JDA,EJ13-2} show that the number of fingers is controlled by a dimensionless surface tension. However, experiments using constant rates of increasing gap widths suggest that the total number of fingers is not only dependent on the dimensionless surface tension but also on the confinement, or aspect ratio of the fluid, $C_0=\frac{R_0}{b_0}$, where $R_0$ is the initial radius of the liquid and $b_0$ is the initial gap width \citep{JDA}; see also Fig. \ref{lncom} in Sec. \ref{sec:numres} where experimental data is plotted together with the results from linear theory and nonlinear simulations. In general, increasing $C_0$ results in an increased number of fingers. In these studies, the rate of increase in the gap width over time is limited, e.g., constant rates of increase. Accordingly, the viscous fingering instability is transient and eventually the interface shrinks as a circle. Interfacial dynamics in the regime where the gap width increases more rapidly in time, which could lead to non-circular vanishing limiting shapes, have been much less explored.

In this paper, we investigate regimes in which the gap width increases rapidly in time. Motivated by linear theory \citep{EJ10,Meng-2}, we consider gaps of the form $\displaystyle b_\mathcal{C}(t)=\left(1-\frac{7}{2}\tau \mathcal{C} t\right)^{-{2}/{7}}$, where $\tau$ is the surface tension and $\mathcal{C}$ is a coefficient. Here, the gap width increases much more rapidly than $b(t)\sim t$ or even $b(t)\sim e^t$ and actually tends to infinity at a finite time, dictated by the surface tension and the coefficient $\mathcal{C}$. If $\mathcal{C}=2(k^2-1)$, with $k$ being the perturbation wavenumber, a $k$-mode perturbation evolves self-similarly (e.g., perturbation size relative to underlying shrinking circle is invariant). Alternatively, if $\mathcal{C}=2(3k^2-1)$ then mode $k$ is the fastest growing mode. In both cases, this leads to non-circular vanishing limiting shapes at least at the level of linear theory. Predictions of the nonlinear dynamics and the emergent interface patterns in the nonlinear regime are very difficult because of nonlocality, strong nonlinearity and rapid evolution. Consequently, until this work the effect of nonlinearity in this special, shrinking regime has not yet been explored. 

Here, we use a recently-developed spectrally-accurate boundary integral method with space-time rescaling  \citep{Meng-2} that enables the accurate simulation of nonlinear interface dynamics in the fast shrinking regime for the first time. In particular, the interface is mapped back to its initial size and time is rescaled such that the speed of the interface in the new frame is prescribed and is slower than the dynamics in the original frame. Thus, a fixed time step in the rescaled frame corresponds to a time step in the original frame that is adaptively and rapidly decreased in time. Together, these features enable us to accurately simulate the fully nonlinear dynamics of the interface at extraordinarily small interface sizes and obtain nonlinear, limiting shapes that have not been previously reported. Other forms of time-space rescaling are also used to study the behavior of bubble extinction in a Hele-Shaw flow \citep{Dallaston13,Dallaston16}.

To validate our methods, we compare with experiments from \citep{JDA}, where the linear gap $b(t)\sim t$ is used. Our numerical predictions for the number of fingers are in quantitative agreement with the experiments when the confinement number $C_0$ is large. This is in contrast with the results of linear theory, which predict fewer fingers and provide a better match to the experimental results at smaller $C_0$ \citep{JDA,EJ13-2}. This illustrates the importance of accounting for nonlinear interactions. 

In the special shrinking regime, where $b(t)=b_\mathcal{C}$ is used, our numerical simulations reveal the existence of strikingly thin, $k$-fold dominant, limiting interfacial shapes. This suggests that these interfaces do not shrink as circles but rather as novel one-dimensional,  web-like networks. Although we do not find evidence of self-similar dynamics in the nonlinear regime, we do find that there is mode selection. We construct a morphology diagram for pattern selection that relates the dominant mode $k$ of the vanishing, limiting interface and the control parameter $\mathcal{C}$.

This paper is organized as follows. In Section \ref{sec:governing equations}, we present the governing equations, the linear stability analysis and the boundary method used to simulate the nonlinear system. In Section \ref{sec:numres}, we present numerical results and in Section \ref{con}, we give conclusions and discuss future work. In the Supplementary Material (SM), we show that if we slightly modify the dynamics of the gap width in time (by making it slightly larger than reported in the experiment), then quantitative agreement between the simulations and experiments can be achieved at small confinement numbers and surface tensions when the gap width increases linearly in time.  Also in the SM, we present additional numerical results in the special-shrinking regime. Finally, in the SM,  we provide a comparison between the nonlinear simulations and new experiments for shrinking interfaces using a gap width that increases nonlinearly in time.

\section{Governing Equations}\label{sec:governing equations}
\begin{figure}
\centering
\includegraphics[scale=0.5]{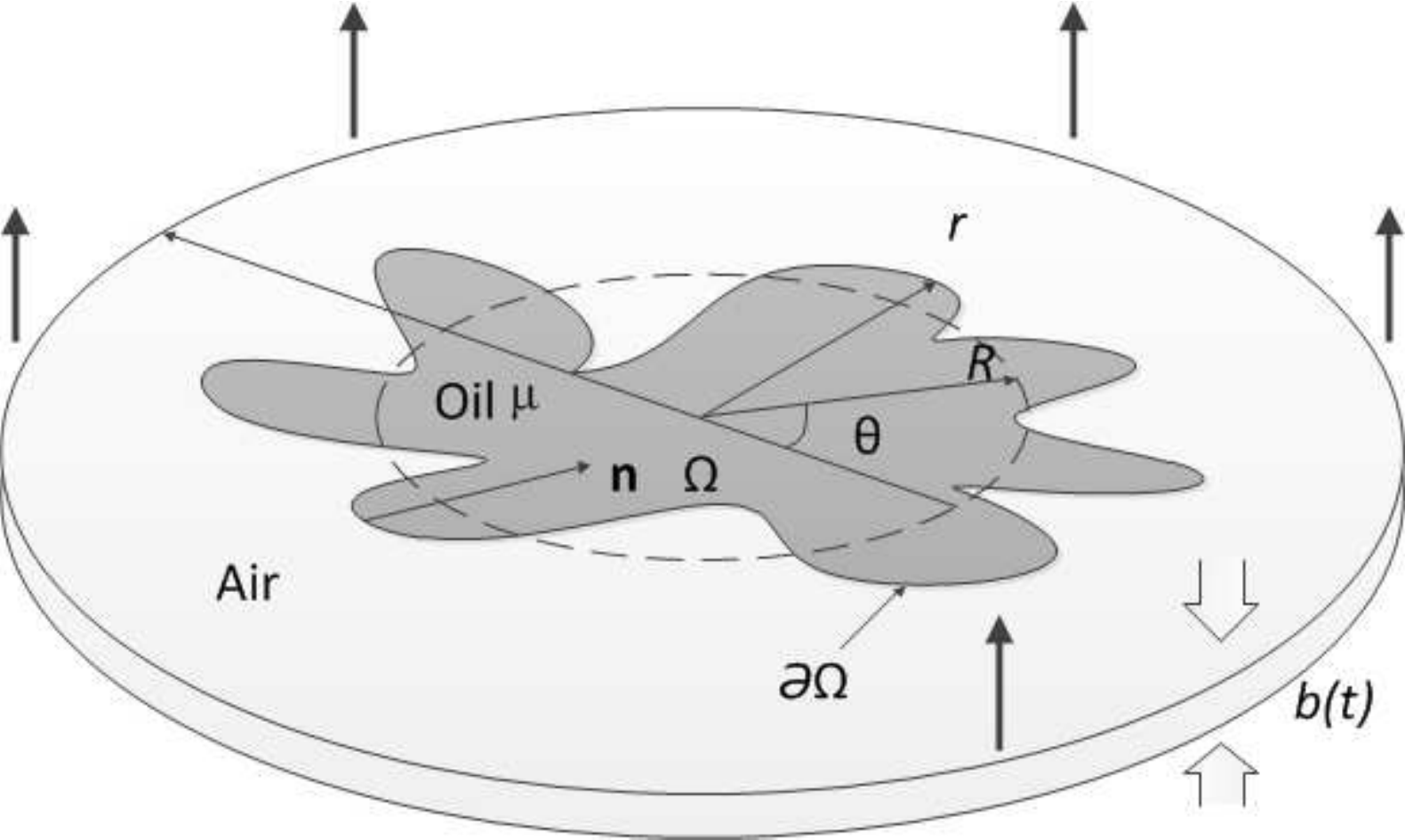}
\caption{Schematic for the lifting plate Hele-Shaw problem. The region $\Omega(t)$ contains oil with viscosity $\mu$. The region exterior to $\Omega(t)$ contains air. $b(t)$ is the time dependent gap width. The normal $\textbf{n}$ to the interface  $\partial\Omega(t)$ points into $\Omega(t)$. $R$ denotes the equivalent radius (radius of a circle with the same enclosed area).}\label{setup}
\end{figure}

We consider a radial Hele-Shaw cell with a time-dependent gap $\tilde b(\tilde t)$, see Fig. \ref{setup} for a schematic. The upper plate is lifted uniformly in space  while the lower plate is stationary. The domain $\tilde\Omega(\tilde t)$ denotes the region containing the viscous fluid (e.g., oil) and $\partial\tilde\Omega(\tilde t)$ denotes its interface. A less viscous fluid (e.g., air) is contained in the exterior of $\tilde\Omega(\tilde t)$. Here, the tildes denote dimensional variables. The nondimensional system, which we analyze and solve numerically, is given below.

\subsection{Governing equations} Following previous studies \citep{MD,ADM,Sinha08,JDA,EJ13-2}, we assume that the motion of the fluid is governed by Darcy's law, which is a 2 dimensional approximation of the Navier-Stokes system obtained by averaging the equations over the narrow gap between the plates. The equations are:
 \begin{equation}
 \tilde \mathbf{u}=-\frac{\tilde b^2(\tilde{t})}{12\mu}\tilde\nabla \tilde P \quad \textnormal{in} \quad\tilde\Omega(\tilde t),\label{vel}
 \end{equation}
 where $\tilde \textbf{u}$ is the velocity,  $\tilde P$ is the pressure, and $\mu$ is the viscosity of the fluid. 
 
From volume conservation, we obtain the gap-averaged incompressibility condition as  
\begin{equation}
\label{dvel}\tilde\nabla\cdot\tilde \mathbf{u}=-\frac{\dot{\tilde b}(\tilde{t})}{\tilde b(\tilde{t})}\quad \textnormal{in} \quad\tilde\Omega(\tilde t),
\end{equation}
where $\displaystyle \dot{\tilde b}(\tilde t)=\frac{d\tilde b(\tilde t)}{d\tilde t}$ is the lifting speed of the upper plate.

The pressure jump $[\tilde P]$ across the interface is given by Laplace-Young condition, which is the product of surface tension $\sigma$ and the curvature of the interface $\tilde\kappa$, 
\begin{equation}
\label{knc}[\tilde P]=\tilde P|_{\partial\Omega^+}-\tilde P|_{\partial\Omega^-}=\sigma\tilde\kappa \quad \textnormal{on}\quad \partial\tilde\Omega(\tilde t),
\end{equation}
where the $+/-$ superscripts denote the limit from the exterior and interior of $\Omega$, respectively.
 The scaled normal derivative  $\frac{1}{\mu}\frac{\partial\tilde P}{\partial\mathbf{n}}$ is continuous across $\partial\Omega$ and the normal velocity of the interface is thus
 \begin{equation}
\tilde V=-\frac{\tilde b^2(\tilde{t})}{12\mu}\frac{\partial \tilde P}{\partial \mathbf{n}}\quad \textnormal{on}\quad \partial\tilde\Omega(\tilde t)\label{norvel},
\end{equation}
where $\textbf{n}$ is the unit normal vector pointing into $\tilde\Omega(\tilde t)$.

We nondimensionalize the system using a characteristic length $L_0$, time  $\displaystyle T=\frac{\tilde b_0}{\dot{\tilde b}_0}$, and pressure  $\displaystyle P_0=\frac{12\mu L_0^2}{T\tilde b_0^2}$, where $L_0$ is a characteristic size of the initial fluid domain $\Omega(0)$ and $\tilde b_0$ and $\dot{\tilde b}_0$ are the initial values of $\tilde b$ and $\dot{\tilde b}$. Further, define the nondimensional modified pressure as $\displaystyle {P}=\tilde P/P_0- \frac{\dot{b}(t)}{4b^3(t)}|\textbf{x}|^2$, where $b(t)=\tilde b(\tilde t)/b_0$, $t=\tilde t/T$, $\displaystyle{\dot b(t)=\frac{db}{dt}}$ and $\mathbf{x}=\tilde\mathbf{x}/L_0$. Then, the nondimensional version of Eqs. (\ref{vel})- (\ref{norvel}) becomes:
\begin{eqnarray}
\nabla^2 {P}&=&0\quad \textnormal{in $\Omega$}\label{pform2},\\
\label{knc2}[{P}]&=&\tau \kappa-\frac{\dot{b}(t)}{4b^3(t)}|\textbf{x}|^2\quad \textnormal{on}\quad \partial\Omega,\\
V&=&-b^2(t)\frac{\partial {P}}{\partial \textbf{n}}+\frac{\dot{b}(t)}{2b(t)}\textbf{x}\cdot\textbf{n}\quad\textnormal{on}\quad \partial\Omega\label{norvel2},
\end{eqnarray}
where $\displaystyle \tau=\frac{\sigma \tilde b_0^3}{12\mu\dot{\tilde b}_0L_0^3}$ is a nondimensional surface tension. Taking $L_0$ to be the equivalent radius of $\Omega(0)$, e.g., the radius of a circle with the same enclosed area, the nondimensional volume of the fluid is $\pi$ since the initial nondimensional gap is $b(0)=1$.

\subsection{Linear theory}
\label{linear theory}
In this section, we briefly review the linear stability analysis in \citep{MS,Meng-2}. We consider the interface to be a slightly perturbed circle, 
\[  r(\alpha,t)=R(t)+\epsilon \delta(t)\cos(k\alpha),\]
where $\epsilon$ $\ll 1$, the perturbation mode $k\geq 2$ is an integer, $\alpha\in[0,2\pi]$ is the polar angle, and $\delta(t)$ is the  amplitude of the perturbation. The shape factor $\displaystyle \frac{\delta}{R}(t)=\frac{\delta(t)}{R(t)}$ can be used to characterize the size of the perturbation relative to the underlying circle \citep{WR}. Then, it can be shown that the shape factor evolves according to
\begin{equation}
\left(\frac{\delta}{R}\right)^{-1}\frac{d}{dt}\left(\frac{\delta}{R}\right)=\frac{\dot{b}k}{2b}- \tau\frac{b^2(k^3-k)}{R^3}.\label{growth}
\end{equation}
Using the relationship $\displaystyle R(t)=\frac{1}{\sqrt{b(t)}}$, which arises from volume conservation at the level of linear theory, the shape factor grows only when 
\begin{equation}
\dot{b}(t)>2\tau (k^2-1)b^{9/2}(t).
\end{equation}
This indicates that the band of modes satisfying $\displaystyle{|k|<\sqrt{{\dot{b}b^{-9/2}}/{2\tau}+1}}$ are unstable and the interface can develop fingering patterns. Note that in the special cases that $\dot{b}=1$, or even $\dot{b}=b$, all modes $|k|>1$ become stable in the long time limit. This occurs because ${\dot{b}b^{-9/2}}\to 0$ as $t\to\infty$. On the other hand, if we consider much faster rates of gap increase, 
\begin{equation}
 b(t)=\left(1-\frac{7}{2}\tau \mathcal{C}t\right)^{-\frac{2}{7}}\label{ssgap},
\end{equation}
where $\mathcal{C}$ is a constant, then the band of unstable modes is fixed in time and depends on $\mathcal{C}$: $\displaystyle{|k|<\sqrt{\mathcal{C}/2+1}}$. Note this gap tends to infinity at the finite time $T_\mathcal{C}=2/(7\tau\mathcal{C})$.

More generally, the fastest growing mode, $k_{max}$, at time $t$ is
\begin{equation}
k_{max}=\sqrt{\frac{\dot{b}}{6\tau b^{9/2}}+\frac{1}{3}},
\end{equation}
and the mode, $k^*$, at which the perturbation is largest at time $t$ is
\begin{equation}
k^*=\sqrt{\frac{\ln{b(t)}}{6\tau \int_0^t b^{7/2}(s)ds}+\frac{1}{3}}.
\end{equation}
Note that because the growth rate of the shape factor depends on time, $k^*$ need not be equal to  $k_{max}$. Further, since
$k_{max}$ and $k^*$ are integers, these have been used to estimate the number of fingers \citep{JDA,EJ13-2}.

When the special gap dynamics in Eq. (\ref{ssgap}) is used, the fastest growing mode $k_{\max}$ can be prescribed by taking $\mathcal{C}=2(3k_{max}^2-1)$. Under this gap dynamics, the mode $k_{max}$ will remain the fastest growing at all times so that $k_{max}=k^*$ and the shape factor is
\begin{equation}
\frac{\delta(t)}{R(t)}= \frac{\delta(0)}{R(0)}~b(t)^{\frac{k_{max}^3}{3k^2_{max}-1}}=\frac{\delta(0)}{R(0)}~R(t)^{-\frac{2k_{max}^3}{3k^2_{max}-1}},
\label{perturbation magnitude}
\end{equation}
which diverges as $t\to T_C$. Alternatively, if $\mathcal{C}=2(\bar k^2-1)$, then the shape factor of mode $\bar k$ does not change in time. In other words, under linear theory, a $\bar k$-mode perturbation of the interface would evolve self-similarly. Both of these conditions suggest that in the special gap regime, there may be mode selection and that perturbations may persist (and even grow) as the fluid domain and interface shrinks.

\subsection{Numerical method} Because of the strong nonlinearity and nonlocality of the lifting plate equations, numerical methods are needed to characterize the nonlinear dynamics. However, the simulations are very challenging because of severe time and space step restrictions introduced by the rapid evolution and shrinking of the interface. To overcome these numerical issues, we have developed a rescaled boundary integral scheme \citep{Meng-2}. The method is briefly described here. The idea is to map the original time and space $({\mathbf x},t)$ into new coordinates $(\bar {\mathbf x},\bar t)$ such that the interface can evolve at an arbitrary speed in the new rescaled frame (see also \citep{ShuwangJCP,Zhao2015,Zhao17}). Introduce a new frame $(\bar{\mathbf{x}},\bar t)$ such that
  \begin{equation}
  \textbf{x}=\bar R(\bar t)\bar{\textbf{x}}(\bar t,\alpha),
  \label{xslp}
  \end{equation}
  \begin{equation}
  \bar t=\int_{0}^t \frac{1}{\rho(t')}dt',
  \label{tslp}
  \end{equation}
  where the space scaling  $\bar R(\bar t)$ captures the size of the interface, $\bar{\textbf{x}}$ is the position vector of the scaled interface, and $\alpha$ parameterizes the interface. The space scaling function maps the interface to its original size and the time scaling function $\rho(t)=\bar{\rho}(\bar{t})$ maps the original time $t$ to the new time $\bar t$, which has to be positive and continuous.  If $\rho(t)<1$, then the evolution in the rescaled frame is slower than that in the original frame. Using mass conversation and requiring the area enclosed by the interface to be fixed in the new frame, we obtain the normal velocity in the new frame
  \[\bar V= \frac{\bar{\rho}}{\bar R}{V}(t).\]
  Representing the pressure $P$ as a double layer potential, with dipole density $\bar \gamma$, Eqs. (\ref{pform2})-(\ref{knc2}) can be written as the following Fredholm integral equation of the second kind
    \begin{equation}
     \bar \gamma(\bar{\textbf{x}})+\frac{1}{\pi}\int_{\partial\bar{\Omega}(\bar t)}\bar \gamma(\bar{\textbf{x}}')\left[\frac{\partial \ln|\bar{\textbf{x}}-\bar{\textbf{x}}'|}{\partial \textbf{n}(\bar{\textbf{x}}')}+\bar R(\bar t)\right]d\bar{s}(\bar{\textbf{x}}')=2\tau \bar{\kappa}-\frac{\dot{b}(t(\bar t))}{2b^3(t(\bar t))}\bar R^{3}|\bar{\textbf{x}}|^2.
     \label{mub}
     \end{equation}
    Once $\bar \gamma$ is determined, the normal velocity in the new frame $\bar V$ can be computed as
     \begin{equation}
     \bar V(\bar{\textbf{x}})=-\frac{b^2(t(\bar t))\bar{\rho}}{2\pi \bar R^{3}}\int_{\partial\bar{\Omega}}\bar \gamma_{\bar{s}}\frac{(\bar{\textbf{x}}'-\bar{\textbf{x}})^{\perp}\cdot{\textbf{n}}(\bar{s})}{|\bar{\textbf{x}}'-\bar{\textbf{x}}|^2}d\bar{s}',
     \label{Vb}
     \end{equation}
     where $\bar{\textbf{x}}^{\perp}=(\bar x_2,-\bar x_1)$ and the interface evolution in the scaled frame is
     \begin{equation}
     \frac{d\bar{\textbf{x}}(\bar t,\alpha)}{d\bar t}\cdot \textbf{n}=\bar V(\bar t,\alpha).
     \label{evolvesc}
     \end{equation}
 Here, we take the time scaling $\rho(t)\propto 1/\dot{b}$ so that in the rescaled frame, the gap increases linearly in time, e.g. $b(\bar t)=1+c\bar t$, where $c$ is a constant. Using a spectrally accurate discretization method in space, a 2nd order accurate semi-implicit scheme in time and the generalized minimum residual scheme (GMRES; \citep{GMRES}) to solve the 2nd kind Fredholm integral equation, the method enables us to accurately compute the dynamics to far longer times than could previously be accomplished. Further details of the numerical method, including convergence studies, can be found in  \citep{Meng-2}.

\section{Numerical Results}\label{sec:numres}
\subsection{Comparison between linear theory, nonlinear simulations and experiments}
\begin{figure}
\includegraphics[scale=0.3]{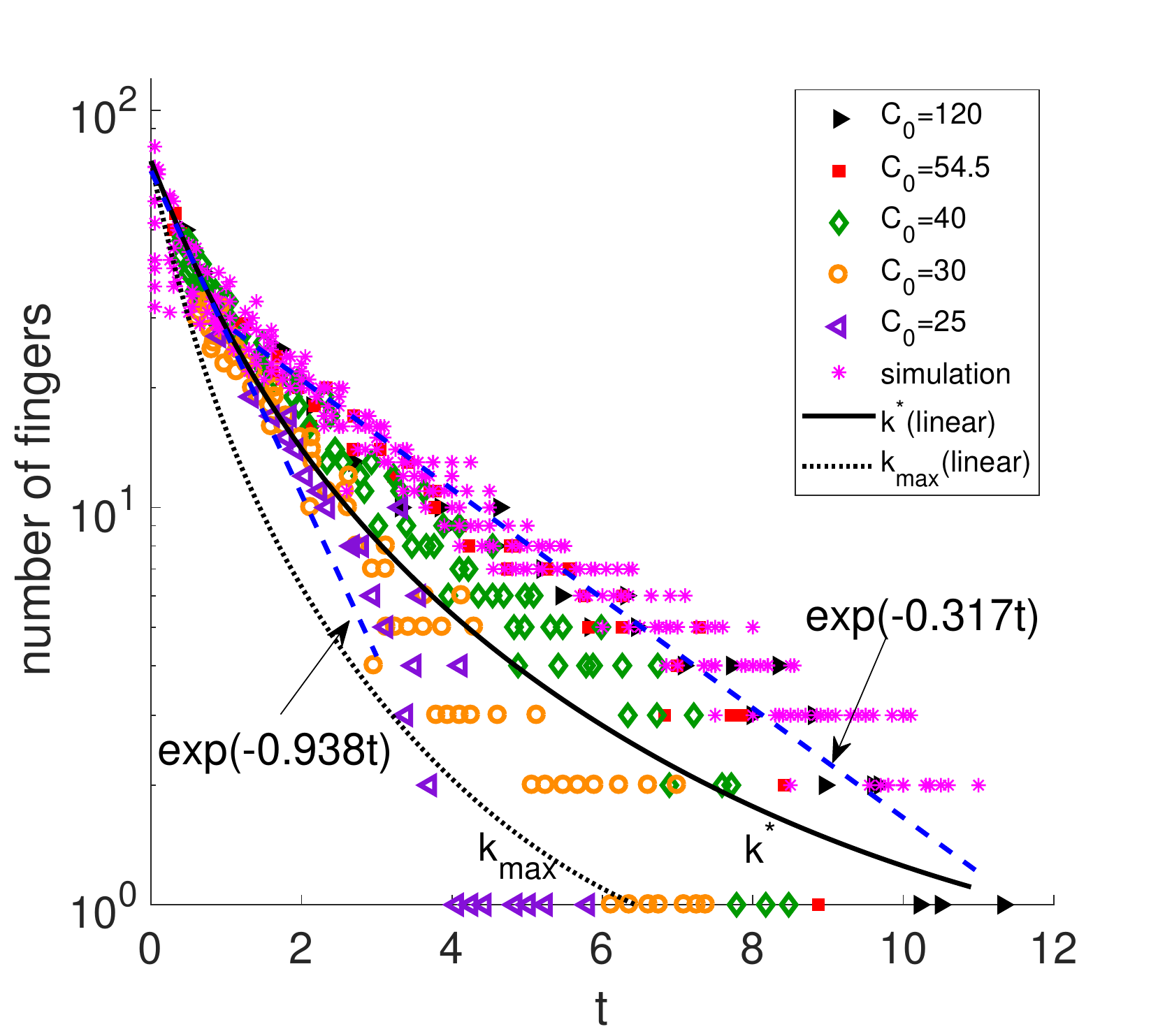}[a1]
\includegraphics[scale=0.3]{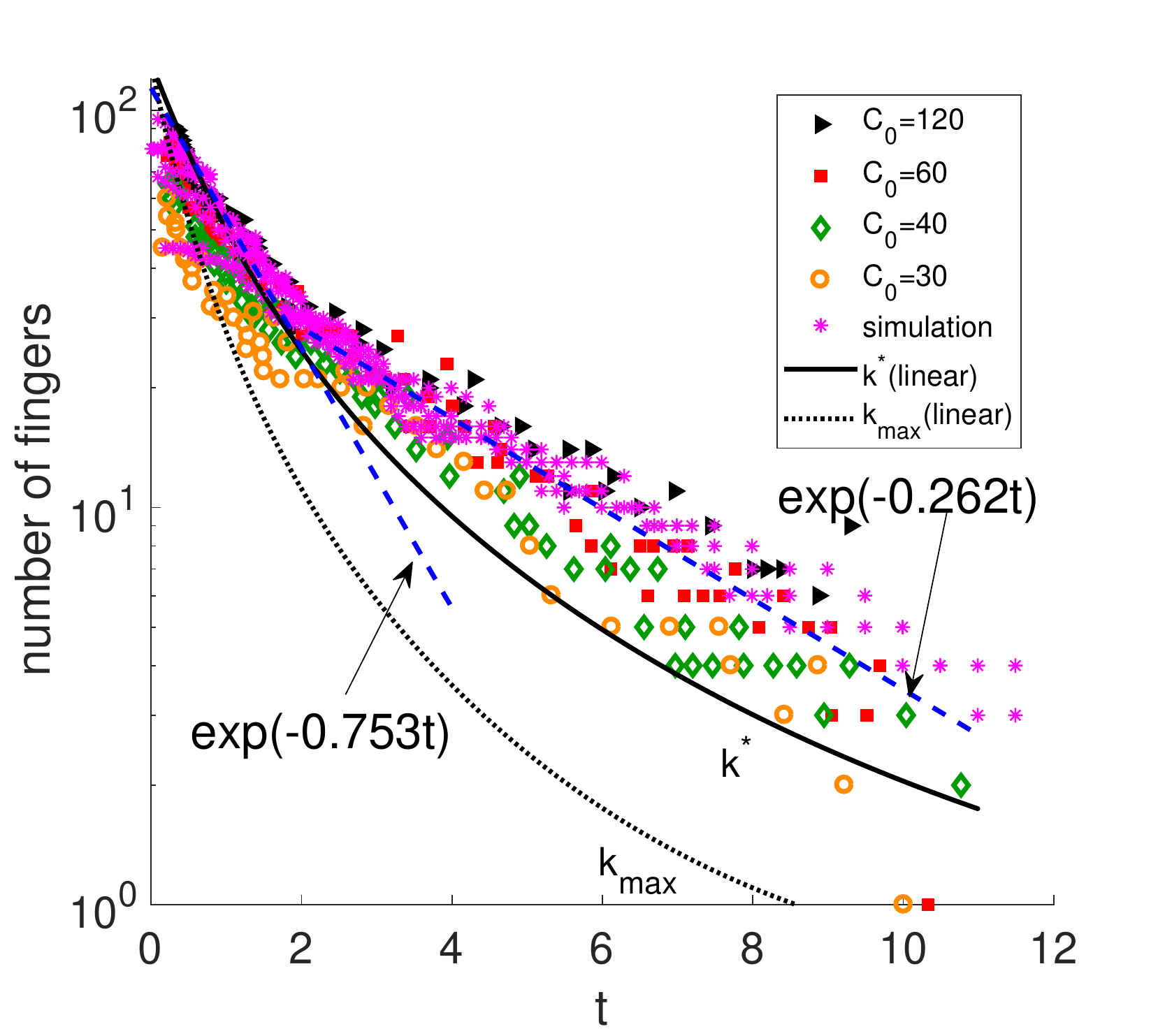}[a2]\\
\includegraphics[scale=0.25]{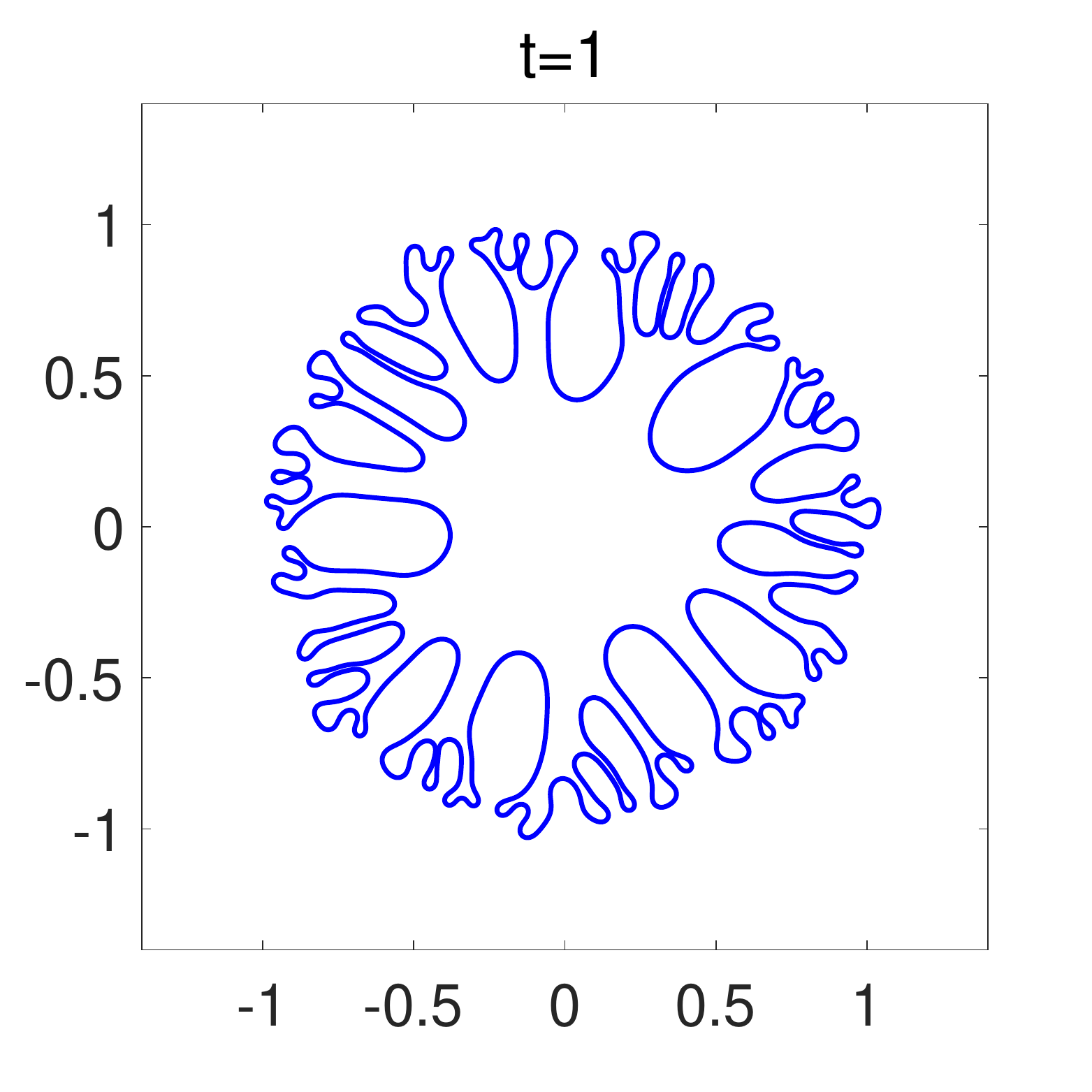} 
\includegraphics[scale=0.25]{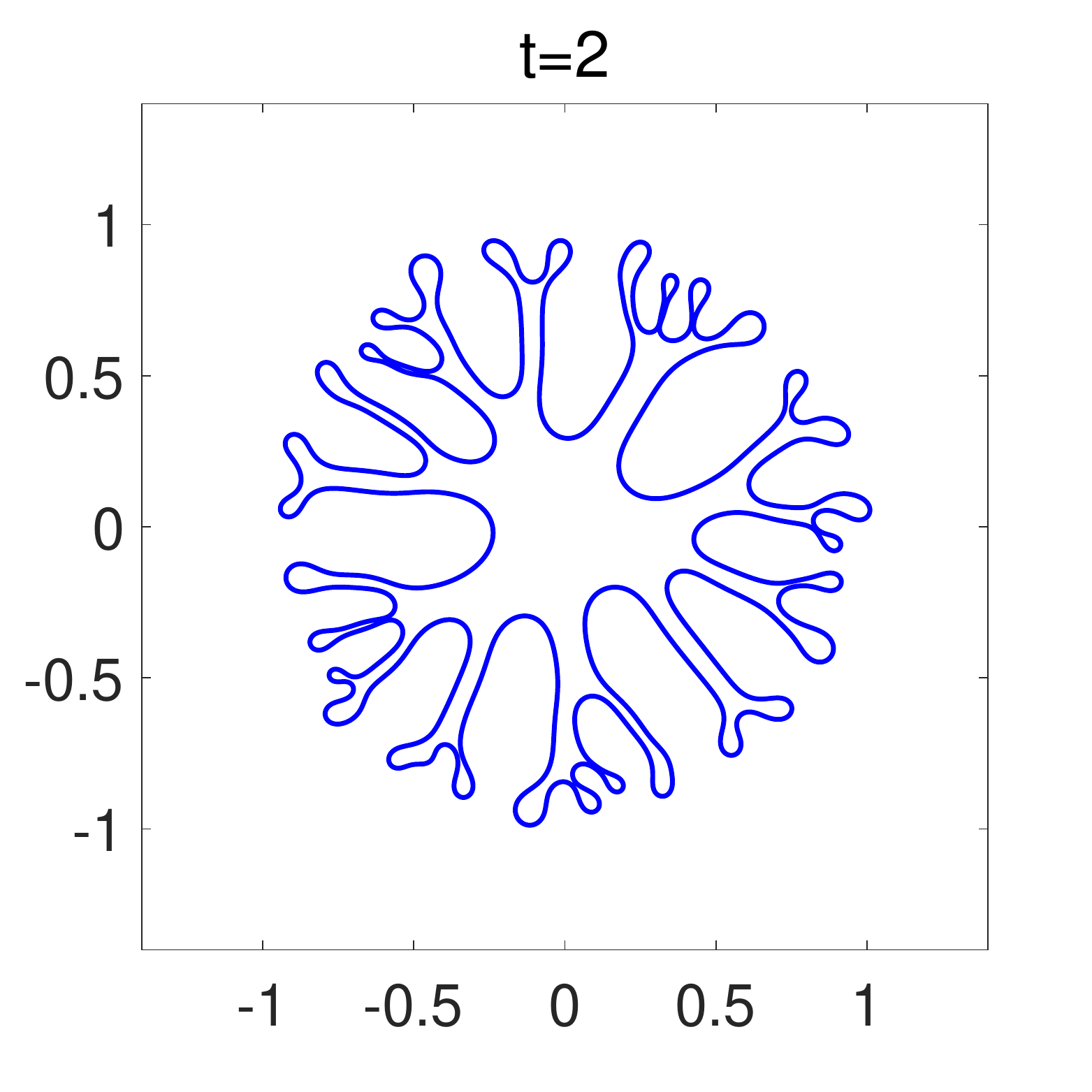}
\includegraphics[scale=0.25]{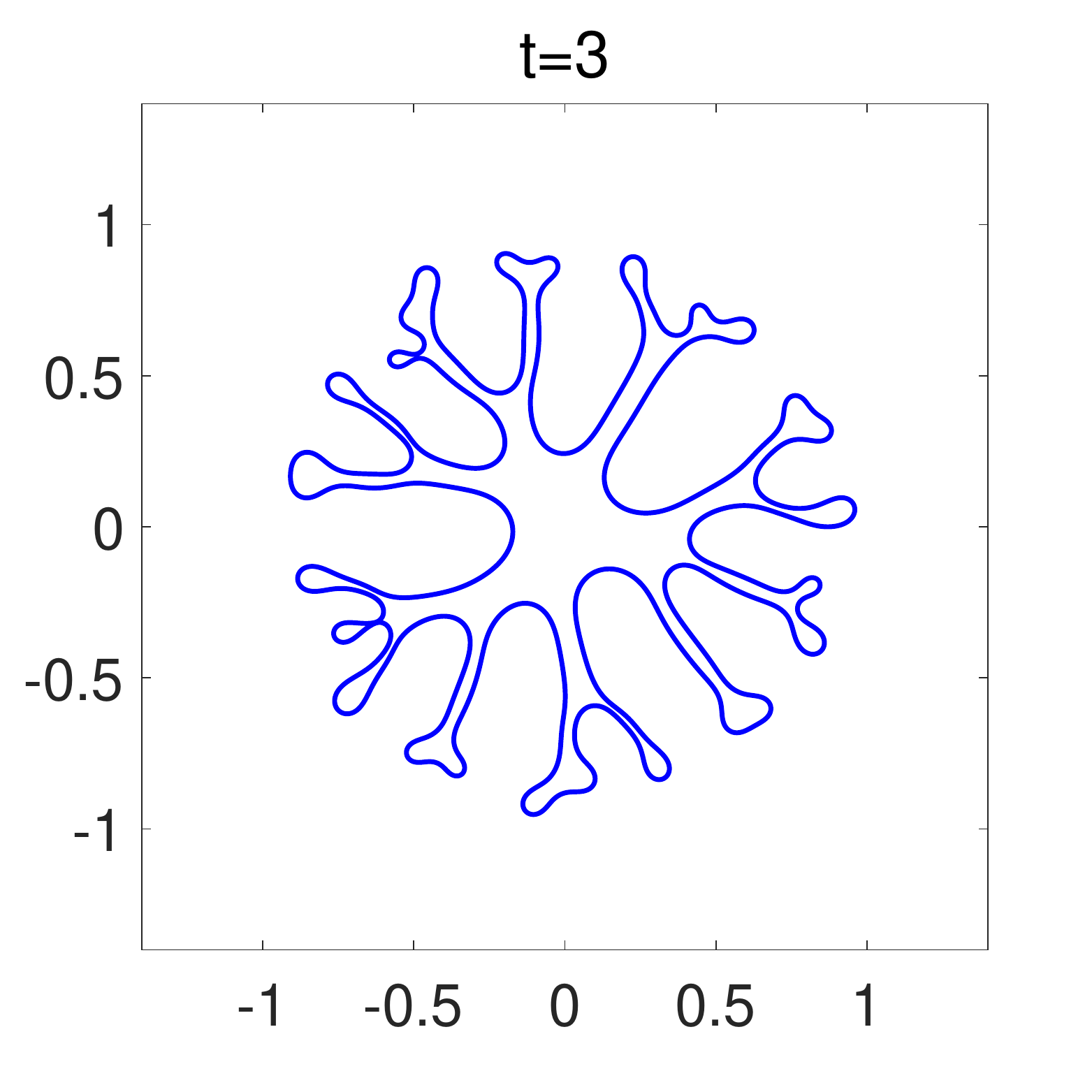}\\
.\hspace{0.15in}
\includegraphics[scale=0.45]{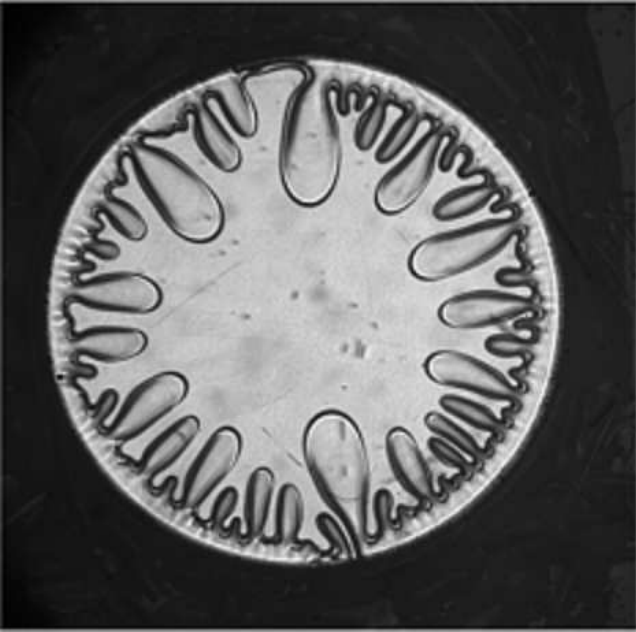}
\hspace{0.38in}
\includegraphics[scale=0.45]{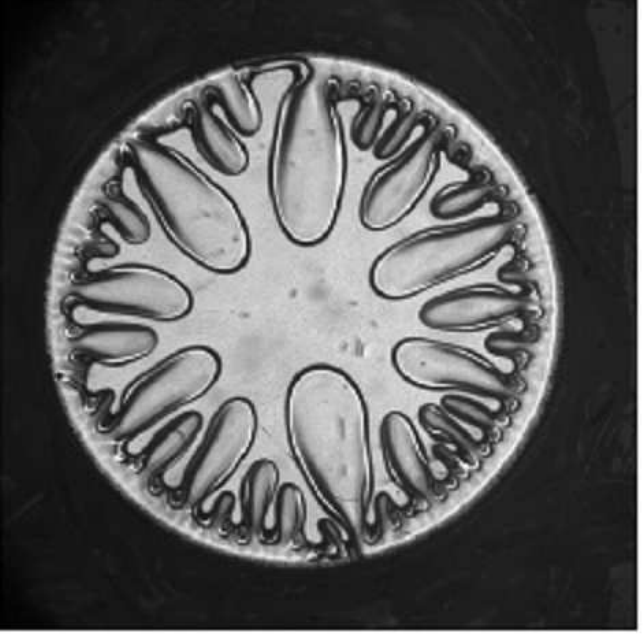}
\hspace{0.38in}
\includegraphics[scale=0.45]{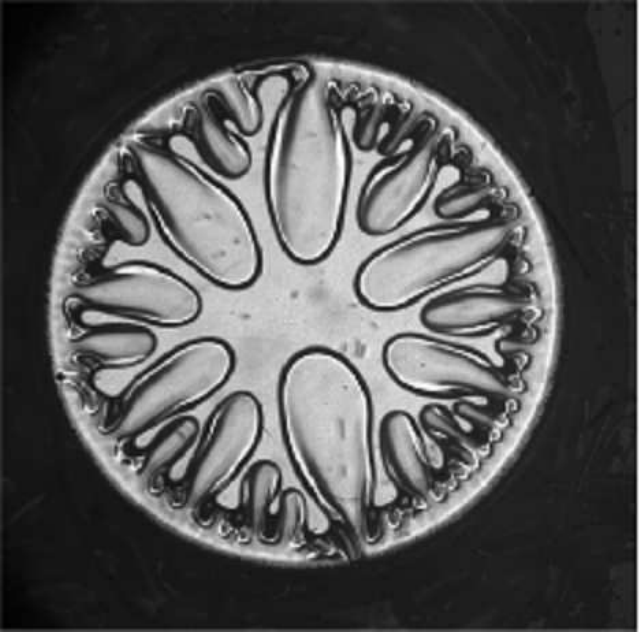}[b]
\caption{Comparison between linear theory, numerical simulations and experiments from \citep{JDA} where the gap width is increased linearly in time. The numbers of fingers over time are shown in [a1] and [a2] predicted by linear theory ($k_{max}$ dotted curves; $k^*$ solid curves), numerical simulations (magenta stars) and experiments with different confinement numbers $C_0$ , ranging from $30$ to $120$ as labeled. The dashed lines are exponential fits to the numerical simulations. In [a1], the nondimensional surface tension is $\tau=3\times 10^{-5}$ while in [a2] $\tau=9.6\times 10^{-6}$. In [b], the simulated interfacial morphologies (top) and those from experiments (bottom), with $\tau=9.6\times 10^{-6}$, are shown at the dimensionless times $t=1,2,$ and $3$. There is an excellent agreement between the numerical simulations and experiments. Experimental results reprinted with permission.} \label{lncom}
\end{figure}

We first compare the experimental results obtained in \citep{JDA} with the results of linear theory \citep{EJ13-2,Meng-2} and nonlinear simulations. In the experiments, a drop of viscous fluid, surrounded by air, was placed in a Hele-Shaw cell and the upper plate was pulled upward at a constant rate of increase, e.g., the nondimensional gap width $b(t)=1+t$, where $t$ is the nondimensional time. High resolution images were used to precisely determine the number of fingers over time. By varying the initial drop radius, the initial plate spacing, the rate of gap increase, and the viscosity, they systematically investigated how the number of fingers depends on the nondimensional surface tension and confinement number (see \citep{JDA} for details). 

The results from two sets of experiments are shown in Fig. \ref{lncom}. In Figs. \ref{lncom}[a1] and [a2] the number of fingers is shown as a function of the nondimensional time for different confinement numbers $C_0$, as labeled. The nondimensional surface tensions are $\tau=3\times 10^{-5}$ [a1] and $\tau=9.6\times 10^{-6}$  [a2].   The solid and dotted curves denote the number of fingers estimated from $k^*$ and $k_{max}$, respectively, obtained from linear theory. The stars denote the results from nonlinear simulations. The number of fingers in the simulations are calculated in exactly the same way as in \citep{JDA} (e.g., number of air fingers penetrating the viscous fluid; see Supplementary Material, Fig. S1[a]). Characteristic drop morphologies using $\tau=9.6\times 10^{-6}$ are shown (at the nondimensional times $t=1$, $2$, $3$) in Fig. \ref{lncom}[b] from simulations (top row) and  experiments (bottom row) with $C_0=60$. The agreement between the experimental and simulation morphologies is striking.

To produce the nonlinear simulation results, we took the initial shape of the drop to be a slightly perturbed circle,
\begin{equation}
r(\alpha,0)=1+\epsilon\Sigma_{k=k_{min}}^{k_N}e^{-\beta k}(a_k\cos(k\alpha)+b_k\sin(k\alpha)),\label{initials}
\end{equation} 
where each $k$, $a_k$ and $b_k$ are chosen randomly from a uniform distribution in the interval $(-1,1)$, and $\alpha$ parametrizes the interface. Varying $\epsilon$, $\beta$, and $k_N$ allows us to vary the amplitude of the perturbation and its modal content. Here, we took $\epsilon=0.05$, $\beta=0.2$, $k_{min}=2$, and $k_N$ is varied between 40 and 100.  Since $b(t)\sim t$, we do not need to rescale time to slow down the evolution and take $\bar{t}=t$. We use N= 4096 mesh points along the interface, the time step $\Delta\bar{t}=1\times 10^{-4}$, and the surface tension $\tau=9.6\times 10^{-6}$ and $3\times 10^{-5}$. 

As can be seen in Fig. \ref{lncom}, the number of fingers decreases in time and the drop eventually shrinks as a circle. The experimental results depend on both the confinement number $C_0$ and the nondimensional surface tension $\tau$. At larger surface tensions [a1], the effect of $C_0$ is more pronounced at later times while at small surface tensions [a2], $C_0$ more significantly influences the early stages of the evolution. Generally, the larger $C_0$ is the more fingers that are observed. 

The number of fingers predicted using $k_{max}$ significantly under-predicts the experimental (and simulation) results while the $k^*$ predictions agree better with the experimental data at small $C_0$. The numerical simulations predict more fingers than either linear estimate and agree best with experimental data at large $C_0$. This illustrates the importance of accounting for nonlinear interactions. Indeed, in \citep{Meng-2}, it was shown that linear theory under-predicts the amplitudes of growing modes in the lifting plate problem.

For both surface tensions, the numerical simulations, and experimental data with large $C_0$, predict a biphasic, exponential decay of the number of fingers over time: rapid decay at early times (e.g., $\approx e^{-0.75t}$ in Fig. \ref{lncom}[a2]) and slower decay over late times (e.g., $\approx e^{-0.26t}$ in Fig. \ref{lncom}[a2]). The experimental data with small $C_0$ seem to predict a single rate of exponential decay (see Supplementary Material, Fig. S2), consistent with the experimental results obtained in \citep{MD}.

It is still not yet well understood why the confinement number influences the number of fingers. Note that the thickness $h$ of the wetting layer on the plates scales as $h \sim O(Ca^{2/3})$ where $Ca=\mu \tilde V/\sigma$ is the Capillary number \citep{Park84,Park,Jackson15}. Estimating $\tilde V\sim (\dot b_0/b_0) R_0= \dot b_0 C_0$ from Eq. (\ref{dvel}) and the definition of $C_0$, we obtain $Ca\sim  \dot b_0 C_0$. In the experiments in \citep{JDA}, the confinement number and lifting rate are changed such that the nondimensional surface tension $\tau=\sigma\tilde b_0^3/(12\mu\dot{\tilde b}_0 R_0^3)$ is fixed. This implies that $\dot b_0\sim C_0^{-3}$ and therefore $Ca\sim  C_0^{-2}$. Thus, the wetting layer thickness $h\sim C_0^{-4/3}$ increases as the confinement number decreases. As a simple test of this, we modified the lifting speed to reflect the fact that more fluid may be left on the plates when the confinement number is decreased, e.g. the lifting speed of the upper plate is increased to reflect the more rapid loss of fluid to the wetting layer. The results are shown in Fig. S1[b] in Supplementary Material and indicate that with this modification, the nonlinear simulations are better able to fit experiments with smaller confinement numbers.

When wetting effects, as well as viscous stresses, are fully included at the level of linear theory, the range of agreement between linear predictions, using the maximum perturbation wavenumber $k^*$, and experiments and be extended to somewhat larger $C_0$, but there is still disagreement between the two at large $C_0$ with linear theory under-predicting the number of fingers, see \citep{EJ13-2}. It is very likely that nonlinear effects also play a key role and the development of a numerical method to accurately account for nonlinearity, viscous stresses, wetting effects and fluid motion in 3D is a subject for future work.

\subsection{Limiting dynamics in the special gap regime}
We next study the interfacial dynamics in the special gap regime using $\displaystyle b(t)=\left(1-\frac{7}{2}\tau \mathcal{C}t\right)^{-\frac{2}{7}}$, which diverges at the finite time $T_C=2/(7\tau\mathcal{C})$. Recall that in this regime, according to linear theory (e.g., see Sec. \ref{linear theory}), the fastest growing mode $k_{max}$ and $\mathcal{C}$ are related by $k_{max}=\sqrt{(1+\mathcal{C}/2)/3}$. Further, $k_{max}=k^*$, the maximum perturbation mode, and the shape factor $\delta/R\propto R^{-2k_{max}^3/(3*k_{max}^2-1)}$, which monotonically increases over time, and diverges at $t=T_C$. Because of the very rapid dynamics, an infinitesmally small time step would be required as $t \to T_C$ if time were not rescaled. This would make it virtually impossible to simulate the drop dynamics in the original frame. Consequently, we rescale time to slow down the evolution and take $\rho(t)=c/\dot b(t)$ so that in the new time scale $\bar t$ we have $b(\bar t)=1+ct$. This makes it possible to study the limiting dynamics in the special gap regime for the first time. 

\subsubsection{Examples of drop dynamics}

\begin{figure}
\includegraphics[width=0.4\textwidth]{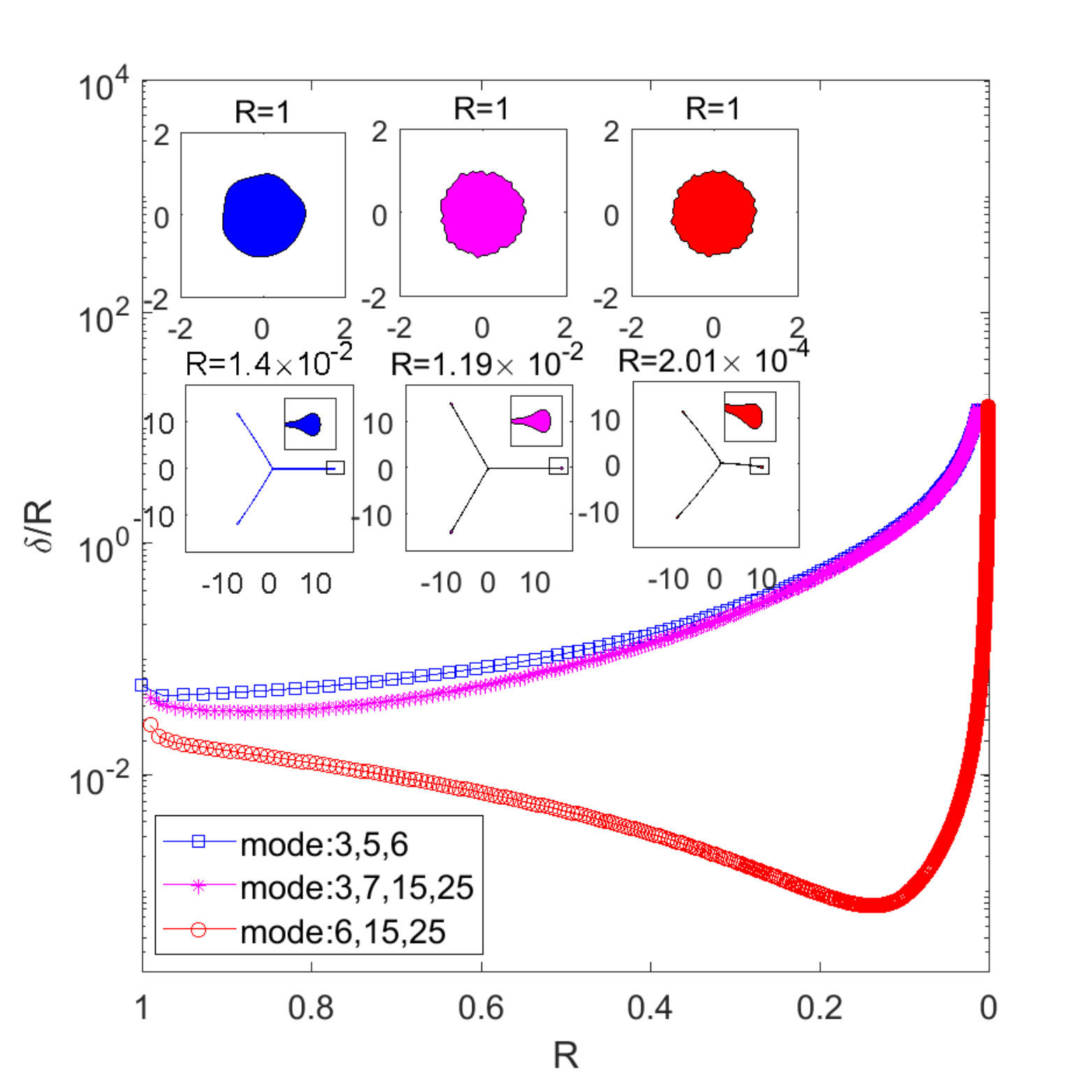}[a]
\includegraphics[width=0.4\textwidth]{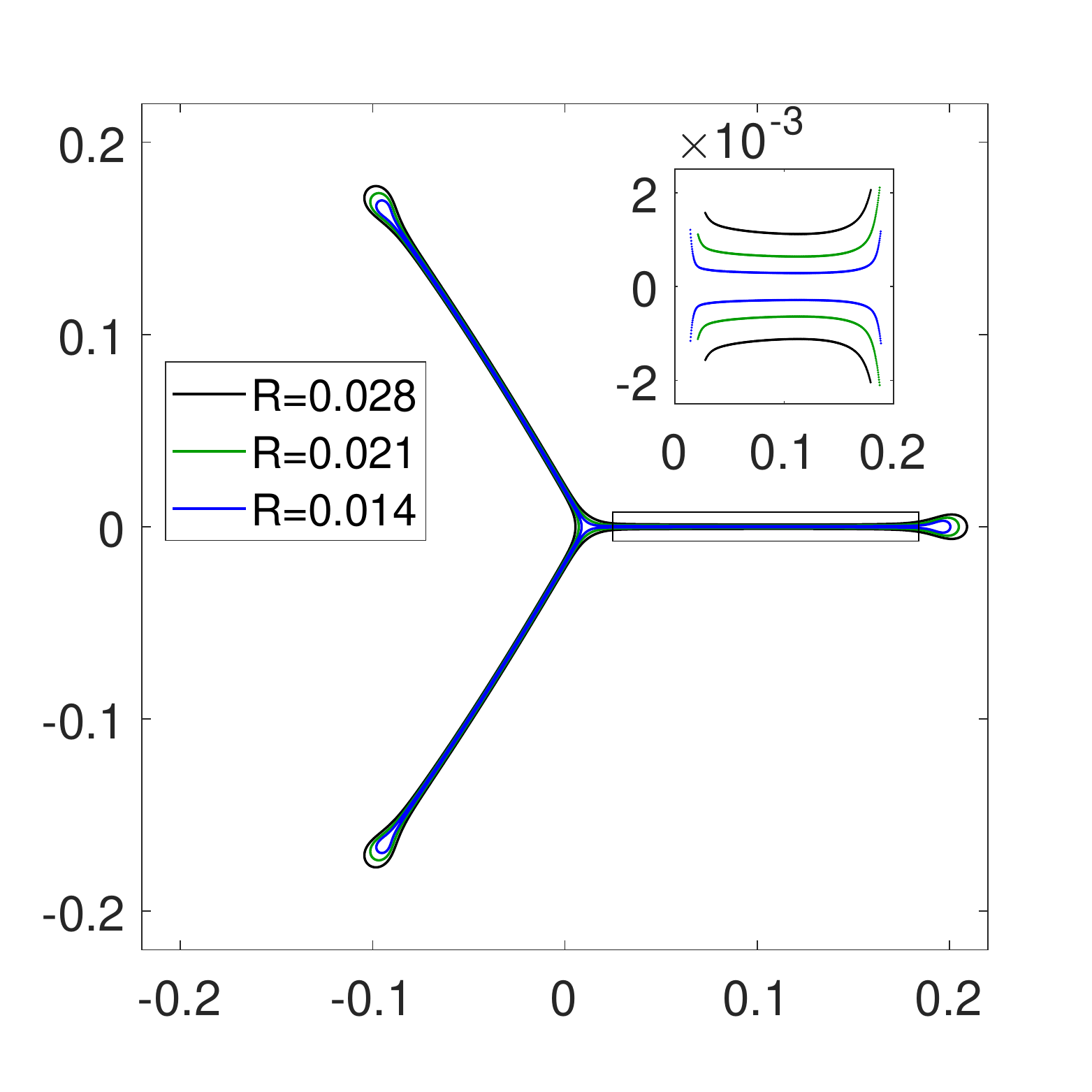}[b]\\
\includegraphics[width=0.4\textwidth]{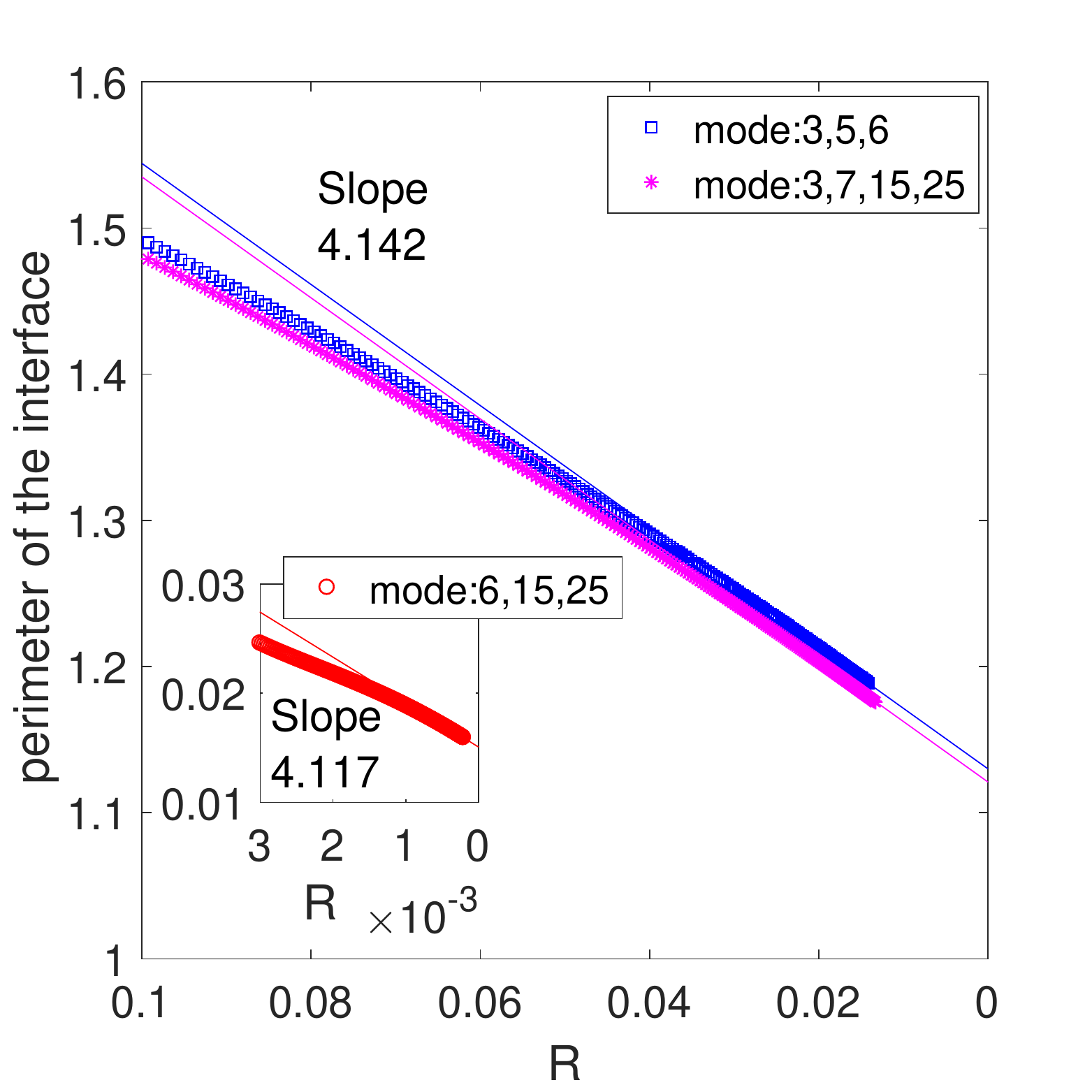}[c]
\includegraphics[width=0.4\textwidth]{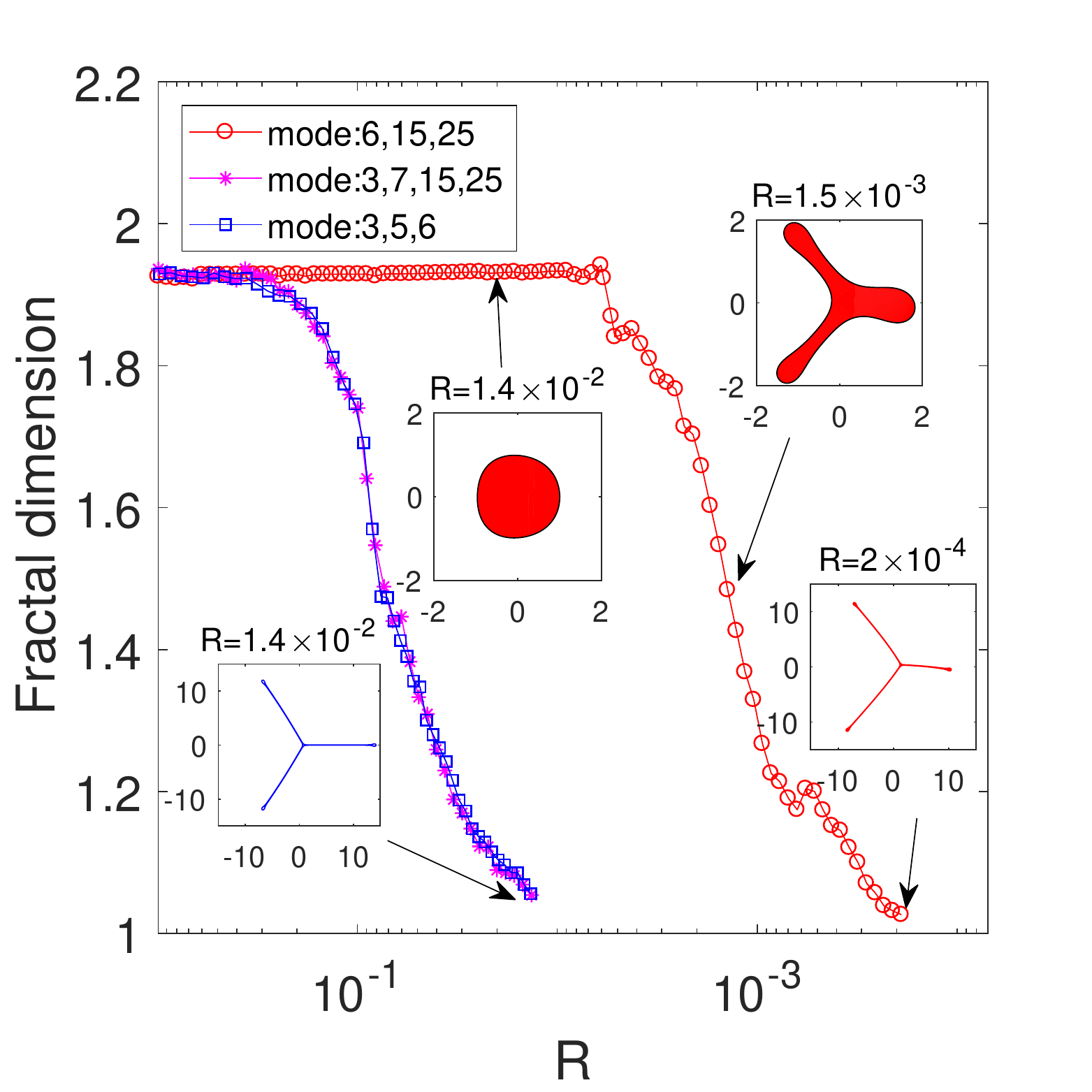}[d]
\caption{Simulations of drop dynamics in the special gap regime with $\displaystyle b_{\mathcal{C}}(t)=\left(1-\frac{7}{2}\tau \mathcal{C} t\right)^{-{2}/{7}}$ with $\mathcal{C}=52$ so that the fastest growing linear mode is $k_{max}=3$. Here $\tau=1\times 10^{-4}$. [a]. The nonlinear shape perturbations for different initial drop shapes are plotted as functions of the equivalent drop radii, as labeled (see text for details). The drop shapes near the vanishing time $T_C=2/(7\tau\mathcal{C})\approx 54.9$ are shown as insets. [b]. The dynamics of the drop with modes 3, 5 and 6 near the vanishing time. [c] The drop perimeters as a function of the effective drop radii as they vanish, together with linear fits near the vanishing time (slopes as labeled). The perimeters tend to a finite number as the drops vanish. [d] The fractal dimensions of the drops are shown as a function of the effective drop radii, together with drop morphologies (insets) at various stages of the evolution. These suggest that the limiting shapes have one-dimensional web-like morphologies.
}\label{lpss3}
\end{figure}

In Fig. \ref{lpss3}, we present results using $\mathcal{C}=52$ and surface tension $\tau=1\times 10^{-4}$, which makes $k_{max}=k^*=3$. We use three different initial drop shapes (see insets in Fig. \ref{lpss3}[a] with $R=1$): $\displaystyle r(\alpha,0)=1+0.02(\cos(3\alpha)+\cos(5\alpha)+\cos(6\alpha))$ (blue); $\displaystyle r(\alpha,0)=1+0.02(\cos(3\alpha)+\sin(7\alpha)+\cos(15\alpha)+\sin(25\alpha))$ (magenta); and $\displaystyle r(\alpha,0)=1+0.02(\sin(6\alpha)+\cos(15\alpha)+\sin(25\alpha))$ (red). Note that in the latter case, mode $k=3$ is not present initially. Here, we take $c=1/2$,  and use the time step $\Delta \bar{t}=1\times 10^{-4}$ and $N=8192$ points along the interface.

\begin{figure}
\includegraphics[scale=0.25]{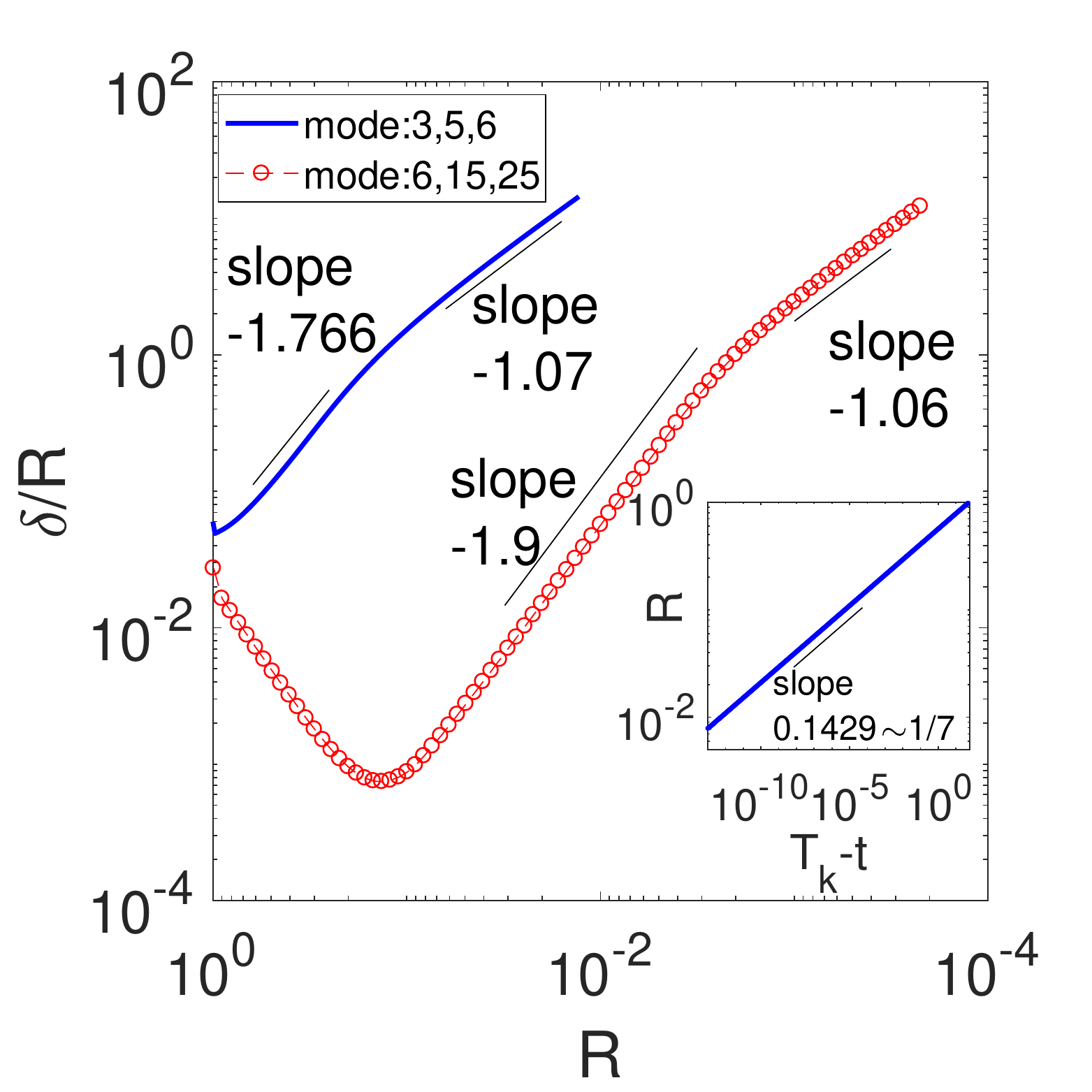}[a]
\includegraphics[scale=0.25]{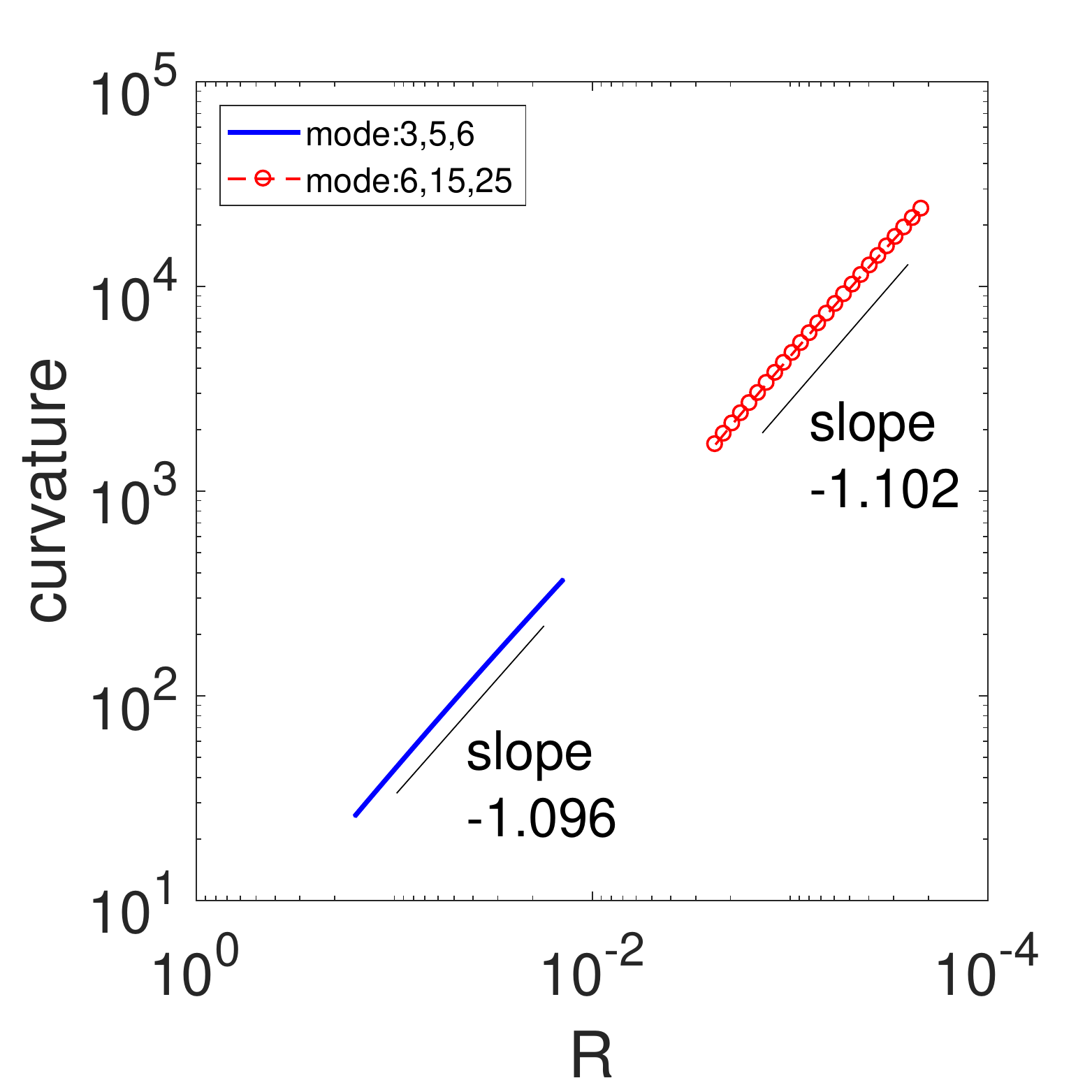}[b]
\includegraphics[scale=0.25]{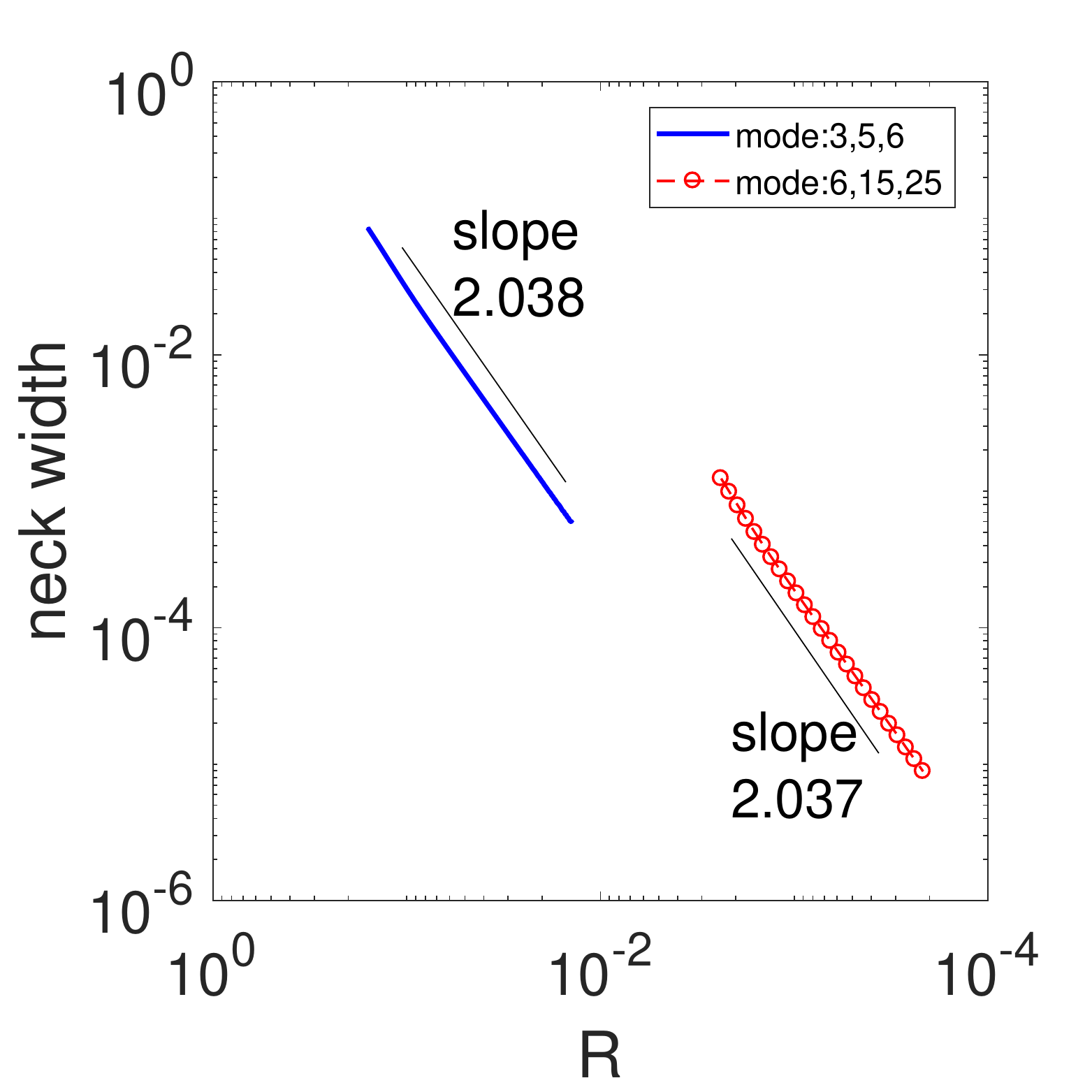}[c]
\caption{Asymptotic behaviors of the drops from Fig. \ref{lpss3}. [a] The nonlinear shape perturbations exhibit a biphasic power-law dependence on the effective drop radii and eventually diverge like $R^{-1}$ as $R\to 0$. [b] The maximum curvature of the drop also diverges like $R^{-1}$ as $R\to 0$. [c] The minimum neck width of the filaments tends to zero like $R^2$ as $R\to 0$.
 }\label{figcd}
\end{figure}

The nonlinear shape factor $\delta/R$ is plotted in Fig. \ref{lpss3}[a] as a function of the effective radius $R(t)$ in the original frame. Here, the nonlinear shape factor is calculated by
$\displaystyle {\delta}/{R}=\max_{\alpha}\left |{|\bar{\mathbf{x}}(\alpha,t)|}/{{\bar R}}-1\right |$,
where $\bar{\mathbf{x}}$ is the position vector measured from the centroid of the shape to the interface, $\displaystyle {\bar R}=\sqrt{{\bar{A}}/{\pi}}$ is the effective radius of the drop in the rescaled frame and $\bar{A}$ is the constant area enclosed by the interface.  Unlike the predictions of linear theory, the dynamics of the nonlinear shape factors are nonmonotone due to nonlinear interactions among the modes and the shape factors even decrease at early times (larger $R$). In particular, when mode $3$ is not present initially (red curve) the shape factor decreases until the drop becomes quite small ($R\approx 0.1$). However, as $R$ continues to decrease, eventually mode 3 starts to dominate and the shape factor grows rapidly. The shape perturbations grow throughout the dynamics (see also Fig. \ref{figcd}[a] below), which suggests that unlike the case for an expanding bubble \citep{ShuwangJCP,ShuwangPRL,Zhao2015}, the evolution does not become self-similar as the drop vanishes. 

The insets in  Fig. \ref{lpss3}[a] show the initial drop morphologies ($R=1$, top) and the final drop morphologies ($R$ as labeled, bottom) in the rescaled frame (the full dynamics can be found in the Supplementary Material Fig. S3) . Generally, the final morphologies are seen to have a 3-fold symmetric, one-dimensional, web-like network structure although the shapes are somewhat different as the drops vanish. The late time (small $R$) evolution in the original frame is plotted in Fig. \ref{lpss3}[b], which shows the morphologies of the drop with initial condition containing modes $3$, $5$ and $6$ (blue drop in Fig. \ref{lpss3}[a]). The drop dynamics in the rescaled frame can be found in the Supplementary Material ({Fig. S4[a]}). As the drop shrinks, we observe that the tips of the three fingers retract while the long filaments connecting the tips become thinner and tend to a finite length, as seen in the inset.The drop dynamics and morphologies when mode 3 is present initially (blue, magenta) are quite similar while in the third case (red) the shape has a less well-developed network structure because it takes some time for nonlinear interactions to generate mode 3 and then for mode 3 to dominate the shape. This is why the dominance of mode 3 emerges at much smaller $R$ than in the other cases (e.g., when the drop is about $\displaystyle {1}/{500}$ of its initial size).

As seen in Fig. \ref{lpss3}[c], the interface perimeter $P\approx P_0+a R$ as $R$ tends to 0, where $a$ is a constant and $P_0$ is a finite number. The slope $a$ depends on the symmetry of the limiting shape ($a$ is a decreasing function of $k_{max}$) and the limiting perimeter $P_0$ depends on the initial shape. See {Fig. S4[b]} in Supplementary Material for fits of the interface perimeter for other values of $k_{max}$. To test whether the limiting shapes are truly one dimensional, we calculate the fractal dimension of the shapes. The fractal dimension $D_0$ can be approximated by a box counting algorithm: cover the pattern with a grid of square boxes of size $\zeta$  and define $N(\zeta)$ to be the total number of boxes of size $\zeta$ to cover the whole pattern \citep{OH}, 
\begin{equation}
D_0=\lim\limits_{\zeta\rightarrow 0}-\frac{\log N(\zeta)}{\log \zeta}.
\end{equation}
Fig. \ref{lpss3}[d] shows the fractal dimensions of the shapes as a function of effective drop radius. At early times, the drops remain compact especially for the case with initial modes 6, 15, and 25, which needs a long time for nonlinear interactions to create mode 3. Later there exists a transition as the fractal dimension decreases from about 2 down to about 1 as the drop vanishes. All together, these results strongly suggest that the limiting shape is not a circle and but instead has a web-like network structure. Although the drop morphologies look similar to patterns of random fractals generated using a large unscreened angle threshold \citep{Kaufman}, the mechanism is different. In \citep{Kaufman}, only the tip region is active, but in our case the whole interface is dynamic. 

In Fig. \ref{figcd} we analyze the properties of the limiting shapes in the original frame as the drop vanishes. In Fig. \ref{figcd}[a], the  nonlinear shape factor is seen to diverge as $R$ tends to $0$. Interestingly, when $R$ is not so small, $\delta/R\sim R^{-2}$, which is consistent with the predictions of linear theory (e.g., see Eq. (\ref{perturbation magnitude}) with $k_{max}=3$). However, as $R$ decreases, nonlinear interactions increasingly dominate the evolution and the shape factor diverges more slowly $\delta/R\sim R^{-1}$. This is due to the curvature of the drop tips, which diverges as $\kappa_{*}\sim R^{-1}$ as seen in Fig. \ref{figcd}[b]. The curvature in the scaled frame $\bar\kappa_{*}$, on the other hand, is bounded and tends to a finite limit as $R\to 0$, see {Fig. S4[c]} in Supplementary Material. The time dependence of the width $w$ of the neck region, shown within the boxed region in Fig. \ref{lpss3}[b], is plotted in Fig. \ref{figcd}[c], which suggests the scaling $w\sim R^{2}$. This can be explained as follows. Let $L$ be the length of the neck region. Then, approximating the filament (neck region and drop tip) as a rectangle with a semi-circular tip with radius $\kappa_{*}^{-1}$, the total area of the drop $\pi R^2\appropto Lw+\pi \kappa_{*}^{-2}$. Since $L$ tends to a finite constant as $R\to 0$, this suggests $w\sim R^2$. Further, taking the same approximation of the filament, the perimeter $P\appropto 2L + 2\pi \kappa_{*}^{-1} \sim P_0+aR$ as suggested in Fig. \ref{lpss3}[c].

\subsubsection{Mode selection and morphology diagram}

\begin{figure}
\includegraphics[scale=0.1]{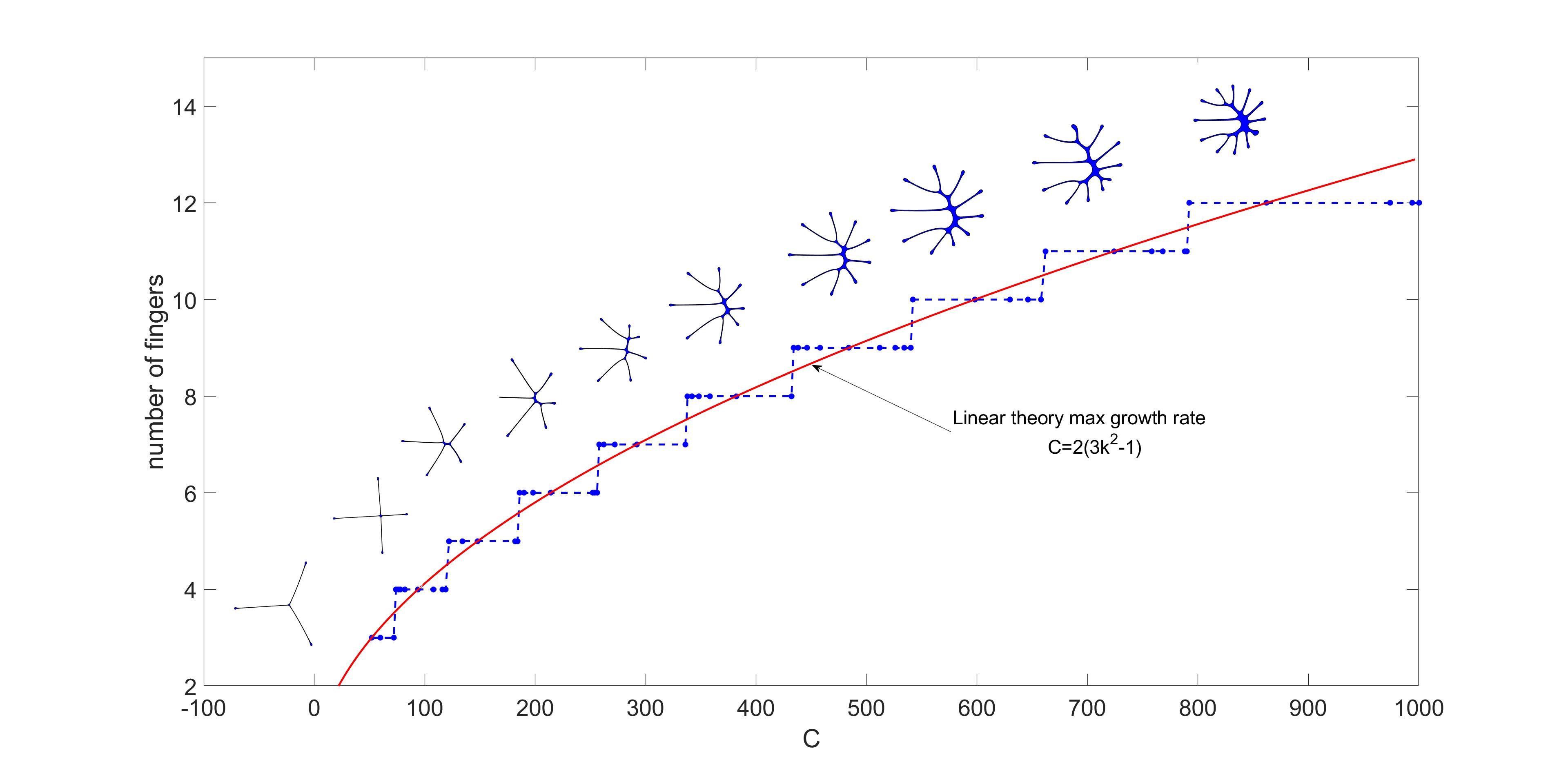}
\caption{A morphology diagram that relates the dominant mode of the limiting shapes (number of fingers) to the constant $\mathcal{C}$ using the special gap dynamics $\displaystyle b_{\mathcal{C}}(t)=\left(1-\frac{7}{2}\tau \mathcal{C}t\right)^{-\frac{2}{7}}$, with $\tau=1\times 10^{-4}$ and $\mathcal{C}$ is varied from 52 to 1000. According to linear theory, the fastest growing mode $k_{max}$ (solid curve) varies from 3 to 12. The dots represent nonlinear numerical simulations using the same initial data (see text for details). The results suggest that the dominant mode of the limiting shapes are selected by $\mathcal{C}$. Linear theory provides a good approximation of the dominant modes, but nonlinear interactions dictate where transitions from $k$-fold to $(k+1)$-fold dominant limiting shapes occur. 
}
\label{diagram}
\end{figure}

Next, we investigate how the limiting shapes are selected by the parameter $\mathcal{C}$ using the special gap dynamics $\displaystyle b_{\mathcal{C}}(t)=\left(1-\frac{7}{2}\tau \mathcal{C}t\right)^{-\frac{2}{7}}$. We consider the dynamics using different initial shapes given by a perturbed circle with the initial radius $r(\alpha,0)=1+2.5\times 10^{-3}\sum\limits_{k=30}^{60}\exp(-0.2k)(a_k\cos(k\alpha)+b_k\sin(k\alpha))$. The coefficients $a_k$ and $b_k$ are randomly selected using a uniform distribution in the interval $[-1,1]$. We generate two such initial shapes and we use the same initial shape for all $\mathcal{C}$, which we vary from 52 to 1000.  According to linear theory, the fastest growing modes correspondingly ranges from $k_{max}=3$ to $12$. By considering initial shapes with these high modes, we guarantee that all the initial modes are decreasing (e.g., only modes $|k|<\sqrt{\mathcal{C}/2+1}$ are growing) and that the fastest growing mode is only generated by nonlinear interactions. The result is a morphology diagram given in Fig. \ref{diagram}, which shows that the dominant mode of the limiting shape (e.g., number of fingers) is an increasing, piecewise constant function of $\mathcal{C}$ where there are sharp transitions from $k$-fold to $(k+1)$-fold dominant limiting shapes. While the morphologies of the limiting shapes are not identical, the dominant mode is solely determined by the constant $\mathcal{C}$. For reference, we also plot the maximum growing mode $k_{max}$ (solid curve). Although linear theory provides a good approximation of the dominant mode of the limiting shape, nonlinear interactions are critical for determining where the transitions from $k$-fold to $(k+1)$-fold dominant limiting shapes occur. Further, range of $\mathcal{C}$ for which the limiting shape is dominated by a particular mode $k$ is an increasing function of $\mathcal{C}$. 

In the Supplementary Material, we present cases where the initial condition contains modes that grow. In these cases, the dominant mode of the limiting shape can be dependent on the initial condition as well as $\mathcal{C}$ (see Figs. {S5}). However, if the initial shape contains $k_{max}$ with magnitude comparable to the other initially growing modes, then the dominant mode of the limiting shape is still given by $k_{max}$, e.g., the limiting shape is still solely selected by $\mathcal{C}$ (see Fig. {S6}).

%

\section{Conclusions}\label{con}

In this paper, we have investigated the fully nonlinear dynamics of viscous drops in Hele-Shaw cells when the upper plate is lifted perpendicularly at a prescribed rate. This action reduces the size of drops in the mid-plane between the plates and increases the extent of the drops in the $z$-direction to preserve volume. As air rushes in, this generates a Saffman-Taylor instability and the drops deform as they shrink. Linear theory predicts that the instability may be transient, if the lifting rate is sufficiently small, or may persist if the rate is sufficiently large. To simulate the nonlinear dynamics of the drops, we used a very efficient, highly accurate boundary integral method that involves space and time rescaling to track the shrinking of the drops in the mid-plane using the Hele-Shaw approximation \citep{Meng-2}. By rescaling time to slow down the rapid evolution of the drops and rescaling space to maintain constant-volume drops in the simulation frame, we can overcome the severe constraints on the time steps and spatial grid sizes that would be encountered in the original frame of reference. This enabled us to study, for the first time, the limiting dynamics as the drops vanish over a wide range of lifting rates. 

We compared the nonlinear simulation results to linear theory and to previously performed experiments with two different surface tensions when the gap is increased linearly in time. Comparisons with a new experiment using a gap width that increases nonlinearly in time are presented in the Supplementary Material ({Figs. S8-10}). In these cases, the instability is transient and the drops eventually shrink like circles. 

When the gap grows linearly in time, we found that nonlinear interactions increase perturbations more rapidly than predicted by linear theory. The nonlinear simulations tend to agree best with experiments at large confinement numbers ($C_0={R_0}/{b_0}$, where $R_0$ is the initial radius of the liquid and $b_0$ is the initial gap) and predict a biphasic exponential decay of the number of fingers over time consistent with experiments. At small $C_0$, the nonlinear simulations over-predict the number of fingers. We suggested that this might be due to an increase in the wetting layer thickness, which scales like $h\sim C_0^{-4/3}$ as the confinement number decreases. By increasing the speed of the lifting plate, to mimic the more rapid loss of fluid to the wetting layer, we found better agreement between the simulations and experiments. Of course, in addition to wetting, other effects such as viscous stresses, inertial forces and three-dimensionality could also play an important role in the drop dynamics.

We also studied the limiting dynamics of drops when the gap is lifted very fast:  $\displaystyle b_{\mathcal{C}}(t)=\left(1-\frac{7}{2}\tau \mathcal{C} t\right)^{-{2}/{7}}$, where $\tau$ is the nondimensional surface tension and $\mathcal{C}$ is a constant. According to linear theory, this gap, which diverges at the finite time $T_C=2/(7\tau\mathcal{C})$, ensures that mode $k=\sqrt{(1+\mathcal{C}/2)/3}$ is the fastest growing mode along the interface at any time. By rescaling in both time and space, we are now able to study the nonlinear limiting shapes as the drops vanish in this regime for the first time. 

Consistent with linear theory, nonlinear simulations predict that the instability is transient when the gap $b(t)$ satisfies
${\dot b(t)}/{b(t)^{9/2}}\to 0$ as $t$ increases, even if the gap $b(t)$ diverges at a finite time. For example, in the Supplementary Material, we show that using the gap $b(t)=\left(1-3\tau \mathcal{C}t\right)^{-1/3}$, which diverges at $T_*=1/(3\tau\mathcal{C})$ even faster than $b_\mathcal{C}(t)$, the interface undergoes a transient instability but ultimately vanishes like a circle since ${\dot b(t)}/{b(t)^{9/2}}\sim (T_*-t)^{1/6}$ as $t\to T_*$ (see Fig. S7). However, when the gap $b_\mathcal{C}(t)$ is used, perturbations may continually grow and the drop morphologies may acquire a striking, one-dimensional web-like shape as the drops shrink. To our knowledge, this limiting behavior has not been reported previously. 

We characterized the limiting drop shapes by generating a morphology diagram that relates the number of fingers to $\mathcal{C}$, independent of initial conditions within a class of interface shapes that contains only high mode interface perturbations so that the most unstable mode is generated solely by nonlinear interactions. While linear theory provides a good approximation of the dominant modes, nonlinear interactions determine where transitions (in $\mathcal{C}$) from $k$-fold to $k+1$-fold dominant limiting shapes occur. We also described the behavior of the system when low mode perturbations are present initially.

A natural question is whether the one-dimensional, web-like shapes we have discovered here when the upper plate is lifted very rapidly are actually achievable in an experiment. Depending on the initial conditions, our simulations suggest that the web-like shapes can be observed when the radius decreases from its initial value by about a factor of 7-10 (Fig. S8). This corresponds to an increase in gap width by a factor of about 50-100. If the initial gap $b_0=50\mu m$, this corresponds to a final gap thickness of about $b\sim 2.5mm-5 mm$. Assuming that the gap width increases by a factor of 100, the corresponding time over which this would occur is $\mathcal{T}\sim \frac{0.29}{\mathcal{C}\tau} T$, where $T$ is a characteristic time scale given by $T\sim \frac{12\mu}{\sigma}C_0^2R_0\tau$ where $\mu$ and $\sigma$ are the dimensional viscosity and surface tension, respectively, and $C_0=R_0/b_0$ is the confinement number with $R_0$ being the initial drop radius. This gives $\mathcal{T}\sim \frac{0.29}{\mathcal{C}} \frac{12\mu}{\sigma}C_0^2R_0$. Using the values of the viscosity and surface tension from \citep{JDA} to be $\mu =10 Pa\cdot s$ and $\sigma=0.02 N/m$, respectively, and the initial radius $R_0=1.5mm$, we obtain $\mathcal{T}\sim  {2349}{\mathcal{C}^{-1}} s$ since $C_0=R_0/b_0=30$. Finally, taking $\mathcal{C}=2(3k^2-1)$ we obtain $\mathcal{T}=\mathcal{T}_k\sim \frac{1200}{3k^2-1}s$. So as the dominant mode $k$ of the limiting shape increases, the time over which the plate needs to be lifted decreases like $k^{-2}$. For example, for 3-mode dominant limiting shapes we obtain $\mathcal{T}_3\sim 46s$ while for 4-mode and 5-mode dominant shapes we obtain $\mathcal{T}_4\sim 25s$ and $\mathcal{T}_5\sim 16s$, respectively. This suggests that it should be possible to access this regime experimentally, at least for limiting shapes dominated by low modes.

Once experiments in the special gap regime are performed, agreement between theory and experimental results may also require accounting for the effects of flow in 3D \citep{MD}, wetting effects \citep{Park,EJ13-2}, and inertia effects \citep{Chevlier06,He11,Anjos17}, which were neglected in our Hele-Shaw formulation. These will be considered in future work. Finally, although we focused on Hele-Shaw problems here, we hope that our findings may help in understanding the selection mechanisms within other pattern forming phenomena such as bacterial colony growth and snowflake formation.

\section*{Acknowledgement}
S. L. and J. L. gratefully acknowledge partial support from the National Science Foundation, Division of Mathematical Sciences through grants NSF-DMS 1719960 (J. L.) and NSF-DMS 1720420 (S. L.).
 J.L. also acknowledges partial support from grants NSF-DMS 1763272 and the Simons Foundation (594598, QN) for the Center for Multiscale Cell Fate Research at UC Irvine. 
\bibliographystyle{jfm}
\bibliography{liftingdiagram}

\end{document}


\maketitle
\section{Modified lifting speed of experiments in \cite{JDA}.}\label{lifting}
In the manuscript, we compared our simulation results, linear theory, and experimental results from \citep{JDA}. In the experiments, the bottom plate of a Hele-Shaw cell is fixed and the top plate is lifted uniformly at a constant speed. A less viscous fluid (air) penetrates a more viscous fluid (silicone oil), forming fingering patterns on the interface. High resolution, high contrast images are used to calculate the number of fingers \citep{JDA}. The finger tip is described as the innermost part of an air finger and the finger base is the outer end of a finger (see Fig. S\ref{jda}). The finger amplitude is the distance between the tip and base of a finger. In \citep{JDA}, the authors
systematically investigate the influence of different parameters on the dynamics of the drop evolution, including how the number of fingers varies as a function of time. To do this, the initial drop size, the initial gap spacing, and the lifting speed are varied. The number of fingers not only depends on the nondimensional surface tension $\tau$ but also depends on the confinement number $\displaystyle C_0=\frac{R_0}{b_0}$, the aspect ratio of the fluid. When the surface tension is large, the effect of $C_0$ is more pronounced at later times while at small surface tensions $C_0$ has a larger effect at early times. See \citep{JDA} for further details.

 To extract the number of fingers over time from the experiments, we use the data presented in Figs. 10 [c] and [e] in \citep{JDA} and use the web application WebPlotDigitizer (https://automeris.io/WebPlotDigitizer/) to extract the data points. In particular, after choosing the scale of the plot, we are able to get the position of each data point on the plots where the number of fingers is usually a decimal close to an integer. Since the number of fingers is an integer, we round to the nearest integer. As pointed out in \citep{JDA} the 
number of fingers depends on both the surface tension and the confinement number. At small confinement numbers $C_0$ (see Fig. S\ref{diffC0}[a][b]), the number of fingers shows a single rate of exponential decay over time, consistent with the experimental results in \citep{MD}. At large confinement numbers $C_0$ (see Fig. S\ref{diffC0}[c][d]), the number of fingers seems to have a biphasic exponential decay in time, similar to our simulation results.
 
In our simulations, we calculate the number of air fingers in exactly the same way as \citep{JDA}. We have found that our simulation results agree very well with the experimental observations at large confinement numbers $C_0$ and predict more fingers than the experimental data at small confinement numbers $C_0$. It is still not well understood how the confinement number influences the behavior of the drop. One possibility might be the volume loss of the drop fluid as the fluid is stuck to the plate. Consequently, the effective shrinking speed of the fluid in experiments might be faster than the constant speed we assumed in our simulations. 

To test this hypothesis, we analyze the dependence of the wetting layer thickness on the confinement number. The thickness of the wetting layer on the plate $h$ is proportional to $\displaystyle Ca^{2/3}$, where $\displaystyle Ca=\mu\tilde V/\sigma$ is the capillary number \citep{Park84,Park,Jackson15}. Using $\displaystyle \tilde V\sim \dot{\tilde b}_0C_0$, we obtain $\displaystyle Ca\sim (\mu \dot{\tilde b}_0/\sigma)C_0$. In the experiments \citep{JDA}, the confinement number and lifting rate are changed such that the nondimensional surface tension $\tau=\sigma\tilde b_0^3/(12\mu\dot{\tilde b}_0 R_0^3)$ is fixed. This gives $\displaystyle Ca\sim C_0^{-2}$ and therefore $h\sim C_0^{-4/3}$. This implies that the wetting layer gets thicker as the confinement number $C_0$ decreases. That is, more fluid is stuck on the plate and the fluid has a faster effective shrinking speed as the confinement number $C_0$ decreases. 

To investigate the increase in wetting layer thickness due to $C_0$, we make a correction to our lifting speed in the simulations and consider $\displaystyle b(t)=1+\frac{1+3e_0t}{1+3t}t$, where $e_0$ is a constant accounting the speed change. This correction is ad-hoc and a simple way to account for the additional volume loss of drop fluid to the wetting layer. Future work should account for the boundary layer dynamics directly. Note that $e_0=1$ corresponds to the original formulation (linear rate of gap increase). However, the larger $e_0$ is, the more fluid is left on the plates and the faster effective shrinking speed is. We take different $e_0=1$, $1.1$, $1.3$, $1.5$, and $1.7$ and perform simulations using this lifting speed. The results are shown in Figs. S\ref{modified gap}[a] and [b]. The morphologies of the interface using $e_0=1.2$ (Fig. S\ref{modified gap}[a]) and $e_0=1.7$ (Fig. S\ref{modified gap}[b]) are shown as insets. At early times, the number of fingers is insensitive to the choice of $e_0$. At later times, the number of fingers decreases as $e_0$ increases. By treating $e_0$ as a fitting parameter, our simulations can fit the experimental data for all confinement numbers $C_0$ using the modified gap dynamics.

\begin{suppfigure}
\center
\includegraphics[scale=0.35]{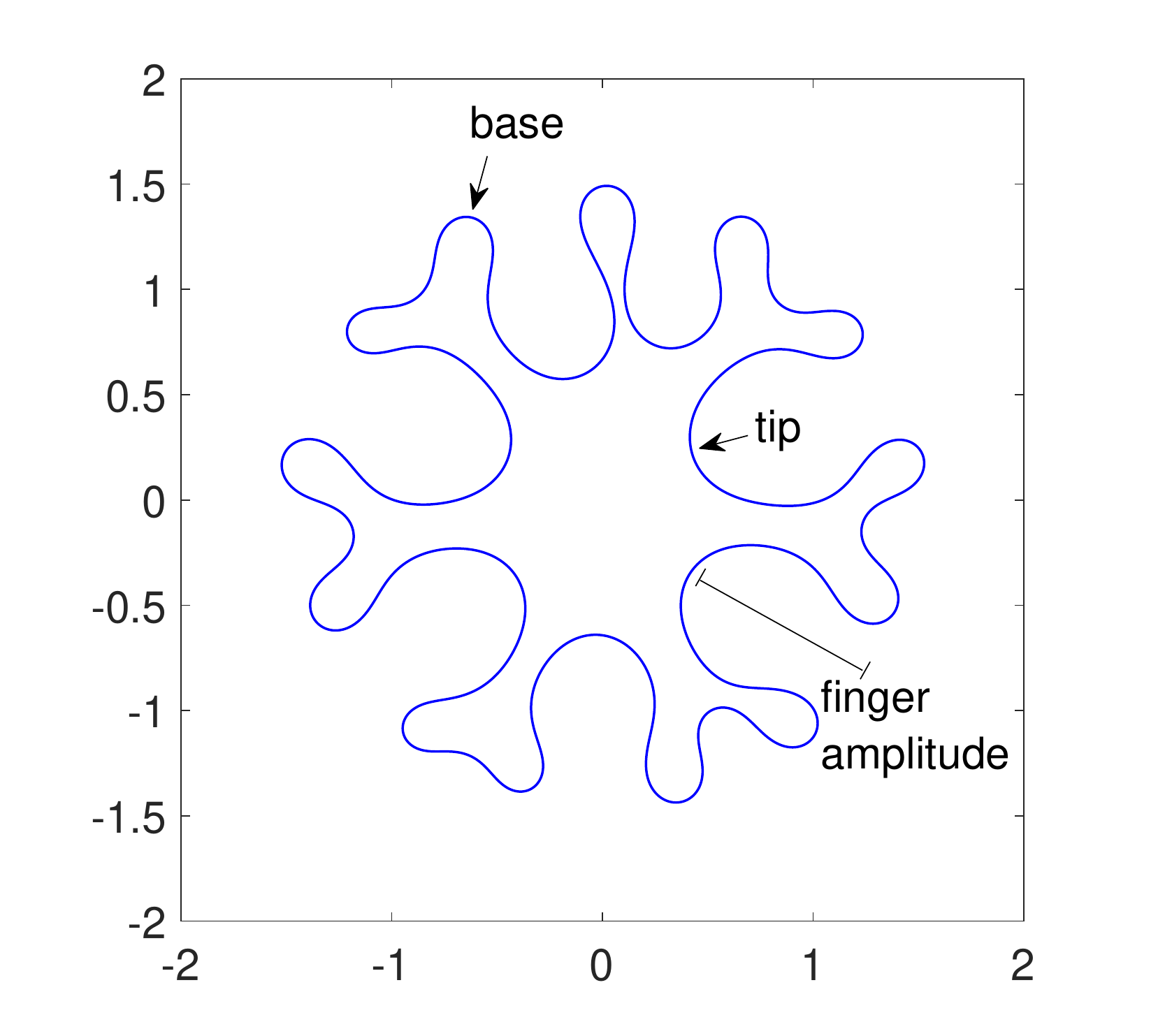}
\caption{A schematic of an interface and definition of the finger tip, finger base, and finger amplitude, following \citep{JDA}. See text for details.}
\label{jda}
\end{suppfigure}

\begin{suppfigure}
\includegraphics[scale=0.3]{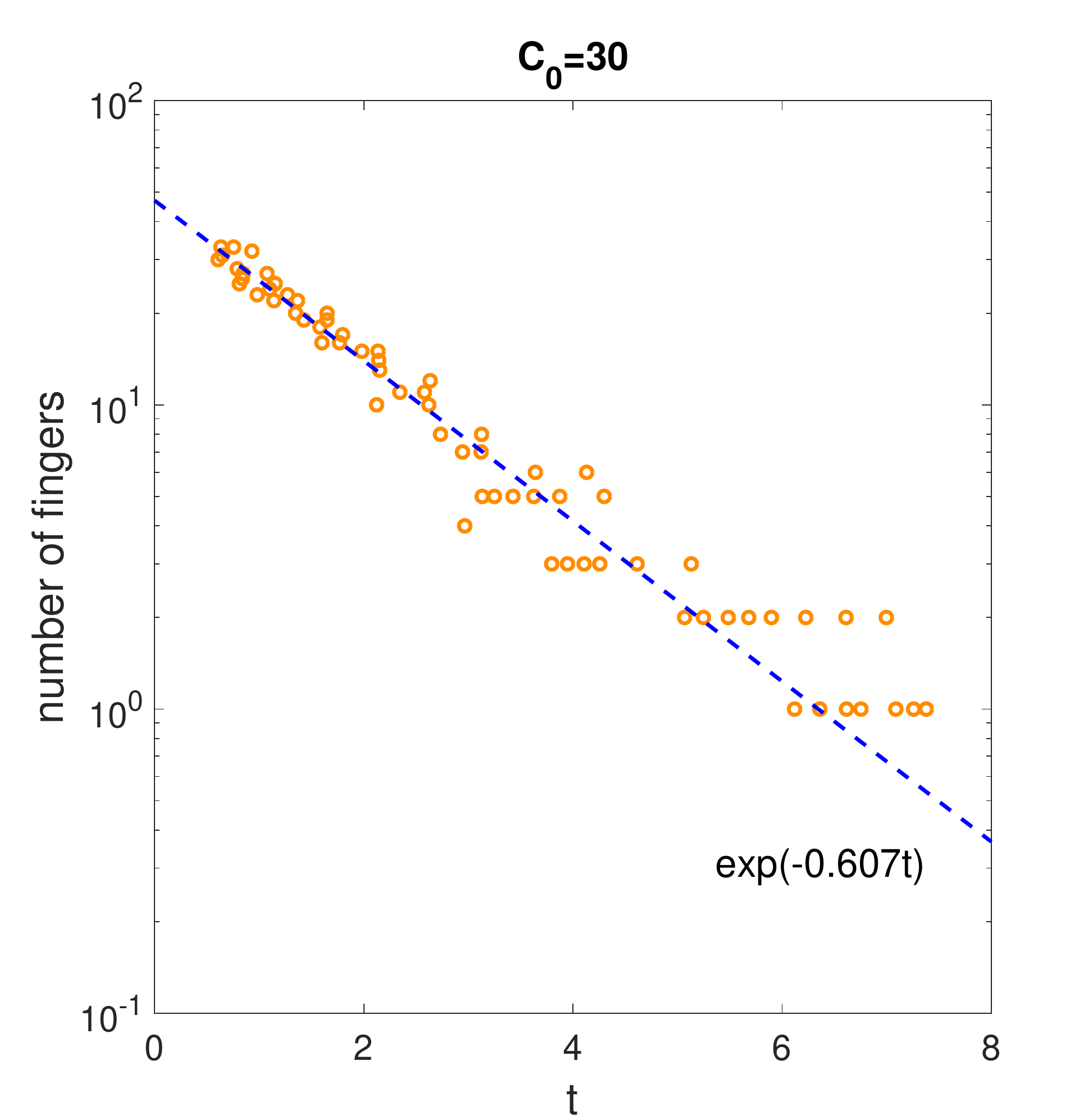}[a]
\includegraphics[scale=0.3]{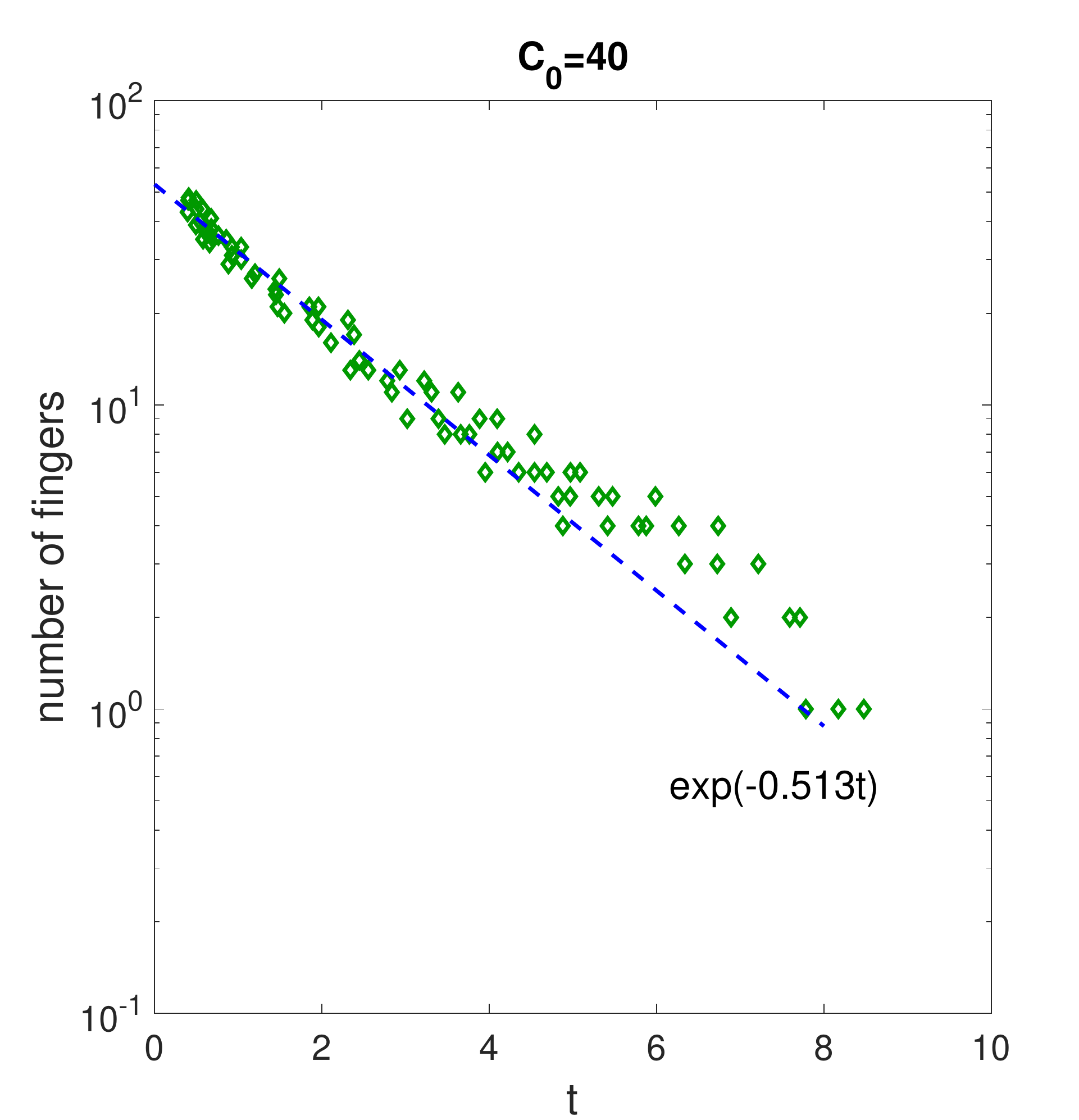}[b]\\
\includegraphics[scale=0.3]{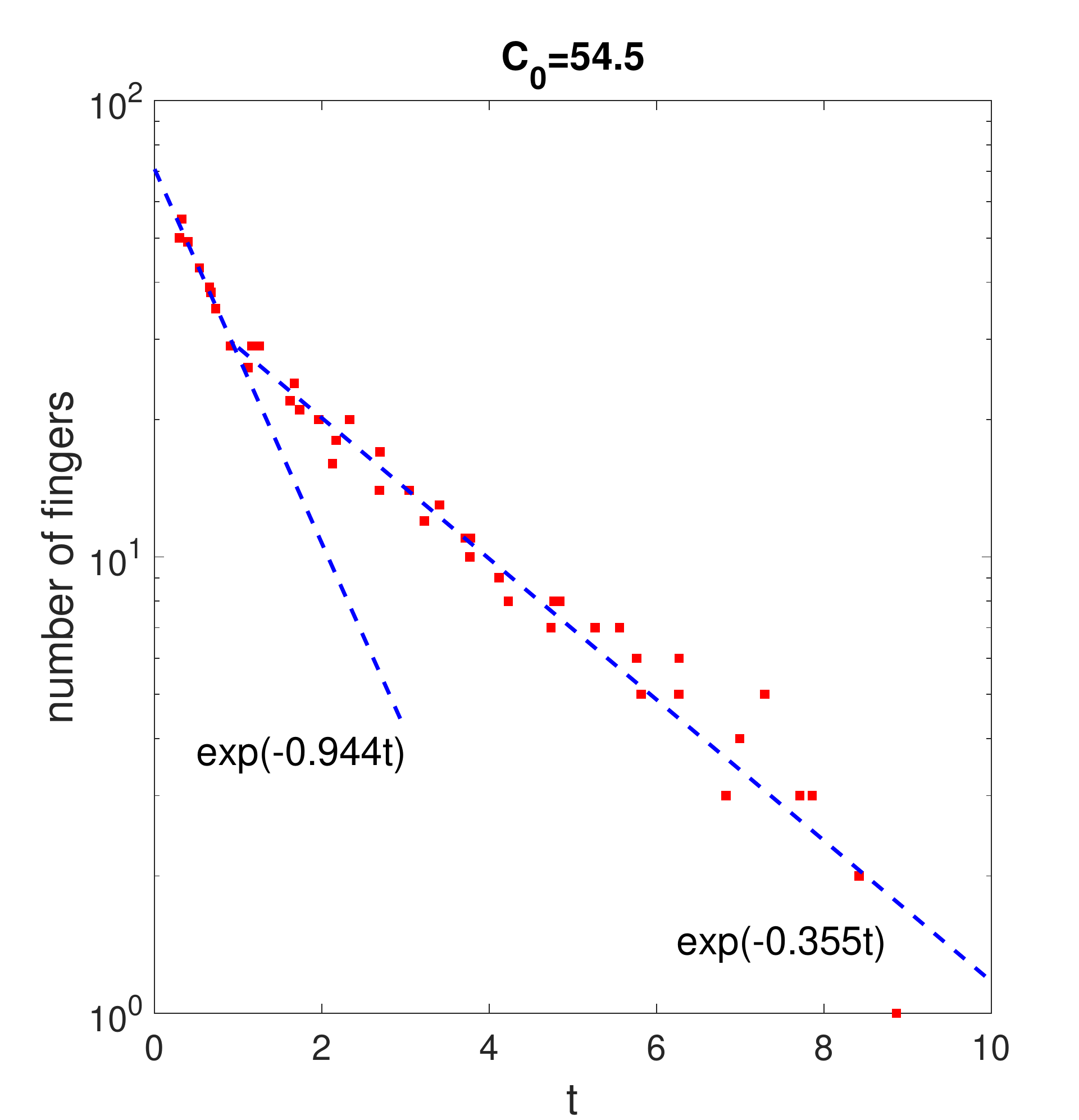}[c]
\includegraphics[scale=0.3]{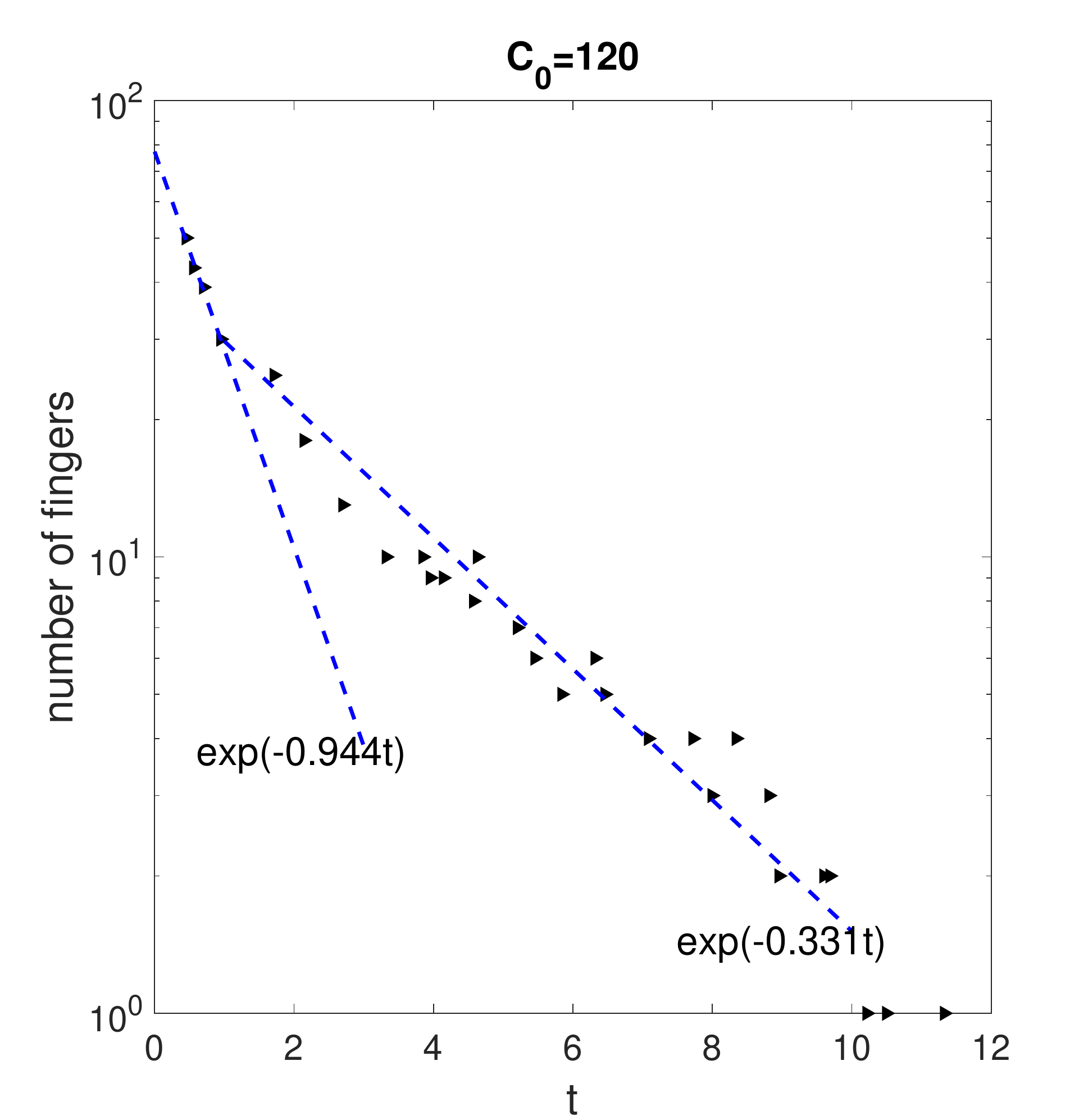}[d]
\caption{The number of fingers in drops with different confinement number $C_0$ from the experiments in \citep{JDA} with surface tension $\tau=3\times 10^{-5}$. At small $C_0$ the number of fingers follows a single rate of exponential decay in time while the data at large $C_0$ suggests that there is a biphasic exponential decay in the number of fingers over time.}
\label{diffC0}
\end{suppfigure}

\begin{suppfigure}
\center
\includegraphics[scale=0.37]{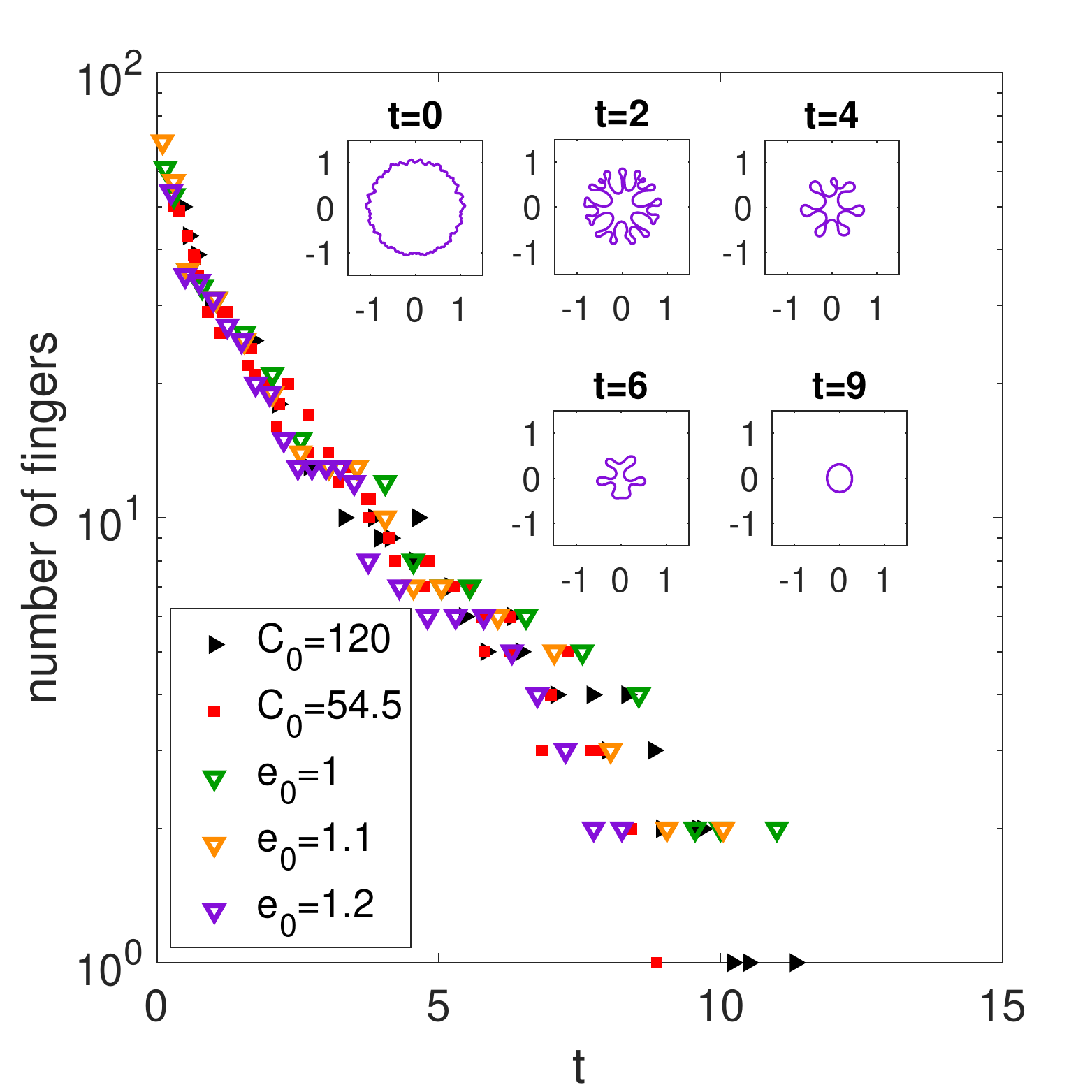}[a]
\includegraphics[scale=0.37]{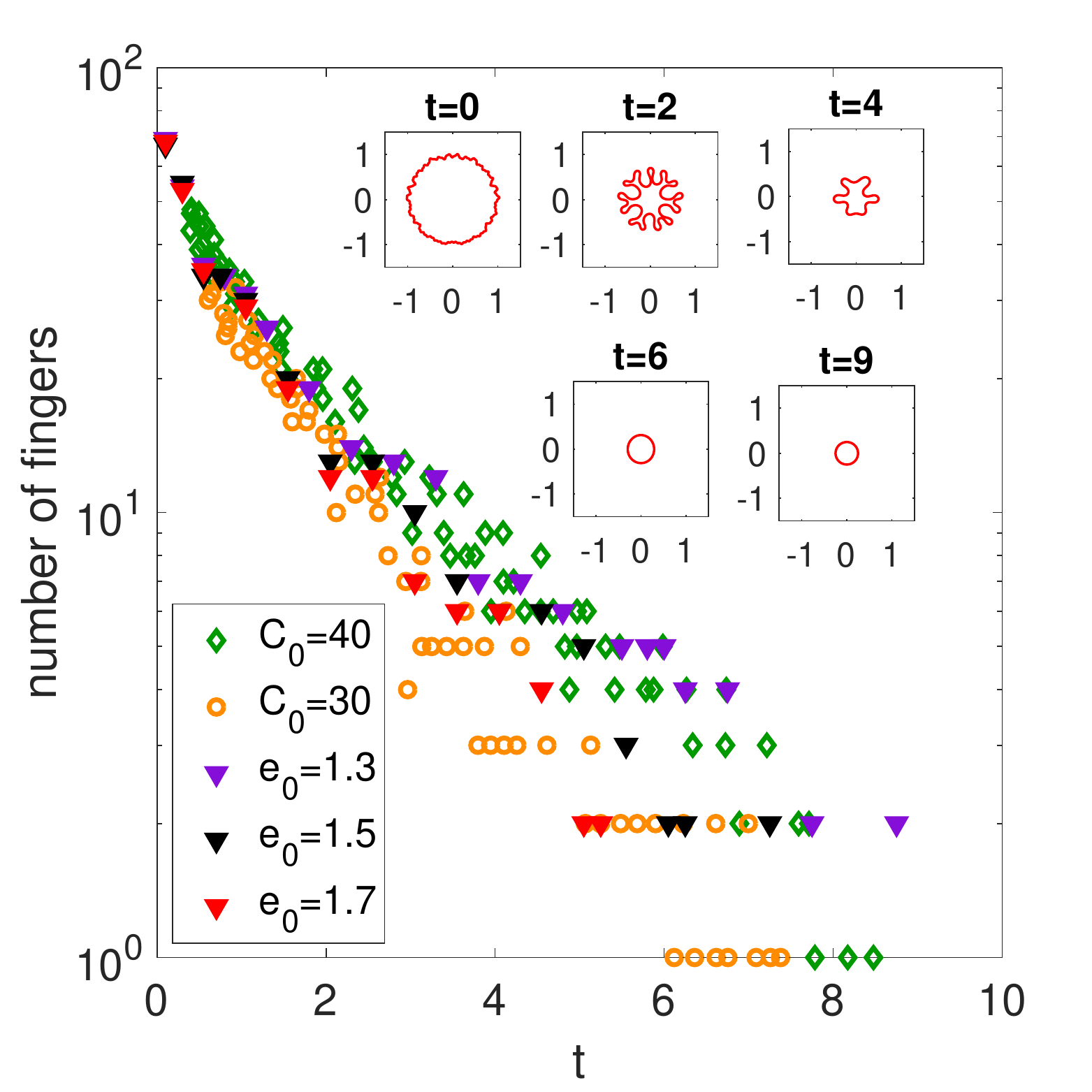}[b]
\caption{Comparisons between the experimental data from \cite{JDA} (with surface tension $\tau=3\times 10^{-5}$; shown also in Fig. \ref{diffC0}) and simulations using the modified gap $\displaystyle b(t)=1+\frac{1+3e_0t}{1+3t}t$ to account for the volume loss of the fluid drop due to changes in the wetting layer thickness due to confinement (see text for details). Increasing the thickness corresponds to increases in  $e_0$ as more fluid is stuck to plates. [a]: Results at large confinement numbers $C_0$ with $e_0$ as labeled. The interface morphologies, with $e_0=1.2$, are shown at different times in the insets. [b] Results at small $C_0$ with $e_0$ as labeled. The interface morphologies, with $e_0=1.7$, are shown at different times in the insets. As $e_0$ increases, the number of fingers decreases. Treating $e_0$ as a fitting parameter, the simulations can fit the experimental data over the whole range of confinement numbers considered.
}
\label{modified gap}
\end{suppfigure}

\section{Examples of dynamics dominated by mode 3}

 In Fig. S\ref{lpss3dyn}, we exhibit the interfacial dynamics in the rescaled under a special gap $\displaystyle b_{\mathcal{C}}=(1-\frac{7}{2}\tau\mathcal{C}t)^{-2/7}$ with $\mathcal{C}=52$ and nondimensional surface tension $\tau=1\times 10^{-4}$. With these parameters, the linear fastest growing mode is 3. We take three different initial shapes: $\displaystyle r(\alpha,0)=1+0.02(\cos(3\alpha)+\cos(5\alpha)+\cos(6\alpha))$ (blue); $\displaystyle r(\alpha,0)=1+0.02(\cos(3\alpha)+\sin(7\alpha)+\cos(15\alpha)+\sin(25\alpha))$ (magenta); $\displaystyle r(\alpha,0)=1+0.02(\sin(6\alpha)+\cos(15\alpha)+\sin(25\alpha))$ (red). All three cases show that after a period of transient morphological changes, the drop morphologies eventually acquire a 
3-fold symmetric, one-dimensional, web-like network structure as they vanish. Since mode 3 is not initially present in the third case (red), it takes some time for nonlinear interactions to generate mode 3 and for mode 3 to dominate the shape. As a result, a longer time has to pass for the the corresponding (red) drop morphologies to acquire 3-fold dominant shapes and as a result the drop is much smaller than the others when this occurs.

\begin{suppfigure}
\includegraphics[scale=0.2]{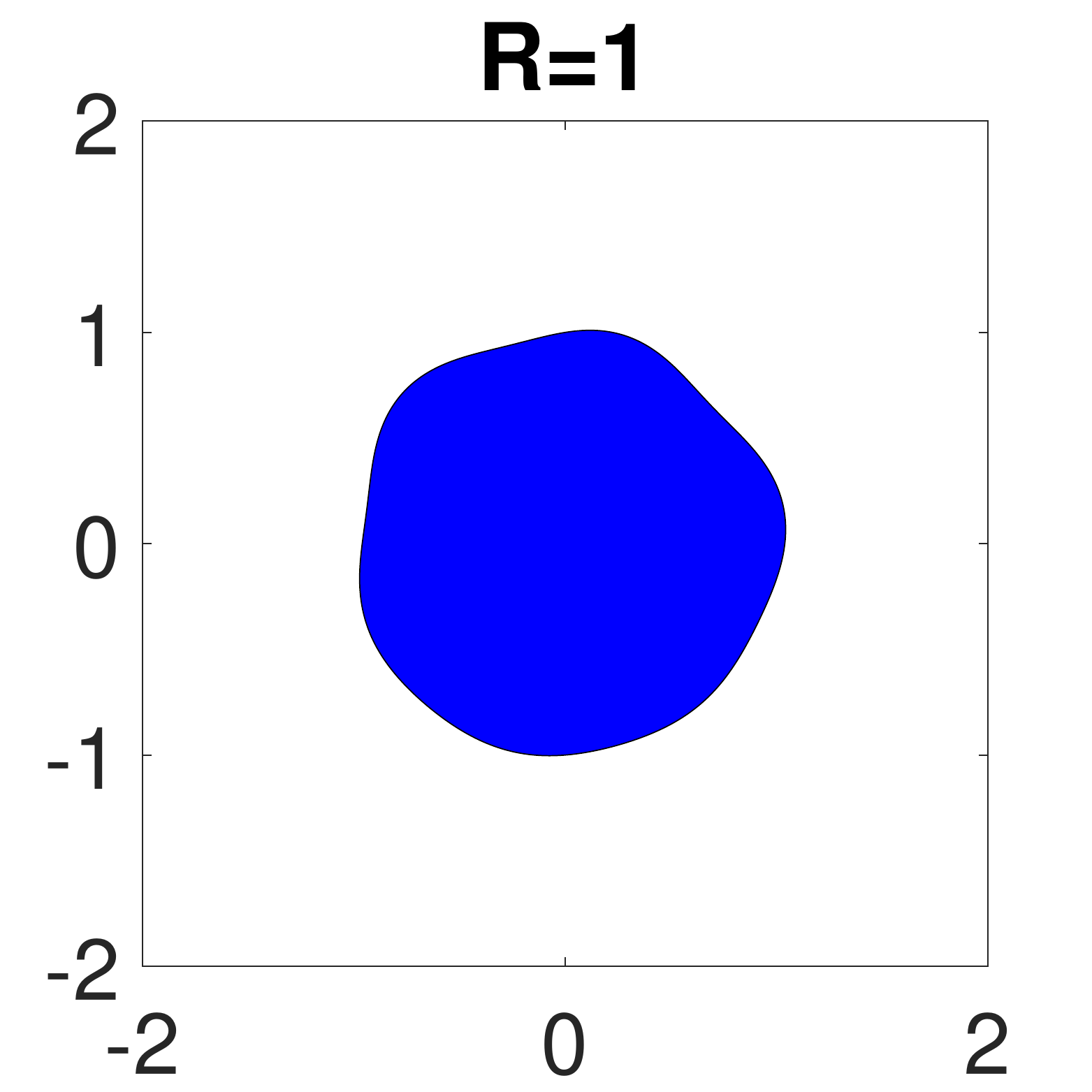}
\includegraphics[scale=0.2]{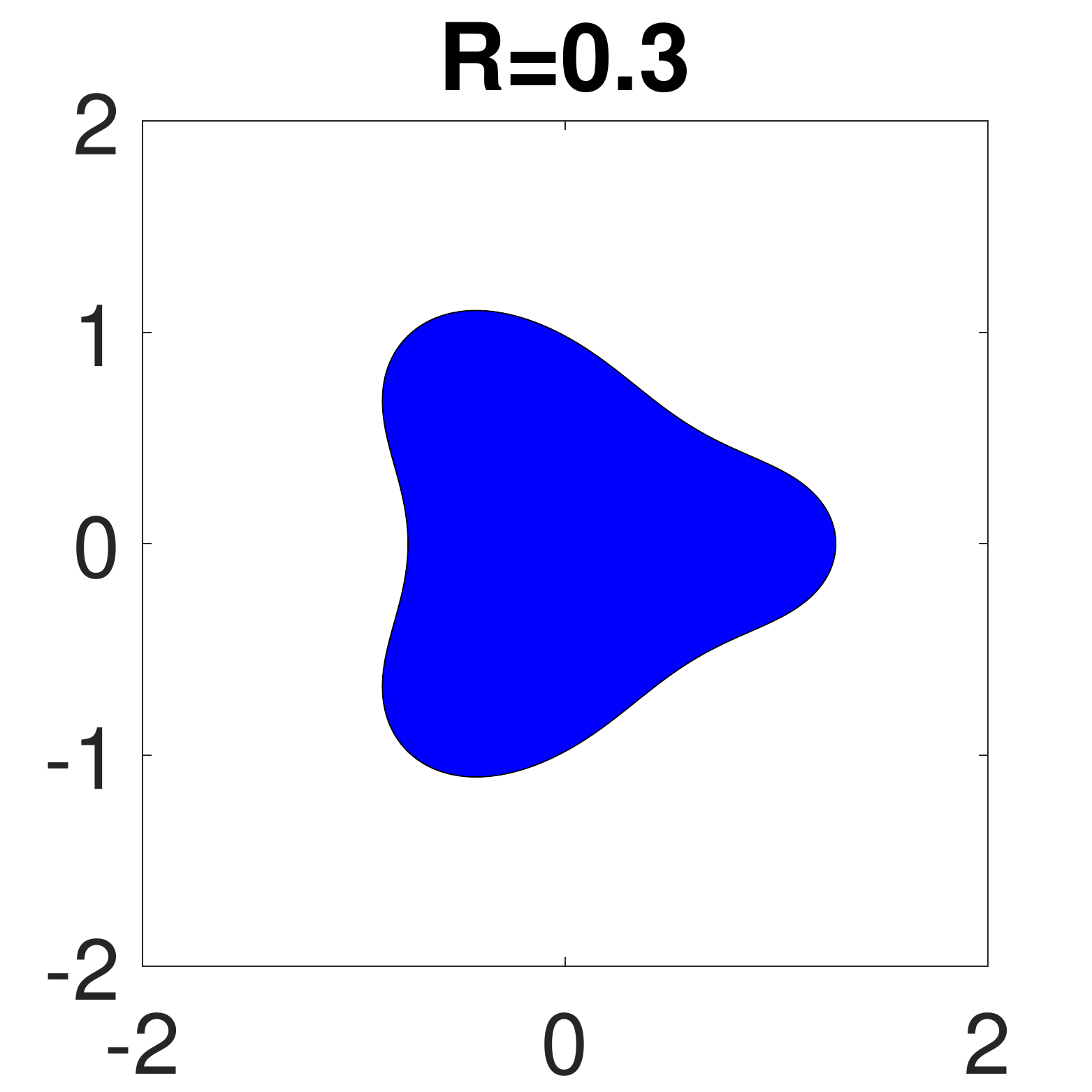}
\includegraphics[scale=0.2]{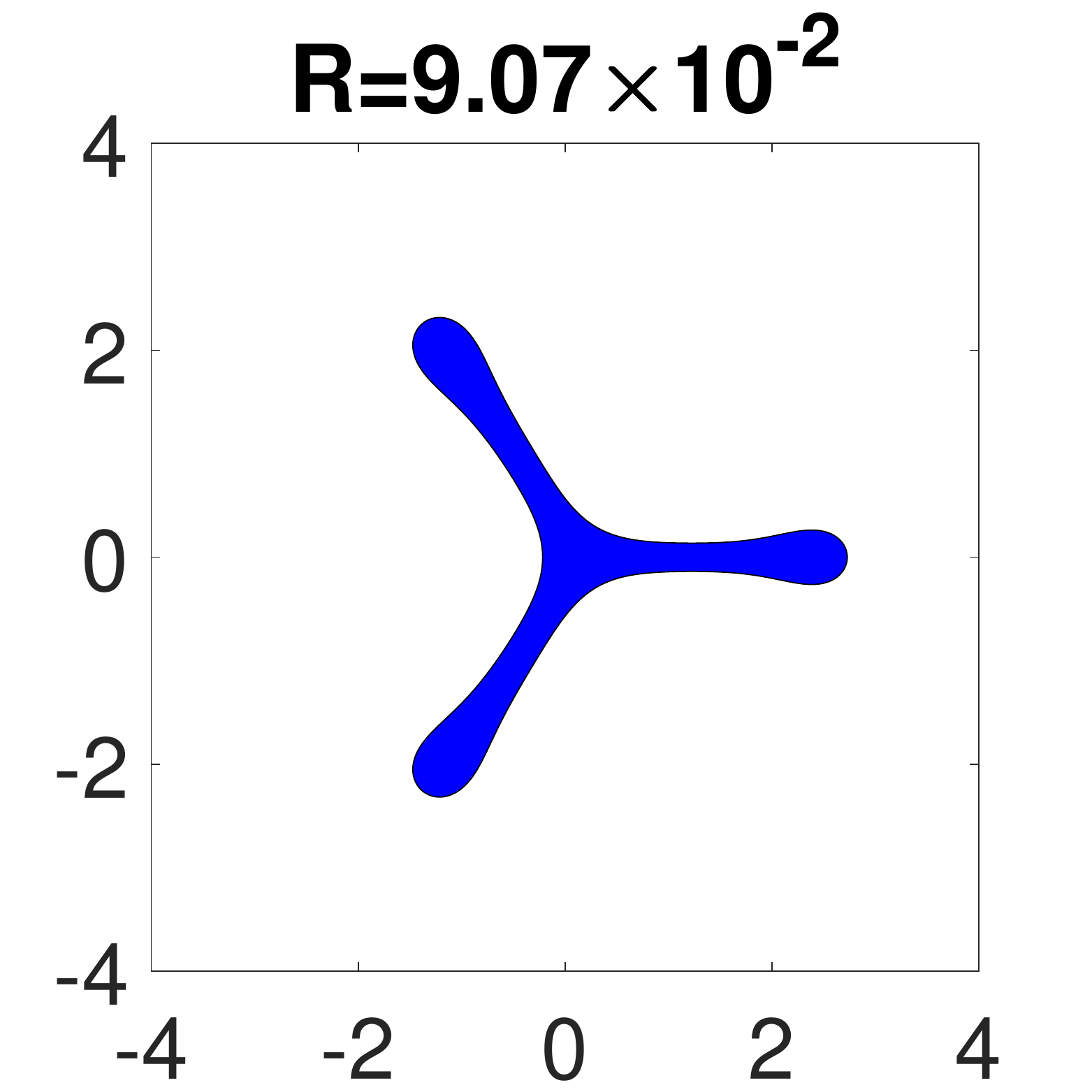}
\includegraphics[scale=0.2]{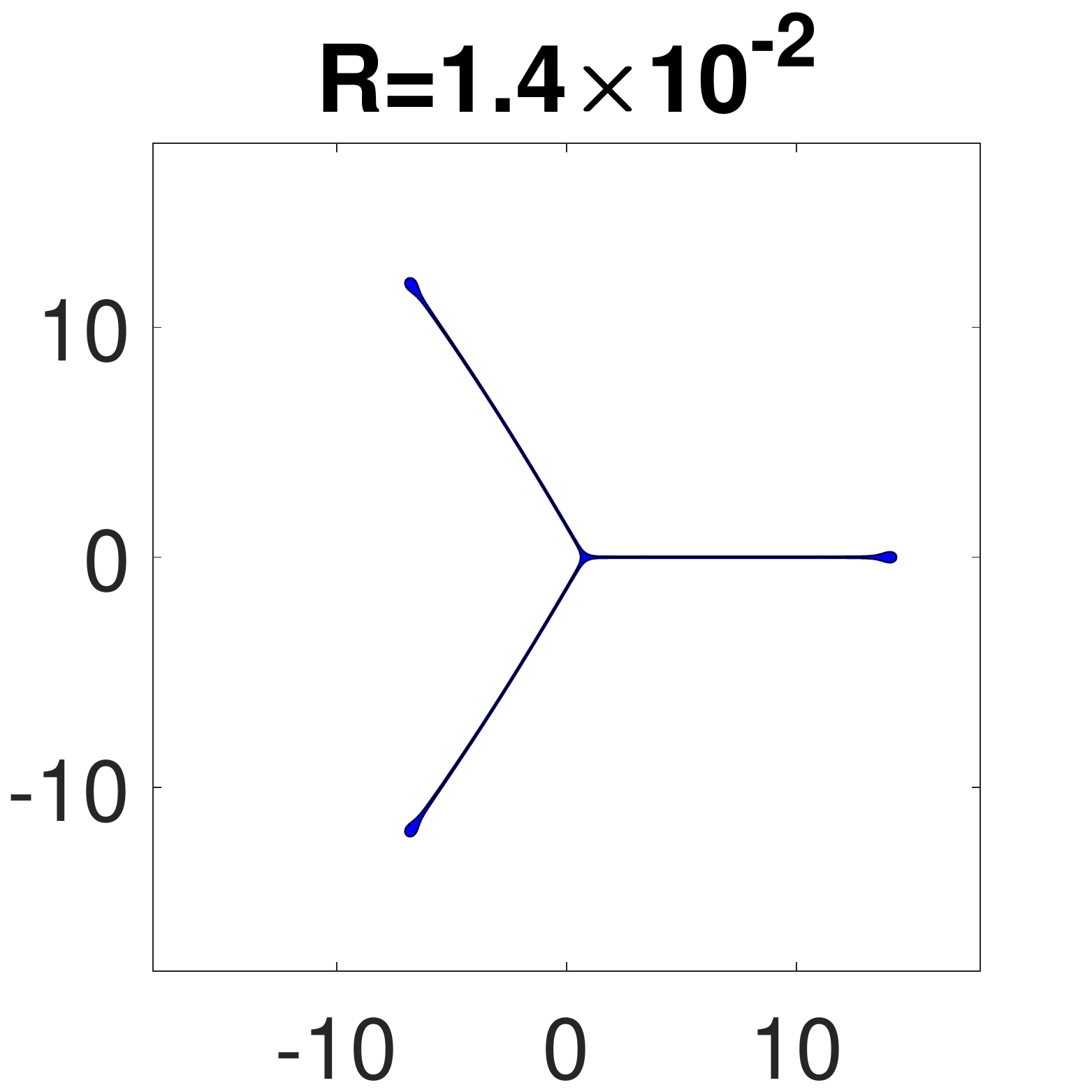}[a]\\
\includegraphics[scale=0.2]{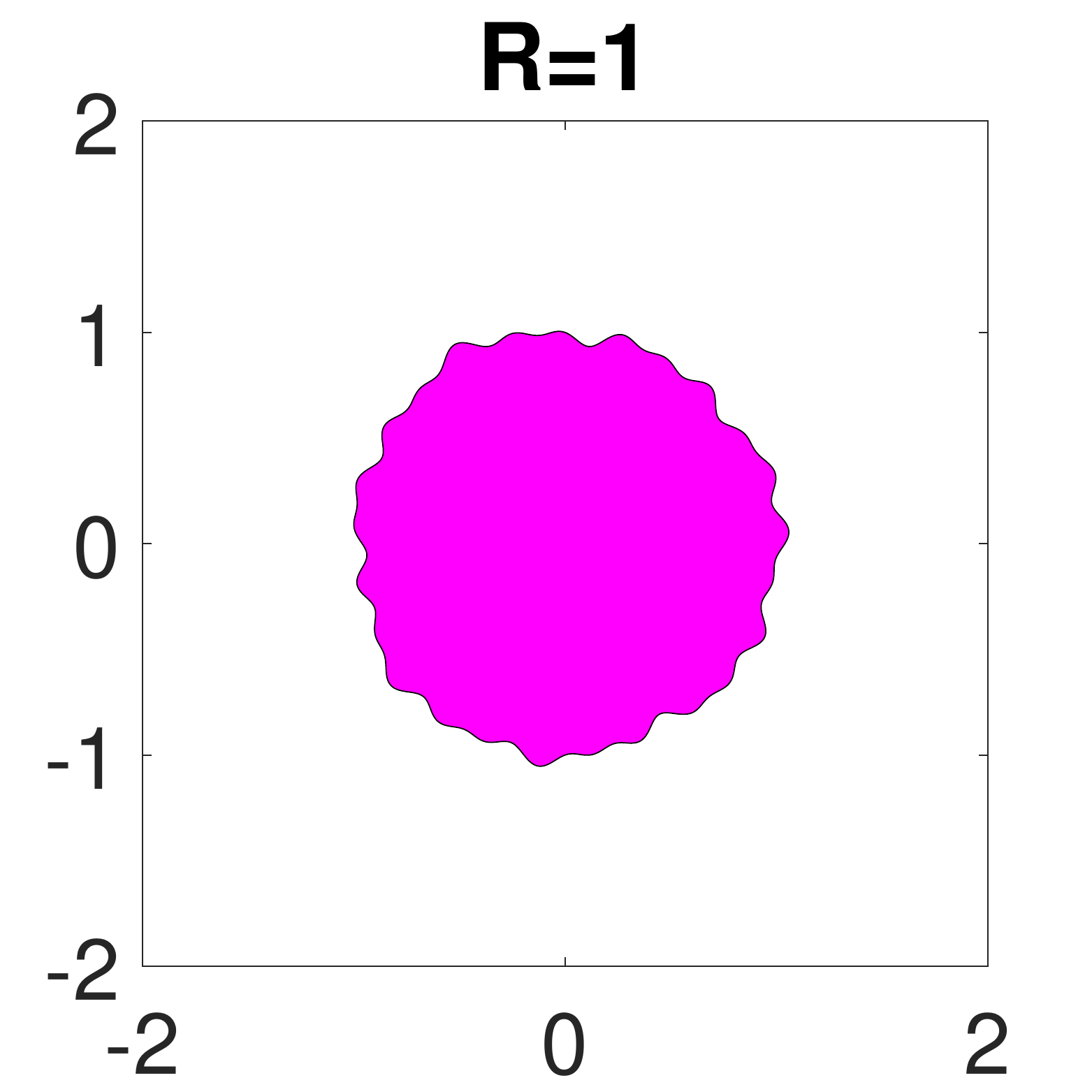}
\includegraphics[scale=0.2]{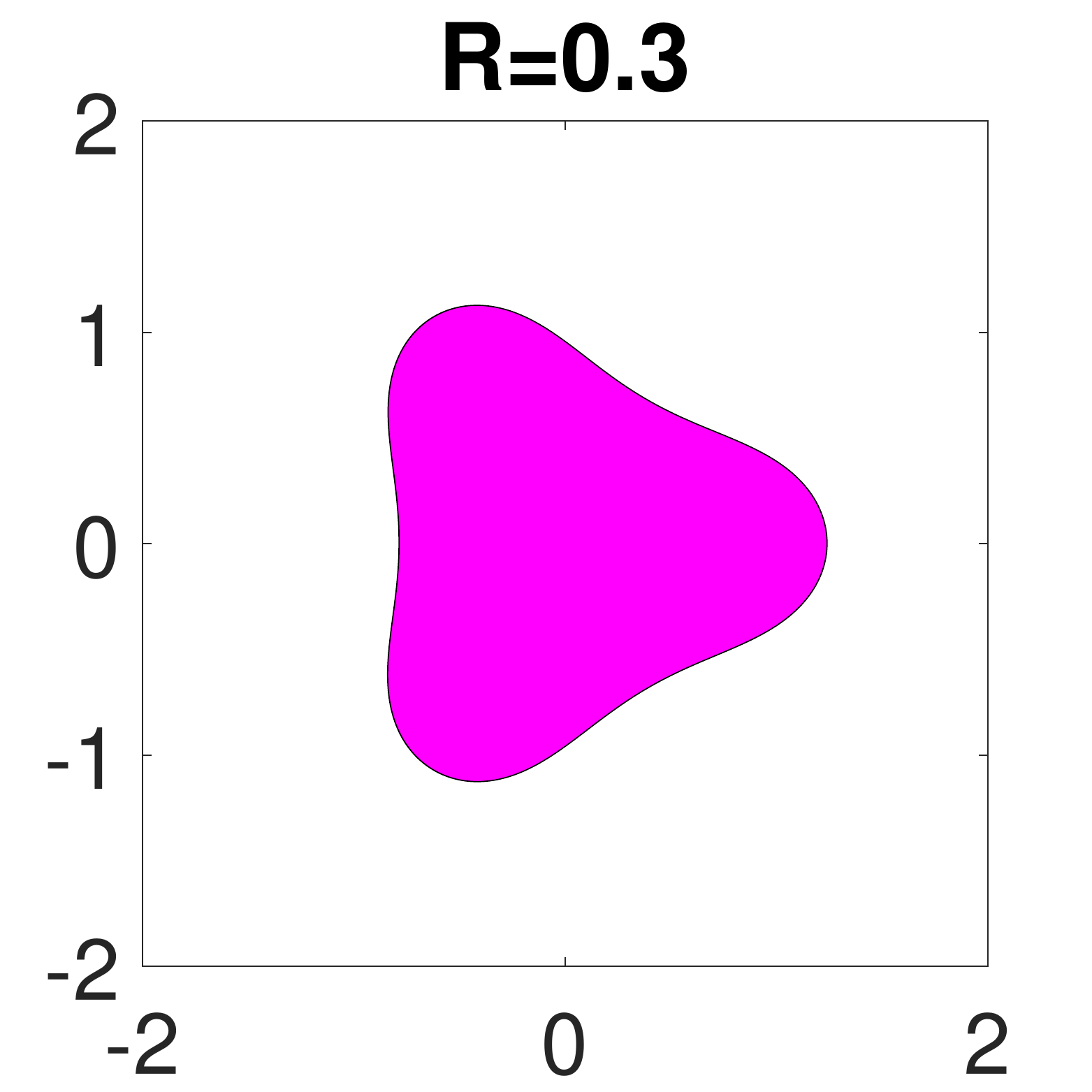}
\includegraphics[scale=0.2]{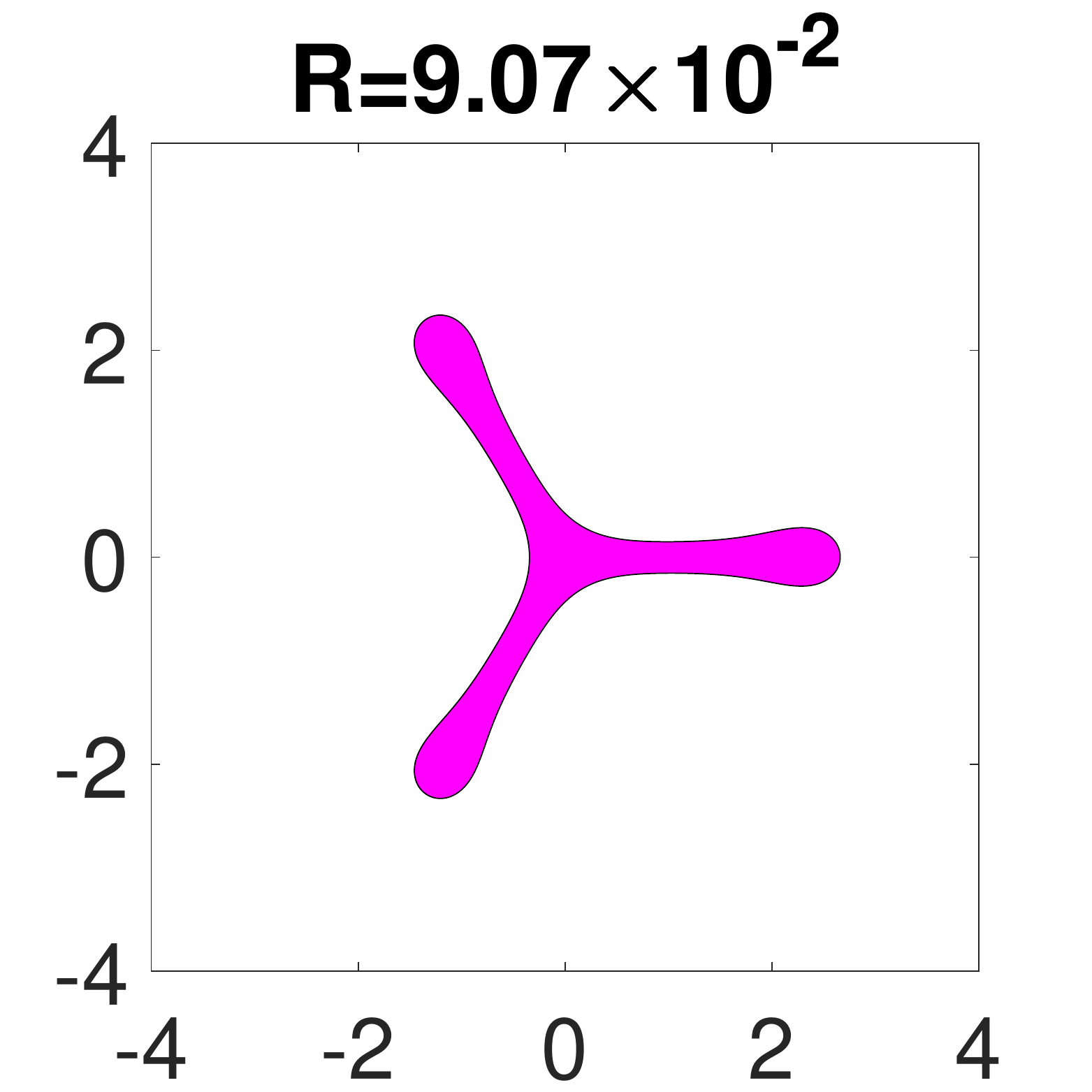}
\includegraphics[scale=0.2]{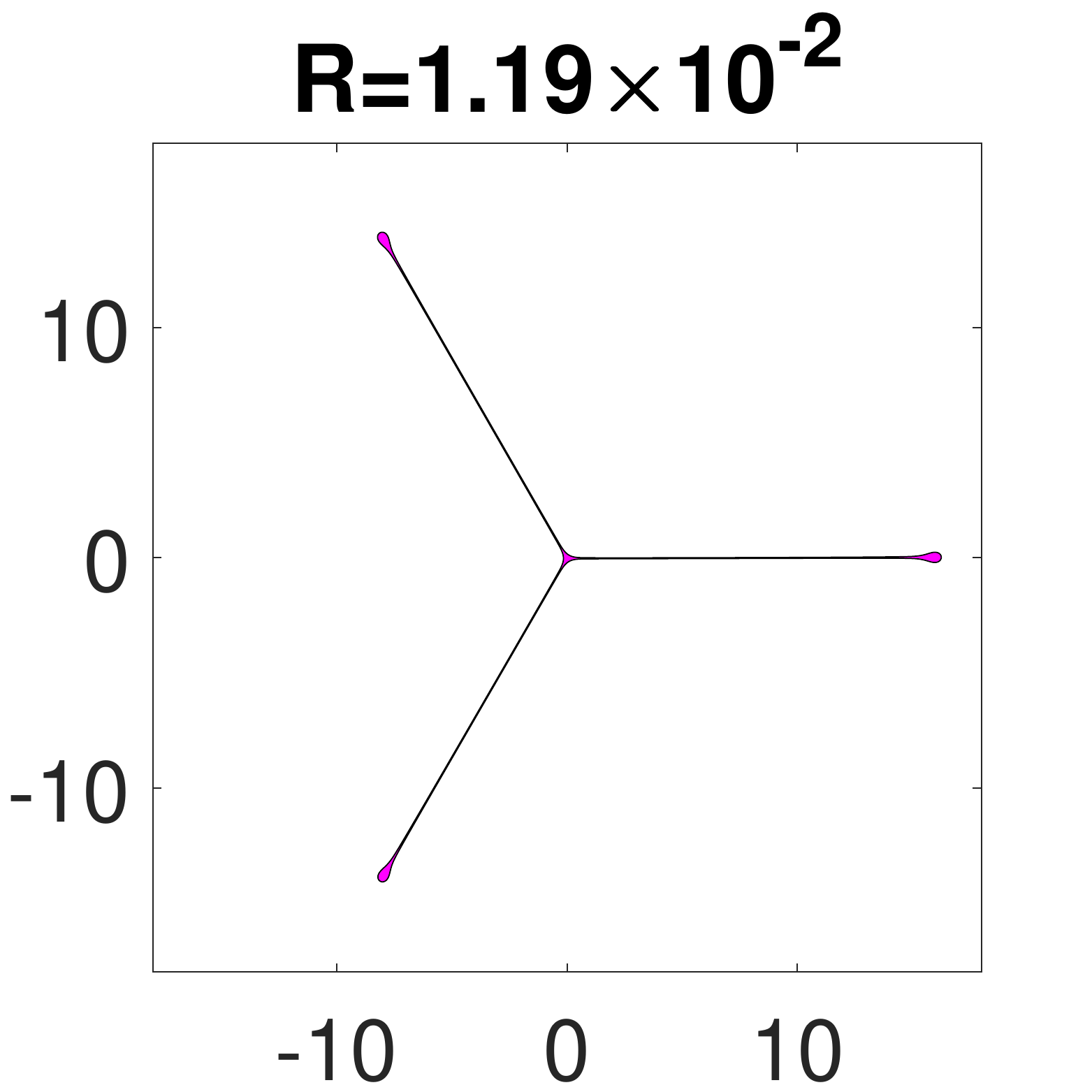}[b]\\
\includegraphics[scale=0.2]{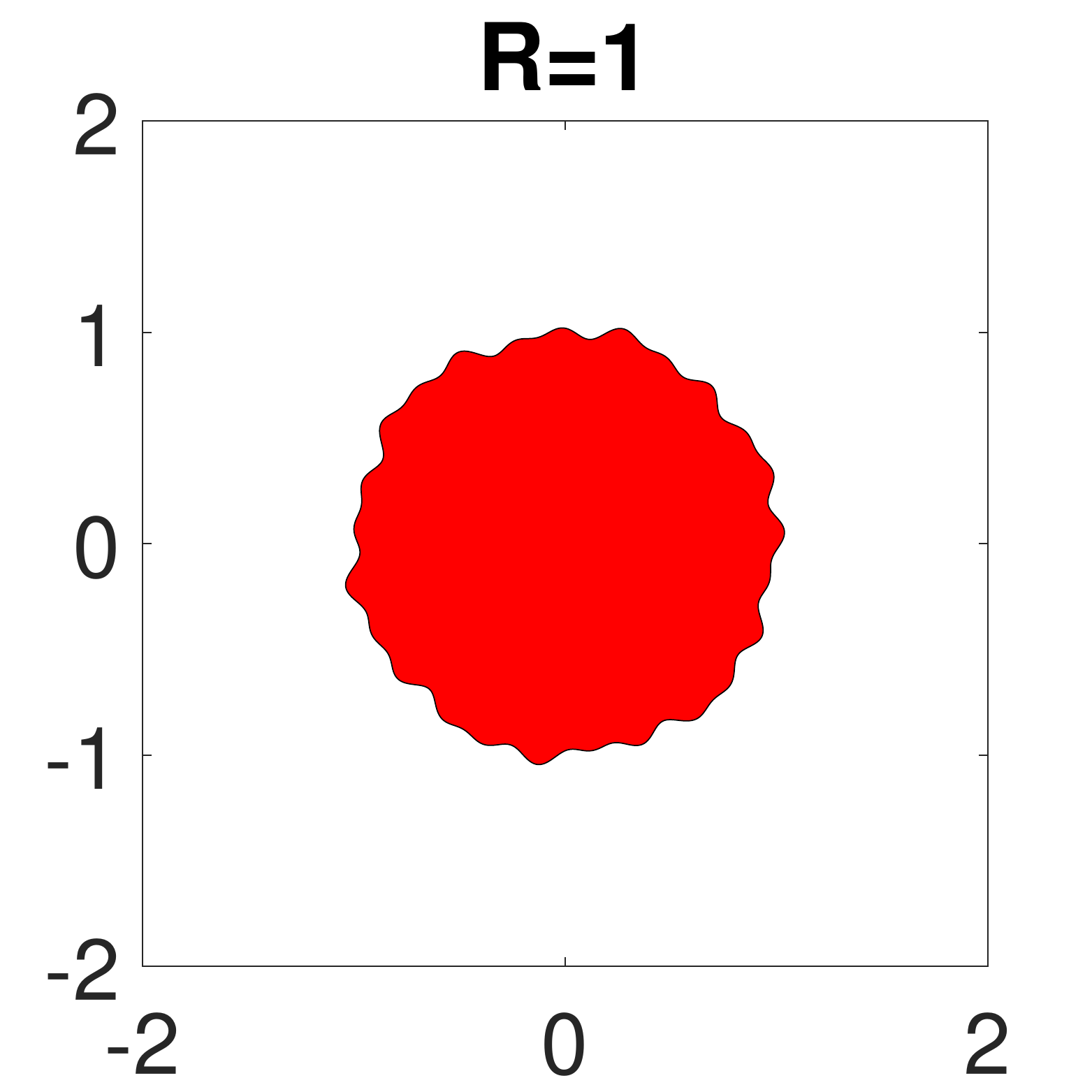}
\includegraphics[scale=0.2]{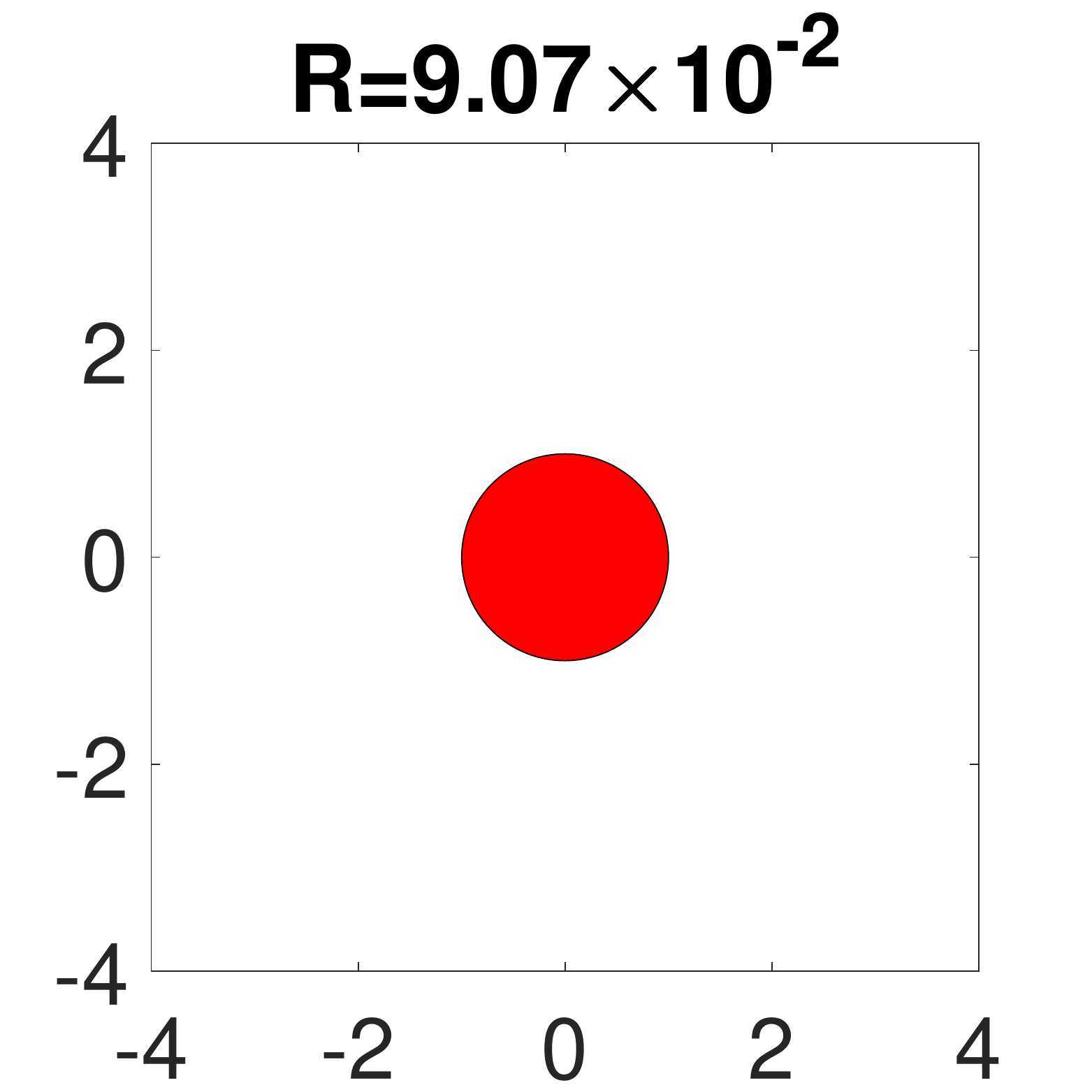}
\includegraphics[scale=0.2]{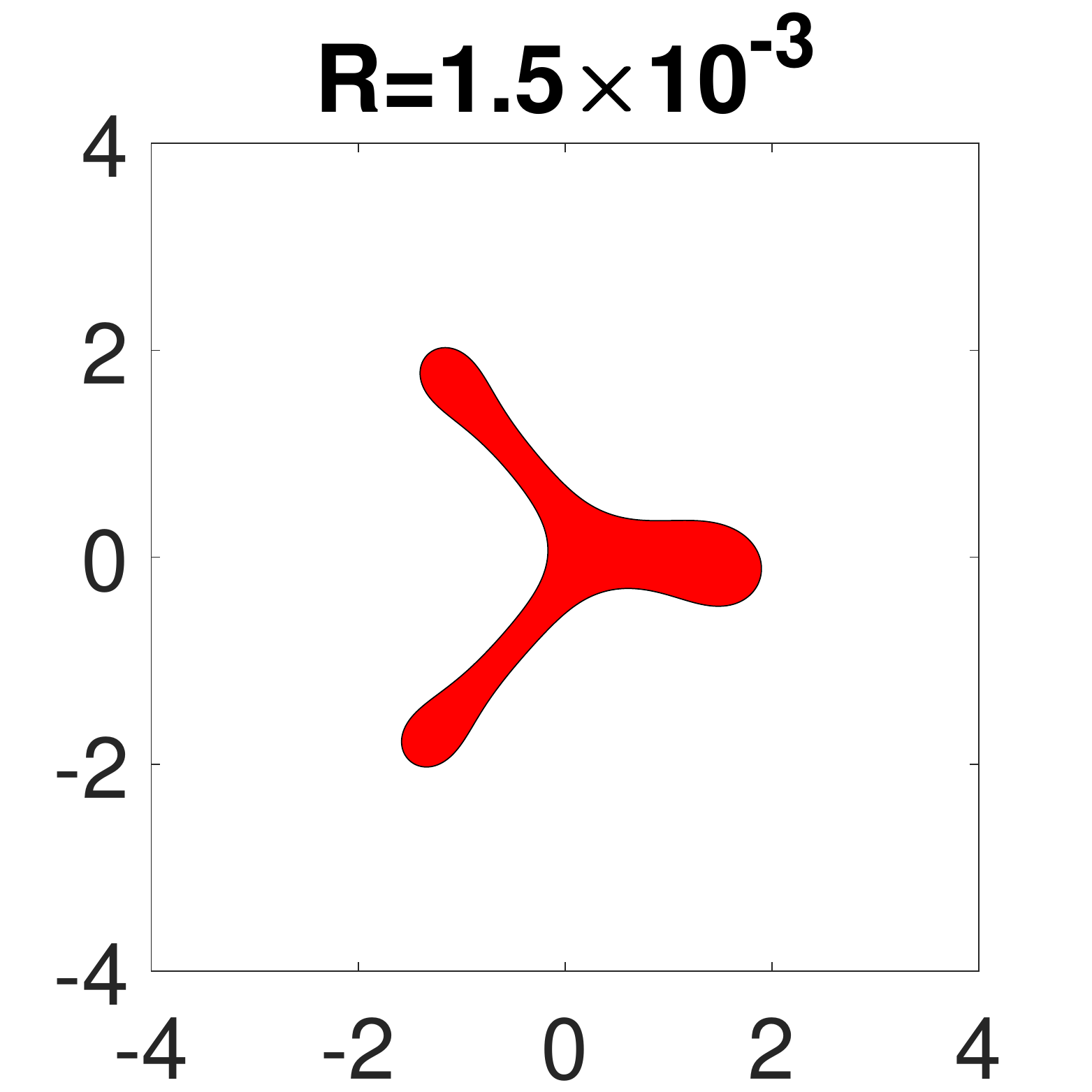}
\includegraphics[scale=0.2]{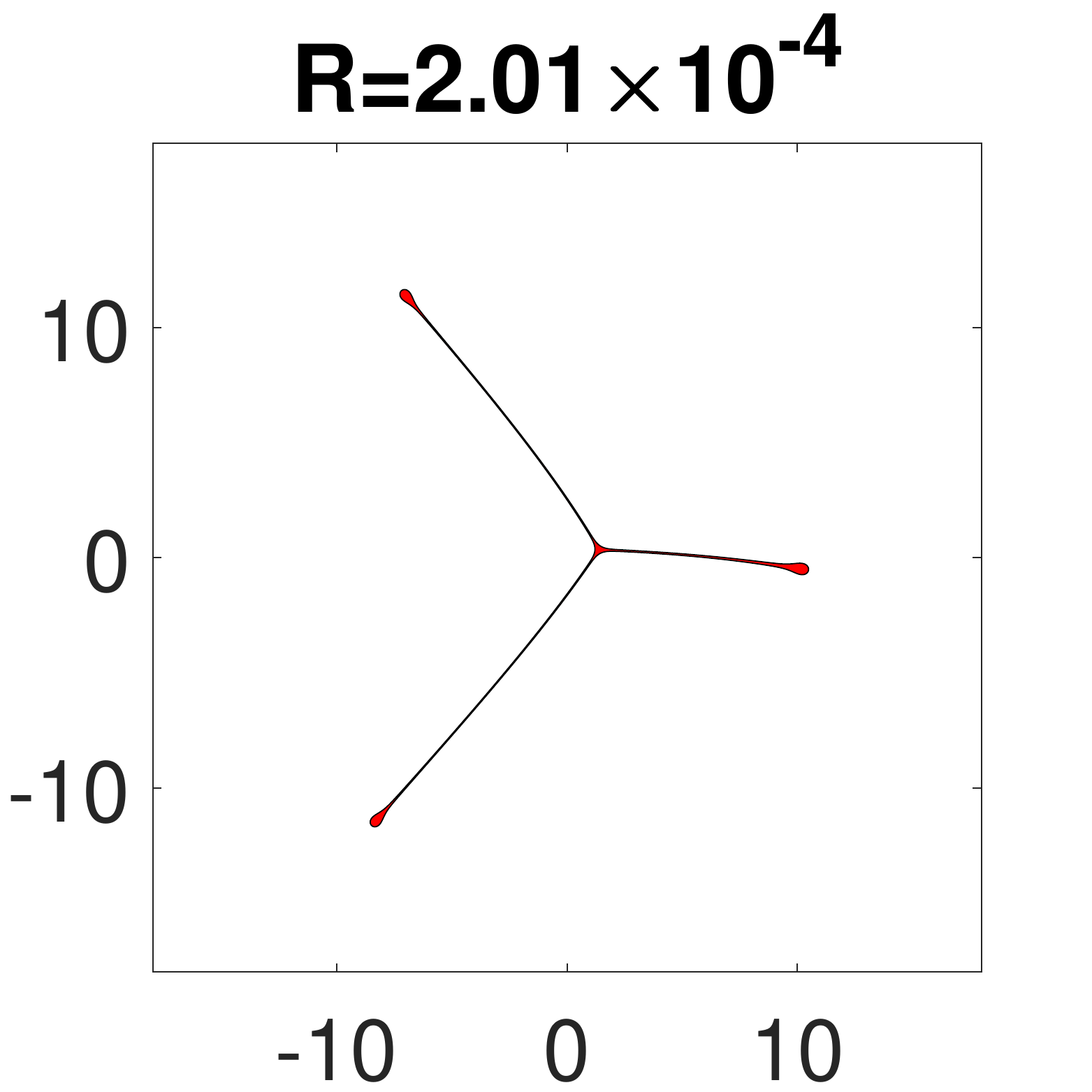}[c]
\caption{The drop dynamics in the rescaled frame under the special gap $\displaystyle b_{\mathcal{C}}=(1-\frac{7}{2}\tau\mathcal{C}t)^{-2/7}$ with $\mathcal{C}=52$, where mode 3 grows fastest according to linear theory. Here we take the nondimensional surface tension $\tau=1\times 10^{-4}$. [a] The evolution under the initial condition  $\displaystyle r(\alpha,0)=1+0.02(\cos(3\alpha)+\cos(5\alpha)+\cos(6\alpha))$; [b] The evolution under the initial condition $\displaystyle r(\alpha,0)=1+0.02(\cos(3\alpha)+\sin(7\alpha)+\cos(15\alpha)+\sin(25\alpha))$; [c] The evolution under the initial condition $\displaystyle r(\alpha,0)=1+0.02(\sin(6\alpha)+\cos(15\alpha)+\sin(25\alpha))$. In all cases, the limiting shape tends to a 3-fold symmetric, one-dimensional web-like structure.}\label{lpss3dyn}
\end{suppfigure}

In the main manuscript, we have shown the length of the neck region tends to be finite. Consequently, the lengths of the long filaments in the rescaled frame increase as the drop sizes decrease. In Fig. S\ref{Exmapleof3}[a], we show the drop dynamics in the rescaled frame from the initial shape $\displaystyle r(\alpha,0)=1+0.02(\cos(3\alpha)+\cos(5\alpha)+\cos(6\alpha))$ and the special gap $\displaystyle b_{\mathcal{C}}=(1-\frac{7}{2}\tau\mathcal{C}t)^{-2/7}$ with $\mathcal{C}=52$, where mode 3 grows fastest linearly. As the drop shrinks, the interface tends to form a bud at the far end and a smooth finger tip at the inner end, connected by a long filament. The inset, which focuses on the boxed region, shows that the filaments get thinner and longer as the size of the drop decreases. We also find the perimeter of the interface $P$ decreases linearly in $R$ to a finite length as $R$ decreases. As the gap width is changed to select other symmetry modes of the limiting shapes, the corresponding interface perimeters still decrease linearly in $R$ but with different slopes (Fig. S\ref{Exmapleof3}[b]). The slopes decrease as the symmetry of the interface increases. In this case, the initial condition for each interface is $\displaystyle r(\alpha,0)=1+0.02\cos(k\alpha)$ where $k=$ 3, 4, 5, 6, and 7. The corresponding (special) gap dynamics are $\displaystyle b_{\mathcal{C}}=(1-\frac{7}{2}\tau\mathcal{C}t)^{-2/7}$ with $\mathcal{C}=2(3k^2-1)$. As expected, the morphology of each drop is dominated by mode $k$. Comparing the slope of the 3 mode symmetry here and the slopes in Fig. 3[c] in the main manuscript, we find that the slopes are close and differences are likely due to the initial condition, which is different. 

In Figs. S\ref{Exmapleof3}[c] and [d] we show the maximum (minimum) curvature at the far end (the inner end) in the rescaled frame for the drop interfaces shown in Fig. \ref{lpss3dyn}. We fit the curvatures as a power function of $R$ (solid line), which indicates the curvatures tend to a finite number as $R\rightarrow 0$. While the minimum curvature at the inner end may not be the same for different initial conditions, the maximum curvature at the far end tends to approach the same limit $\bar{\kappa}_*\approx 5.8$. This suggests that the buds at the far end may acquire a universal shape that is independent of the initial conditions. From Fig. 3[b] and Fig. 4[a] in the main manuscript, we see that the length of the filaments approaches a finite number. Approximating the filaments as a rectangle with a circular tip with radius $\bar{\kappa}_*/R$, the perimeter $P\sim P_0+aR$, where $a$ is a constant related to $\bar{\kappa}_*$ and $P_0$ is a finite number depending on the initial shape. This explains why the perimeter decreases as a linear function of $R$, as seen in Fig. \ref{Exmapleof3}[b].

\begin{suppfigure}
\includegraphics[scale=0.4]{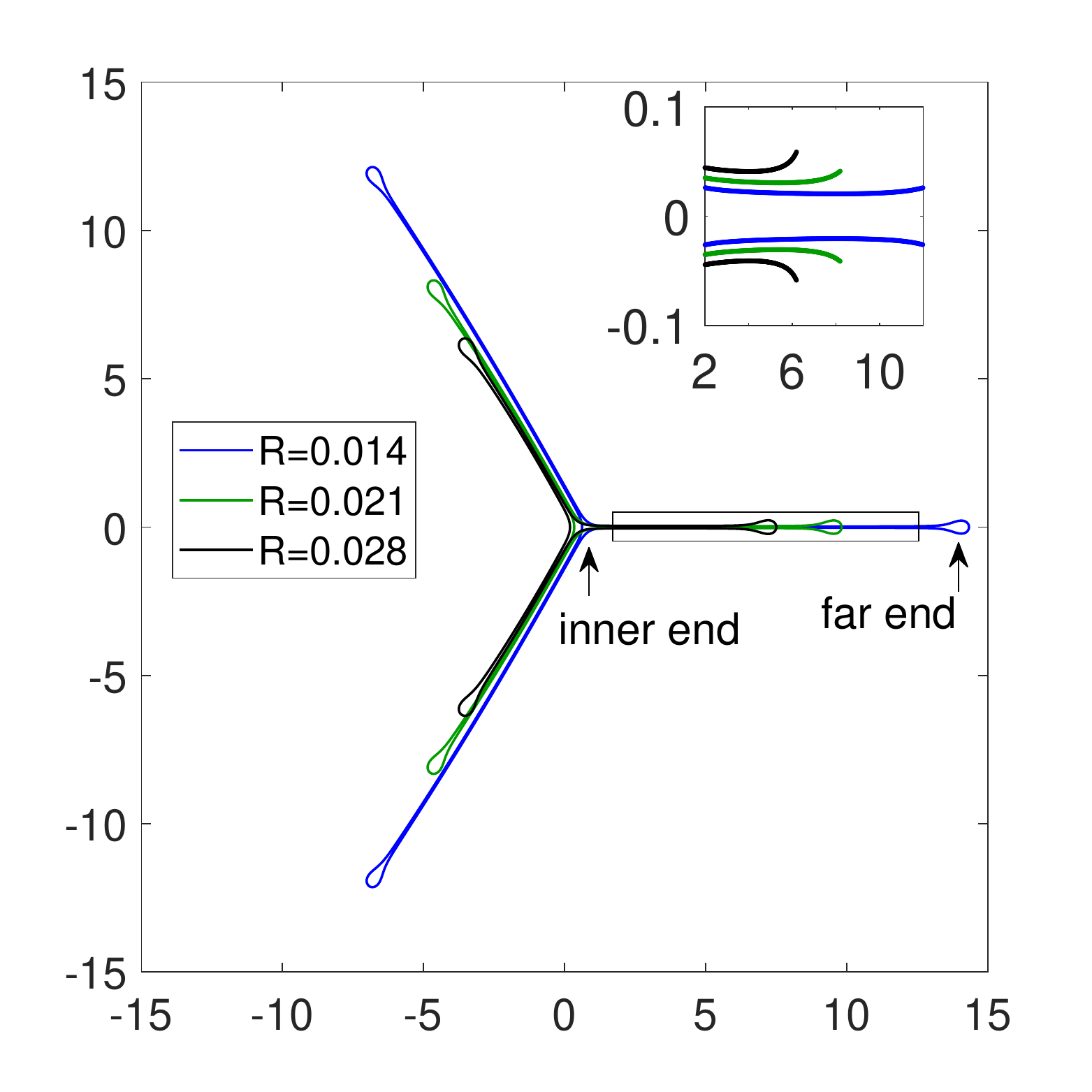}[a]
\includegraphics[scale=0.4]{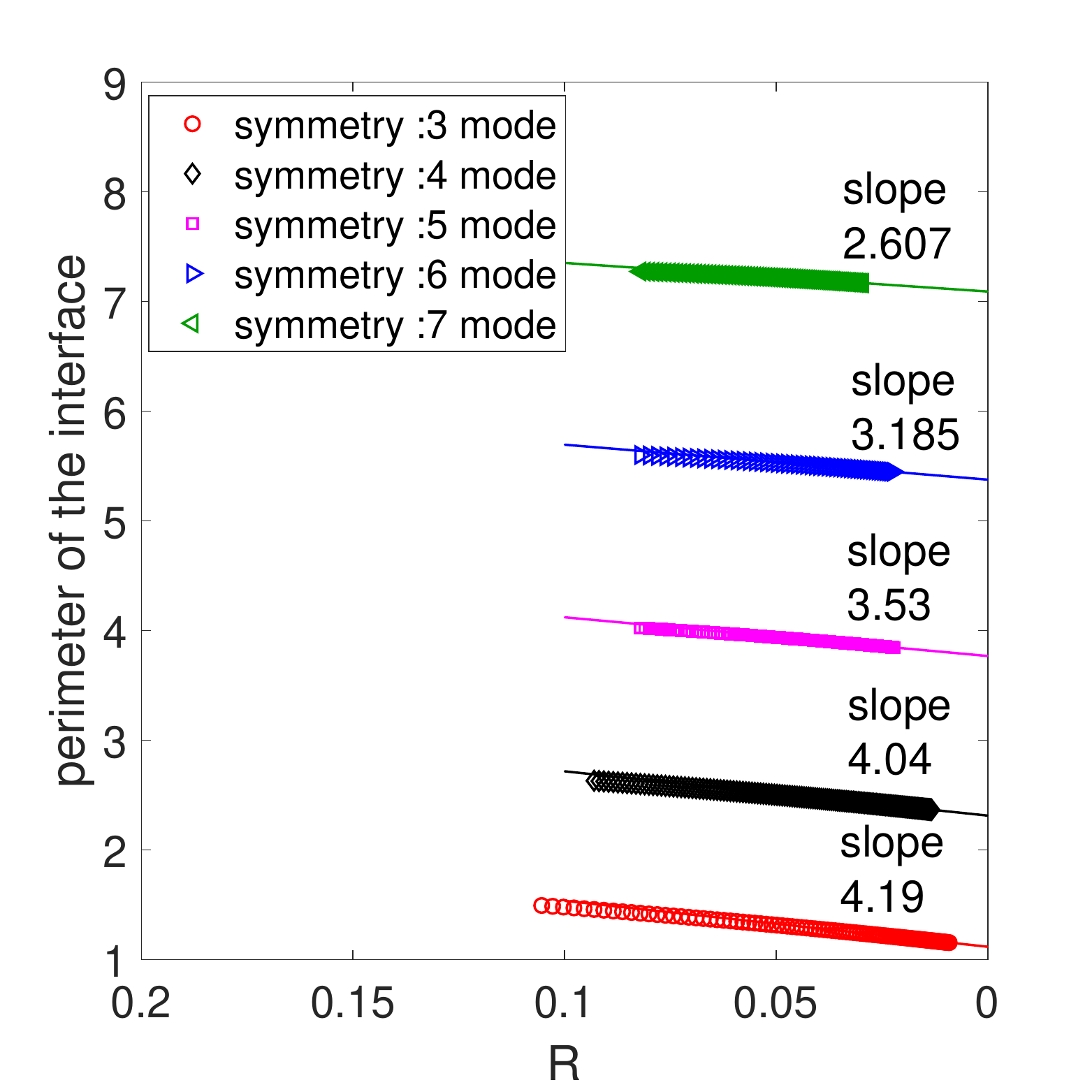}[b]\\
\includegraphics[scale=0.4]{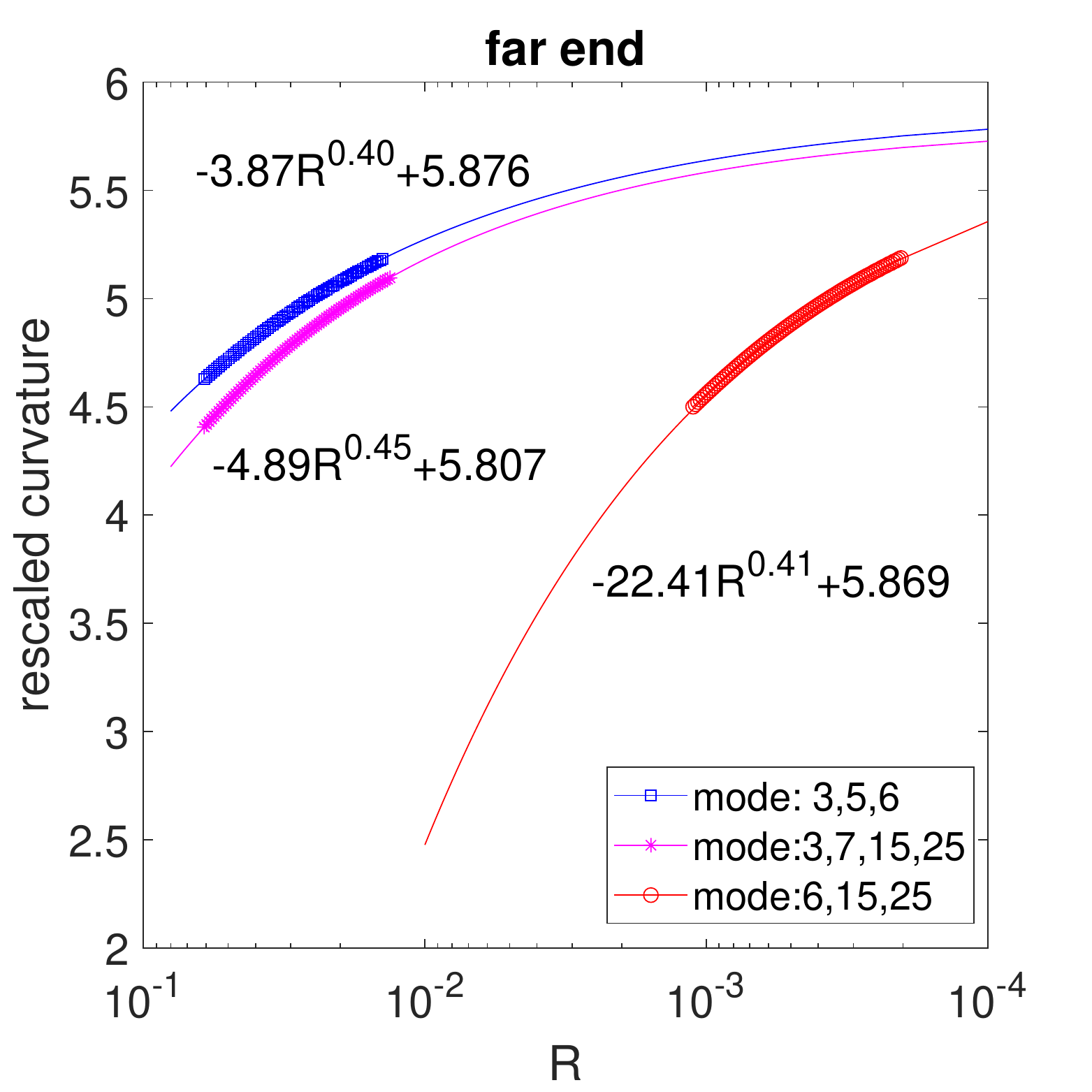}[c]
\includegraphics[scale=0.4]{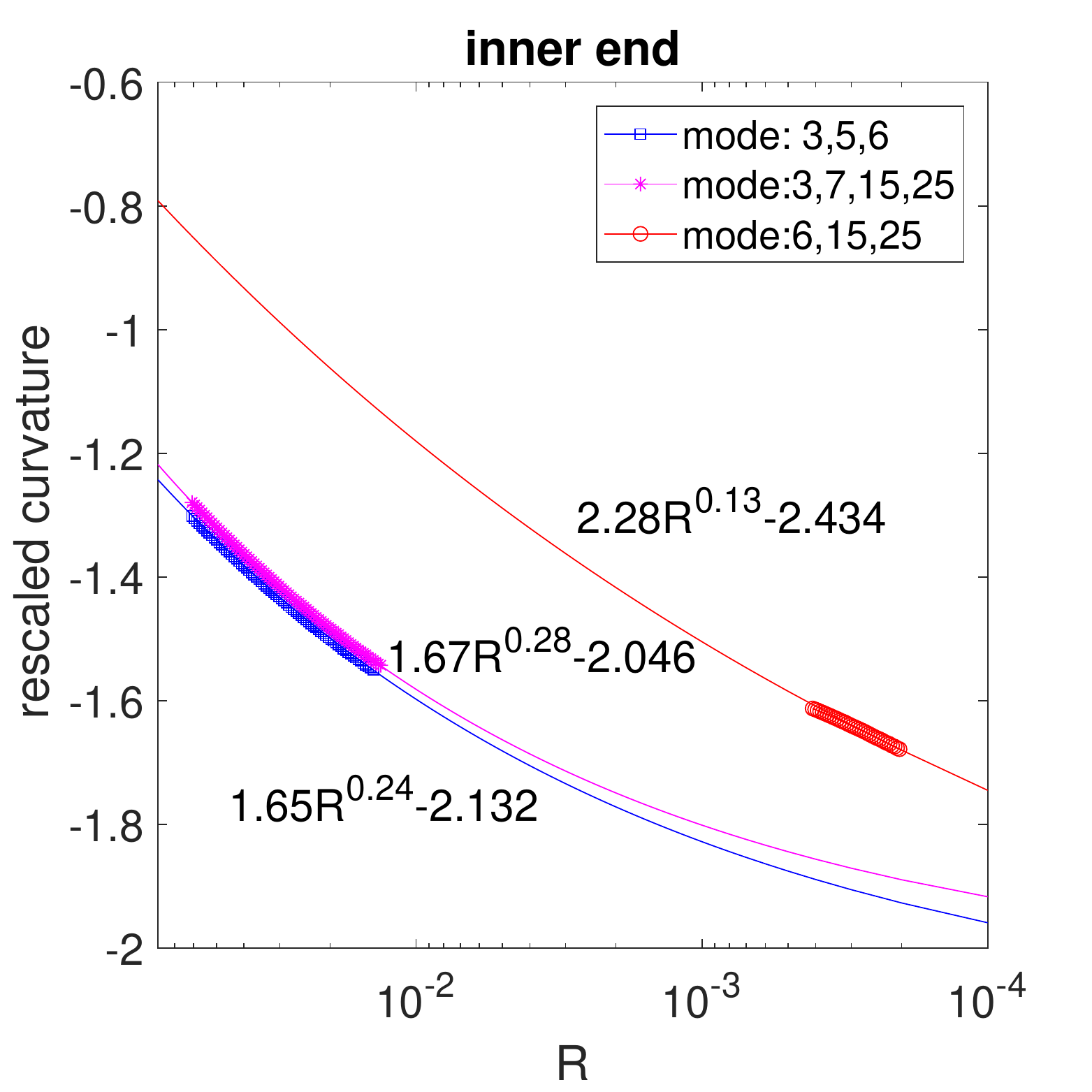}[d]
\caption{Details of the drop morphologies. [a]  The drop dynamics in the rescaled frame near the vanishing time and corresponds to Fig. 3[b] in the main text, which shows the drop morphologies in the original frame. The inset shows a blow-up of the neck region in the boxed region. [b] The relation between the perimeter of the interface $P$ and the effective radius $R$ for different symmetric limiting shapes (see text for details) using the special gap $\displaystyle b_{\mathcal{C}}=(1-\frac{7}{2}\tau\mathcal{C}t)^{-2/7}$, where $\tau=1\times 10^{-4}$, $\mathcal{C}=2(3k^2-1)$ and $k$ varies as 3, 4, 5, 6, and 7. [c] The maximum curvature of the far end in the rescaled frame with respect to $R$. [d] The minimum curvature of the inner end in the rescaled frame with respect to $R$.}\label{Exmapleof3}
\end{suppfigure}

\section{Active lower modes under the special gap dynamics $b_{\mathcal{C}}(t)=(1-\frac{7}{2}\tau \mathcal{C}t)^{-\frac{2}{7}}$}
As we have shown in the manuscript, the dominant mode of the morphology is solely determined by the control parameter $C$ if the initial shape only contains high modes,
\begin{equation}
r(\alpha,0)=1+\epsilon\Sigma_{k=k_{min}}^{k_N}e^{-\beta k}(a_k\cos(k\alpha)+b_k\sin(k\alpha)),
\label{IC}
\end{equation}
where $\epsilon=2.5\times 10^{-3}$, $\beta=0.2$, $k_{min}=30$, $k_{max}=60$, and $a_k$ and $b_k$ are uniformly distributed in $(-1, 1)$. 
 All these high modes are stable and tend to vanish. However, lower modes are created by nonlinear interactions and depending on $\mathcal{C}$, a particular mode $k_{max}$ will have the fastest growth rate and will finally dominate the morphology of drop. This is still true if lower modes are present in the initial condition. These modes may be active in the sense that they may grow due to the Saffman-Taylor instability. In Fig. S\ref{lowermode}, we report the results with lower modes present in the initial condition. Using the gap as $b_{\mathcal{C}}(t)=(1-\frac{7}{2}\tau \mathcal{C}t)^{-\frac{2}{7}}$, where $\mathcal{C}=484$, linear theory suggests mode 9 grows fastest and our simulations show a 9-fold shape if the initial shape only contains modes 30 to 60 (first row). If the initial conditions contain low modes, the results can be different. For example, simulations show that a 13-fold shape emerges when modes from 13 to 60 are present initially(Fig. S\ref{lowermode}, second row) and a 5-fold shape emerges when the initial data contains modes from 5 to 60 (S\ref{lowermode}, third row). The reason is these low modes are active all the time and their initial perturbations are much larger than the perturbation of mode 9 because of the form of the initial condition in Eq. (\ref{IC}). For instance, in the second row the initial magnitude of mode 13 is about $1.857\times 10^{-4}$ while mode 9 does not appear in the initial shape. For the third row the initial magnitude of mode 5 is about $9.19\times 10^{-4}$, which is about as twice as the magnitude of mode 9 initially. Even though mode 9 grows faster than all the other modes, there is not enough time for mode 9 to dominate the shape. To demonstrate this, we add mode 9 to these initial shapes such that mode 9 has the same magnitude as that of the smallest mode. Figure S\ref{lowermode2} shows that in this case, mode 9 is selected and morphologies tend to a 9-fold dominant shape.
\begin{suppfigure}
\includegraphics[scale=0.5]{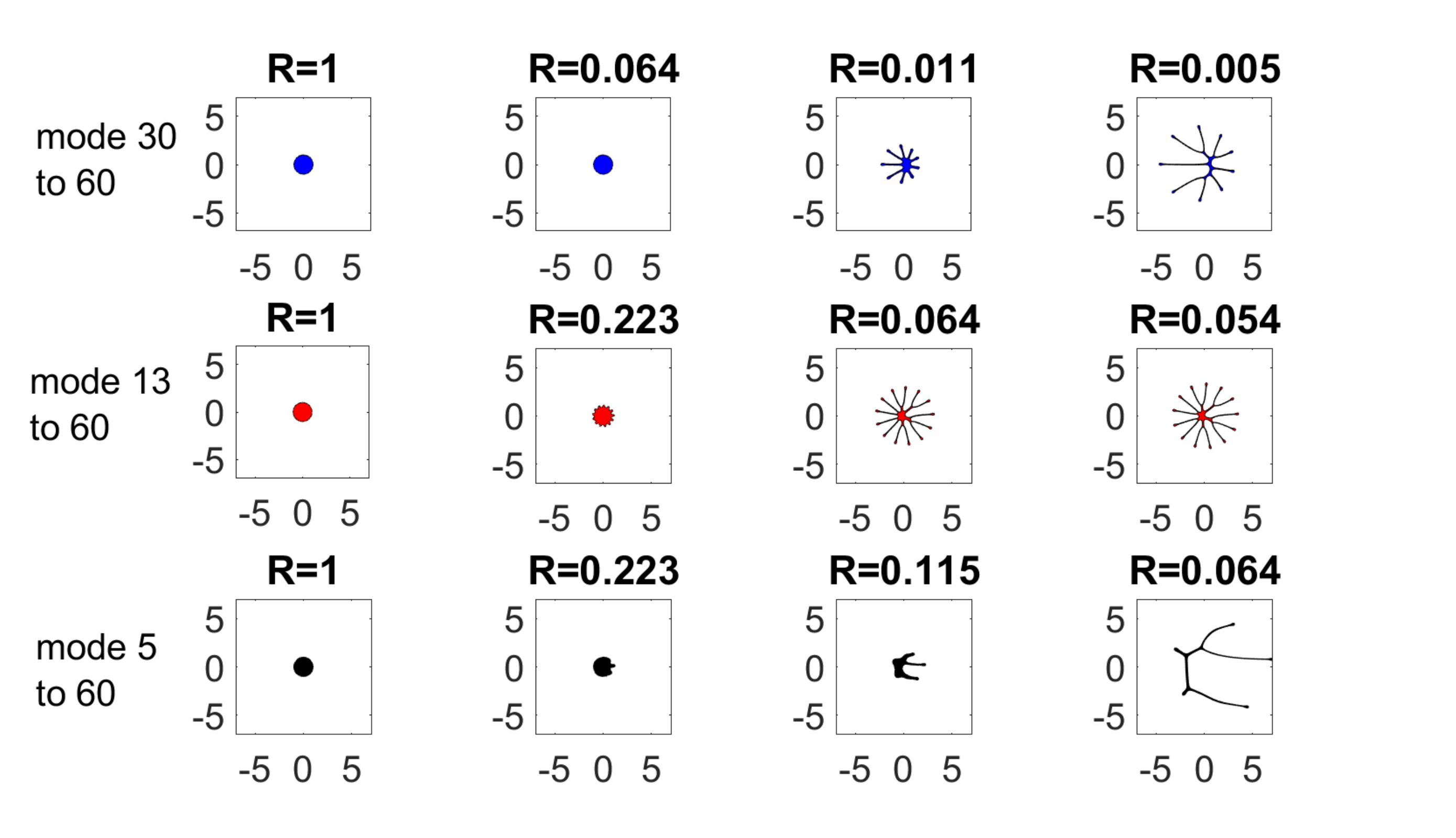}
\caption{The drop morphologies using the special gap $b_{\mathcal{C}}(t)=(1-\frac{7}{2}\tau \mathcal{C}t)^{-\frac{2}{7}}$ with
$\mathcal{C}=484$, which makes mode 9 the fastest growing mode from linear theory. When the initial shape does not contain lower unstable (active) modes (e.g., modes 30 to 60), the interface develops a 9-fold dominant shape (top row). When there exist lower active modes, the interfaces develop a 13-fold shape (when modes 13 to 60 are present in the initial condition) and a 5-fold shape (when modes 5 to 60 are contained in the initial condition). See text for details.
}\label{lowermode}
\end{suppfigure}

\begin{suppfigure}
\includegraphics[scale=0.6]{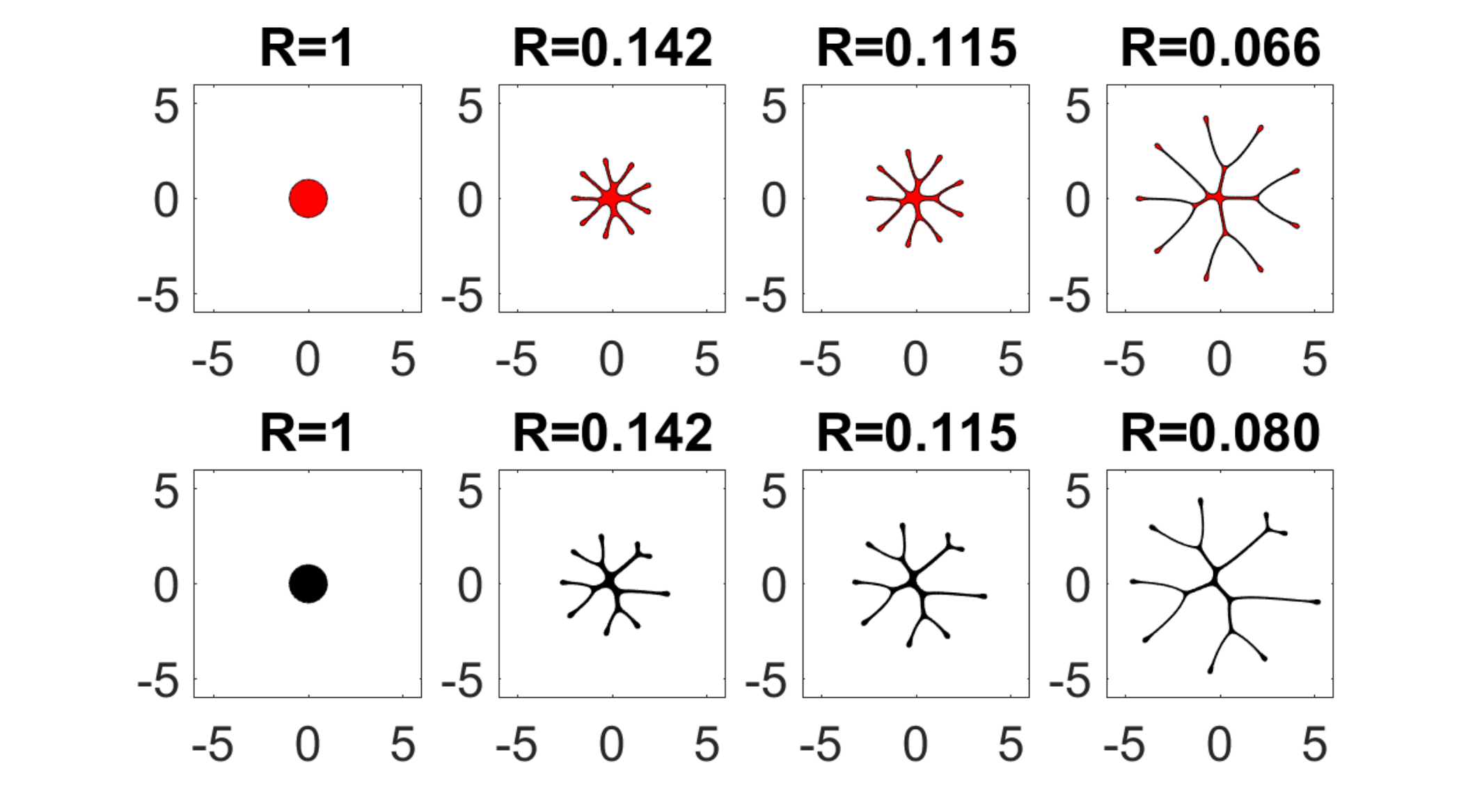}
\caption{The drop morphologies using the special gap $b_{\mathcal{C}}(t)=(1-\frac{7}{2}\tau \mathcal{C}t)^{-\frac{2}{7}}$ with
$\mathcal{C}=484$ (mode 9 is the fastest linearly growing mode) but with initial conditions that are the same as in Fig. S\ref{lowermode}(second row) and (third row) except that in each case, mode 9 has been added into the initial condition with the same magnitude as modes 13 (first row) and 5 (second row), respectively. In both cases the morphologies are dominated by mode 9 as the drop shrinks. }\label{lowermode2}
\end{suppfigure}

\section{Other choices of gap dynamics}

According to linear theory given in the main text), the critical gap width is determined by the relation that $\dot b(t)/b(t)^{9/2}$ is a constant in time (see Eq. ({2.9}). Therefore, if $\dot b(t)/b(t)^{9/2} \to 0$ as time increases, the drop should eventually shrink like a circle, even if the gap width blows up at a finite time. For example, $\dot b(t)/b(t)^{9/2} \to 0$ when $b(t)=1+t$, $b(t)=e^t$ or even $\displaystyle b_\mathcal{D}(t)=(1-3\tau\mathcal{D}t)^{-1/3}$, for any $\mathcal{D}$. In the latter case, the rate of blow up is actually larger than that for the special gap $b_\mathcal{C}(t)$. To confirm this behavior, we perform nonlinear simulations using $b_\mathcal{D}(t)$ for different choices of $\mathcal{D}$. The results are shown in Fig. S\ref{slowerss}. We observe that for each case considered, the drop eventually shrinks like a circle after a transient instability, which confirms the predictions of linear theory. The initial condition for each simulation is $\displaystyle r(\alpha,0)=1+0.02(\cos(3\alpha)+\cos(5\alpha)+\cos(6\alpha))$ and the drop morphologies using $\mathcal{D}=100$ are shown as insets. 


\begin{suppfigure}
\center
\includegraphics[scale=0.4]{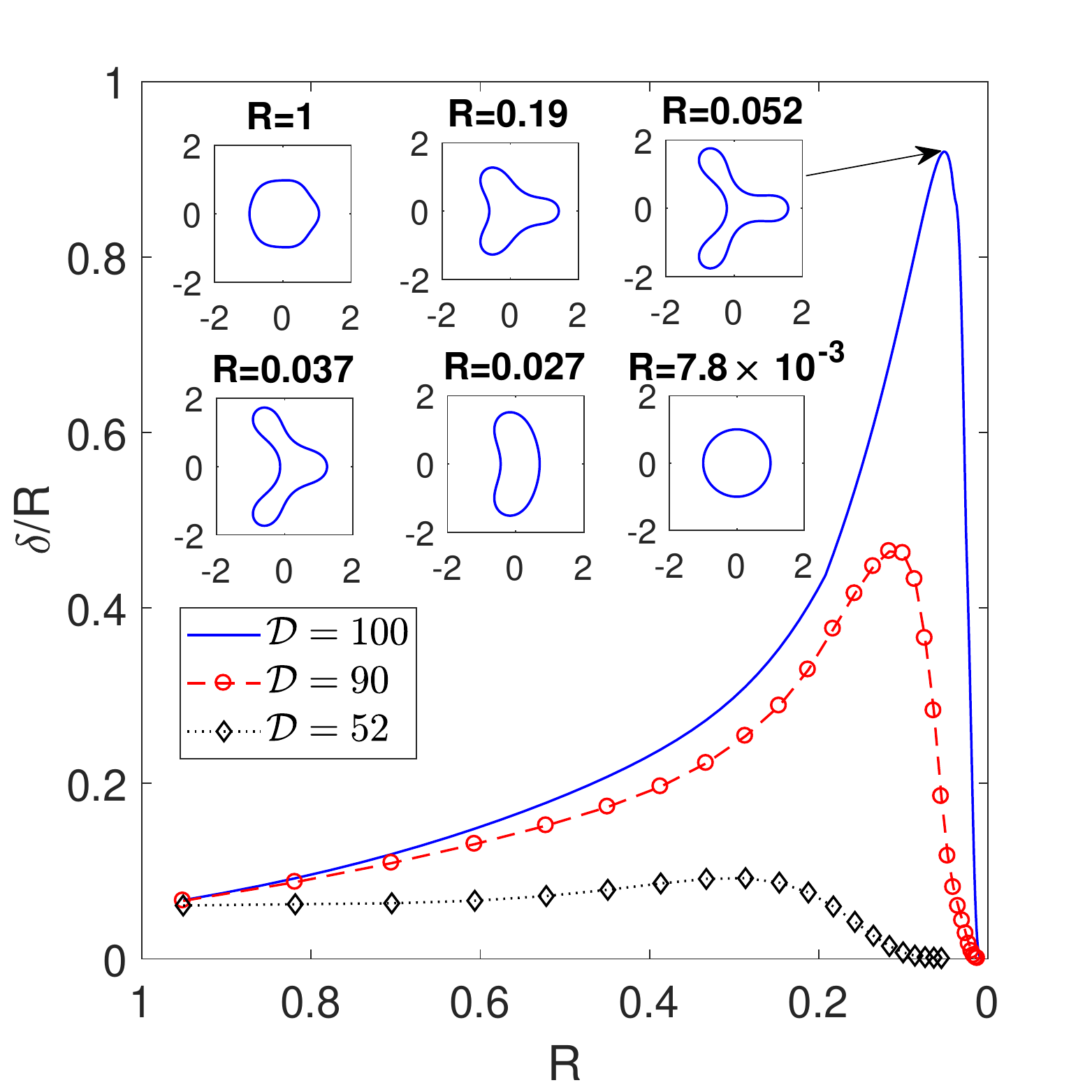}
\caption{The relative perturbation magnitude of drop interfaces, as a function of $R$, using the gap  $\displaystyle b_\mathcal{D}(t)=(1-3\tau\mathcal{D}t)^{-1/3}$ with $\mathcal{D}$ as labeled. Insets show the drop interfaces with $\mathcal{D}=100$ and $\tau=1\times 10^{-4}$. In all cases, the drop shrinks like a circle after a transient instability.
}\label{slowerss}
\end{suppfigure}

\section{Comparison with experimental data under a nonlinear gap width}
We next compare the simulation results with an experiment in which the gap is increased nonlinearly in time.
In the experiment, a thin layer of Canola oil is confined between two parallel plates. We use clear acrylic plates $12'' \times 12'' \times 1/2''$ with an initial gap $b_0=100$ $\mu m$. The upper plate is lifted at two diagonal corners manually. The Canola oil purchased from Great Value has viscosity $\mu=65$ $cP$ and surface tension $\sigma=31.3$ $mN/m$. We use video images to capture the air-oil interface. The contour of the air-oil contact line is found by using IMAGEJ and MATLAB (see Fig. S\ref{lncom4}). However, since the plate is lifted by hand, the gap width may not be uniform in space. Nevertheless, we will assume the gap is spatially uniform in our simulations.

Because the upper plate is lifted manually, the time-dependent gap is unknown and nonlinear, but we are able to reconstruct the gap width from the area enclosed by the contact line $A(\tilde{t})$, where $\tilde t$ is the dimensional time. We find the area of the oil $A(\tilde t)$, fit the data using a smooth spline in MATLAB, and compute $A'(\tilde{t})$ using the spline (See Fig. \ref{simudata}[a]). Since there is fluid leftover on the plate due to a thick wetting layer, we assume that the rate of volume loss of the drop fluid is proportional to the area change
 \begin{equation}
 \frac{1}{V_{ol}}\frac{d V_{ol}}{d \tilde{t}}=e_1A'(\tilde{t}),
 \end{equation}
 where $V_{ol}$ is the volume of the drop and $e_1$ is a positive constant.  The gap is then given as 
 \begin{equation}
 b(\tilde t)=\frac{b_0A(0)}{A(\tilde t)}\exp(e_1(A(\tilde{t})-A(0))).
 \end{equation}
 We have constructed the gap width with different parameters $e_1$ in Fig. S\ref{simudata}[b], where increasing $e_1$ decreases the gap width. 

Taking the time scale to be $\displaystyle T=-\frac{A(0)}{A'(0)}=0.3243$, the nondimensional time is $t=\frac{\tilde{t}}{T}$.
At early times, the oil/air interface is nearly a circle as Fig. S\ref{lncom4}[c]. We find the initial equivalent radius of the interface is $R_0=2.349$ cm and the nondimensional surface tension is $\displaystyle \tau=\frac{\sigma b_0^2}{12\mu R_0^3}T=1.0273\times 10^{-5}$. The experimental results are shown in Fig. S\ref{comparison} (top (first) row) and simulations using different values of $e_1$ are shown in Fig. S\ref{comparison} in rows 2-4. In both the experiments and simulations, 
there is a transient instability as fingers are generated and grow towards the center of the drop at early times before eventually decaying away at later times.
Generally, there is good agreement between the simulations and experiments.
\begin{suppfigure}
\includegraphics[width=0.3\textwidth]{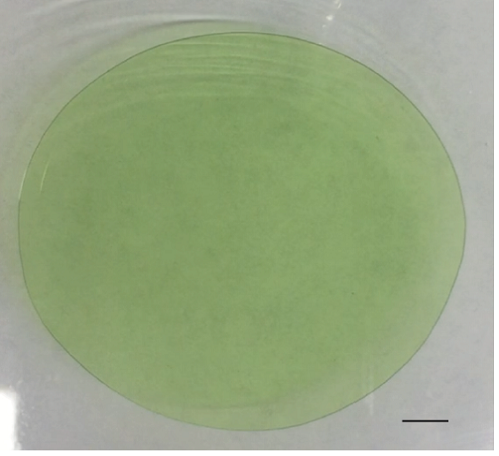}[a]
\includegraphics[width=0.3\textwidth]{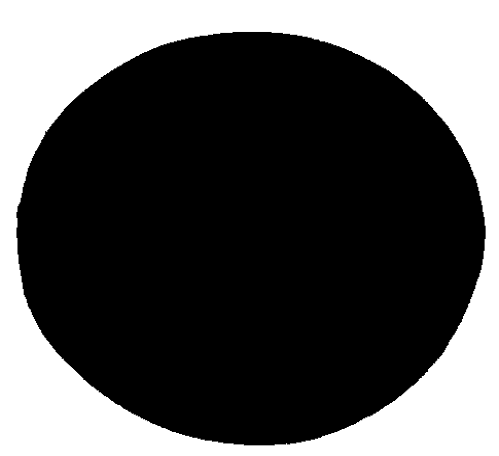}[b]
\includegraphics[width=0.3\textwidth]{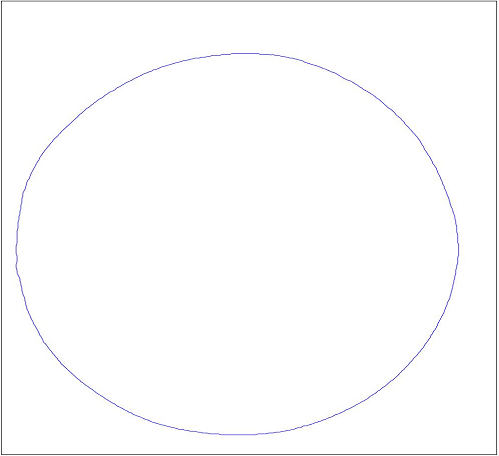}[c]
\caption{Image processing in the manual lifting plate experiment. [a] Top view of the drop in the Hele-Shaw cell (the scale bar is 
$5$mm); [b] The fluid domain after applying thresholding in IMAGEJ; [c]  The contour of the air-oil contact line extracted in MATLAB.
}\label{lncom4}
\end{suppfigure}

\begin{suppfigure}
\centering
\includegraphics[scale=0.39]{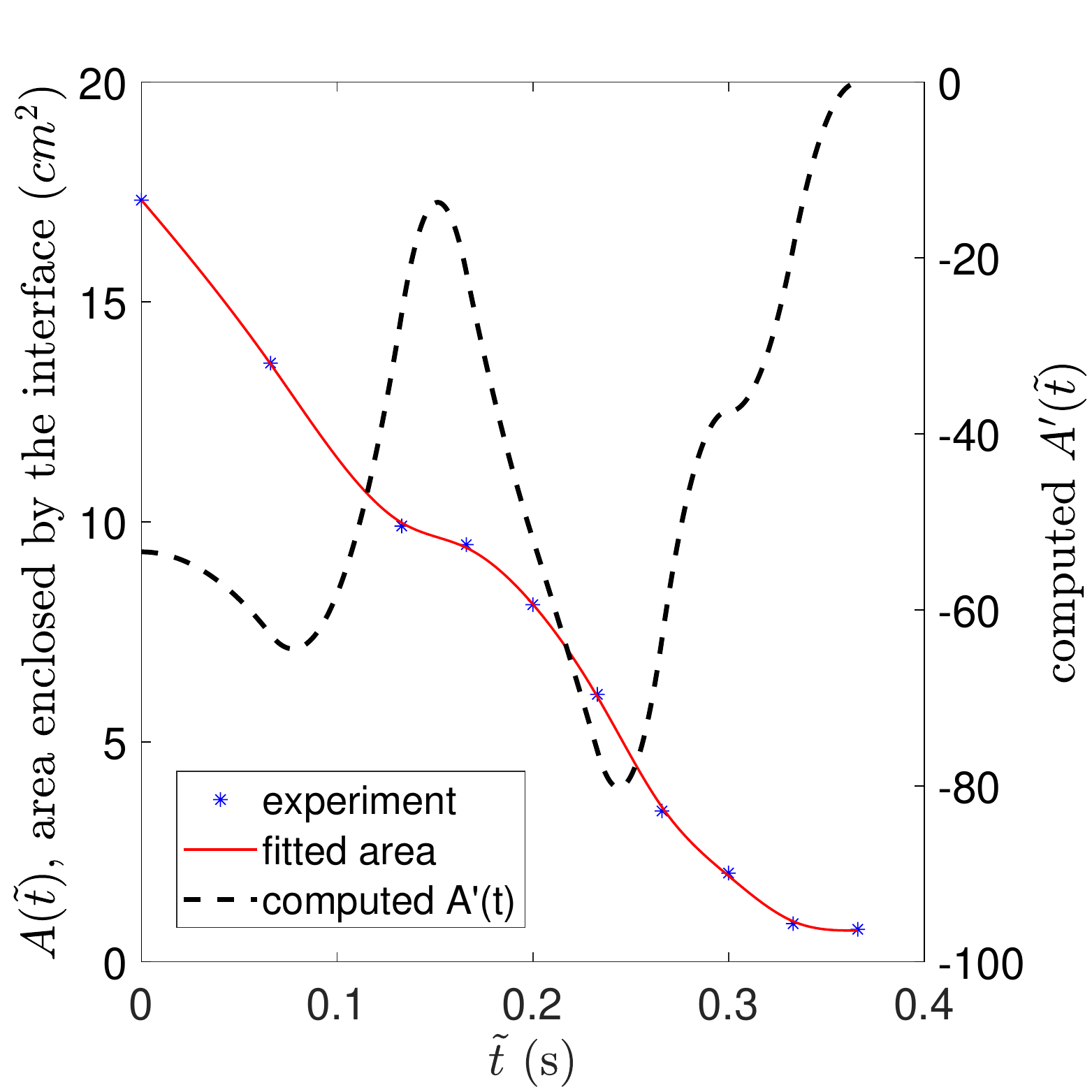}[a]
\includegraphics[scale=0.39]{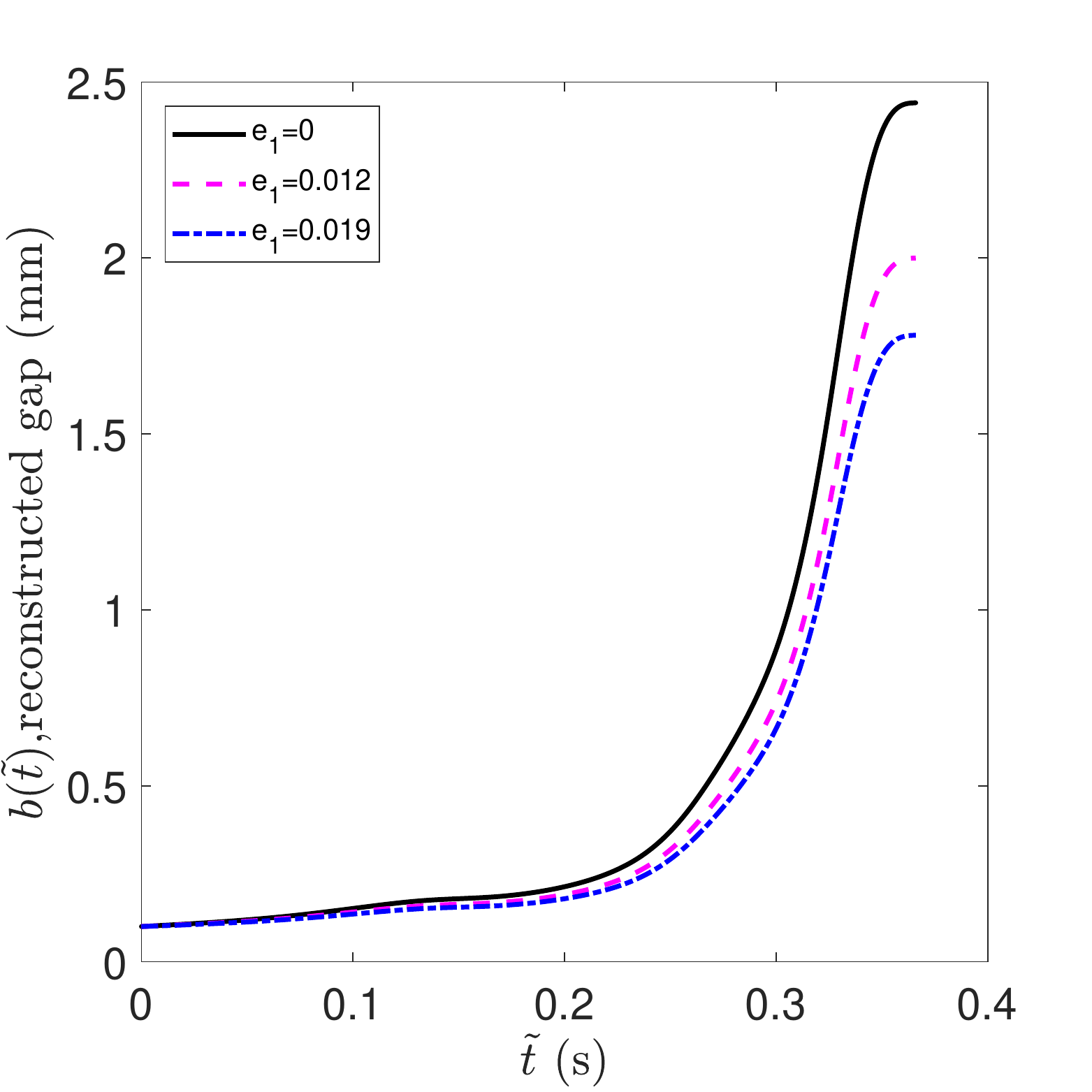}[b]
\caption{The drop areas and gap widths for experiments with nonlinear dynamics of the gap width. 
[a] The drop areas (*), the fitted area (solid line) and the computed $A'$ (dashed line) in dimensional time $\tilde t$. [b]  The gap widths reconstructed from the area $A(\tilde{t})$ as $\displaystyle b(\tilde t)=\frac{b_0A(0)}{A(\tilde t)}\exp(e_1(A(\tilde{t})-A(0)))$ under different $e_1$ as labeled. See text for details.}\label{simudata}
\end{suppfigure}

\begin{suppfigure}
\hspace{0.03in}
\includegraphics[scale=0.49]{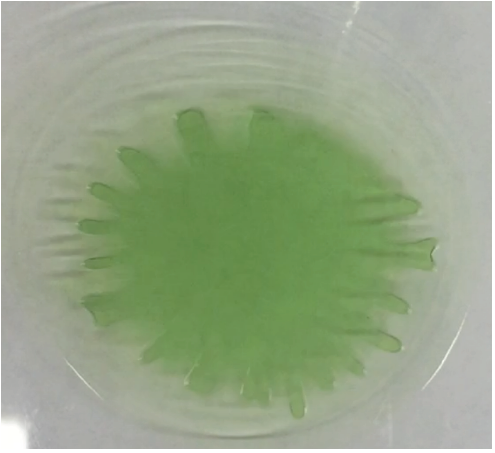}
\hspace{0.01in}
\includegraphics[scale=0.49]{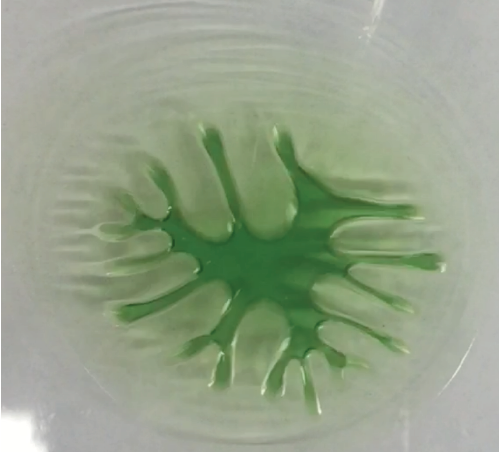}
\hspace{0.01in}
\includegraphics[scale=0.49]{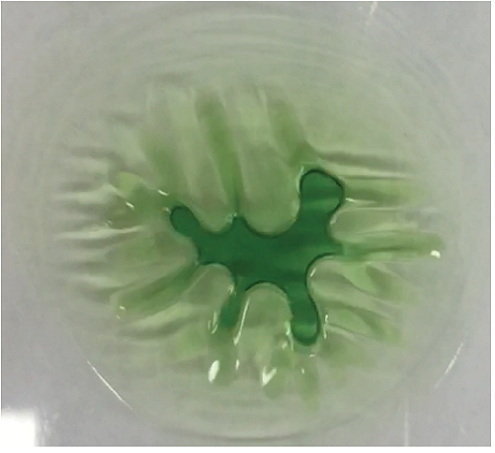}\\
\includegraphics[scale=0.25]{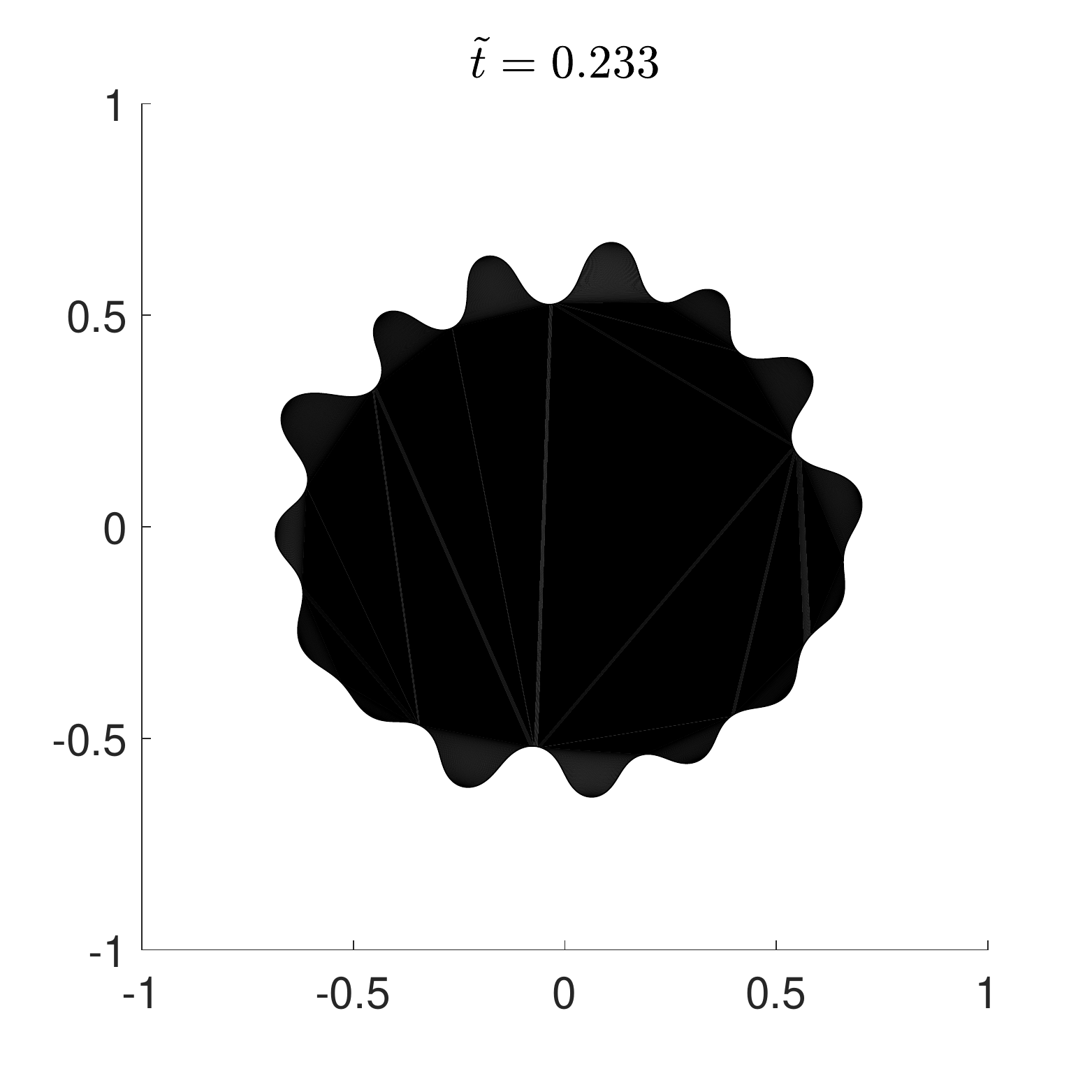}
\includegraphics[scale=0.25]{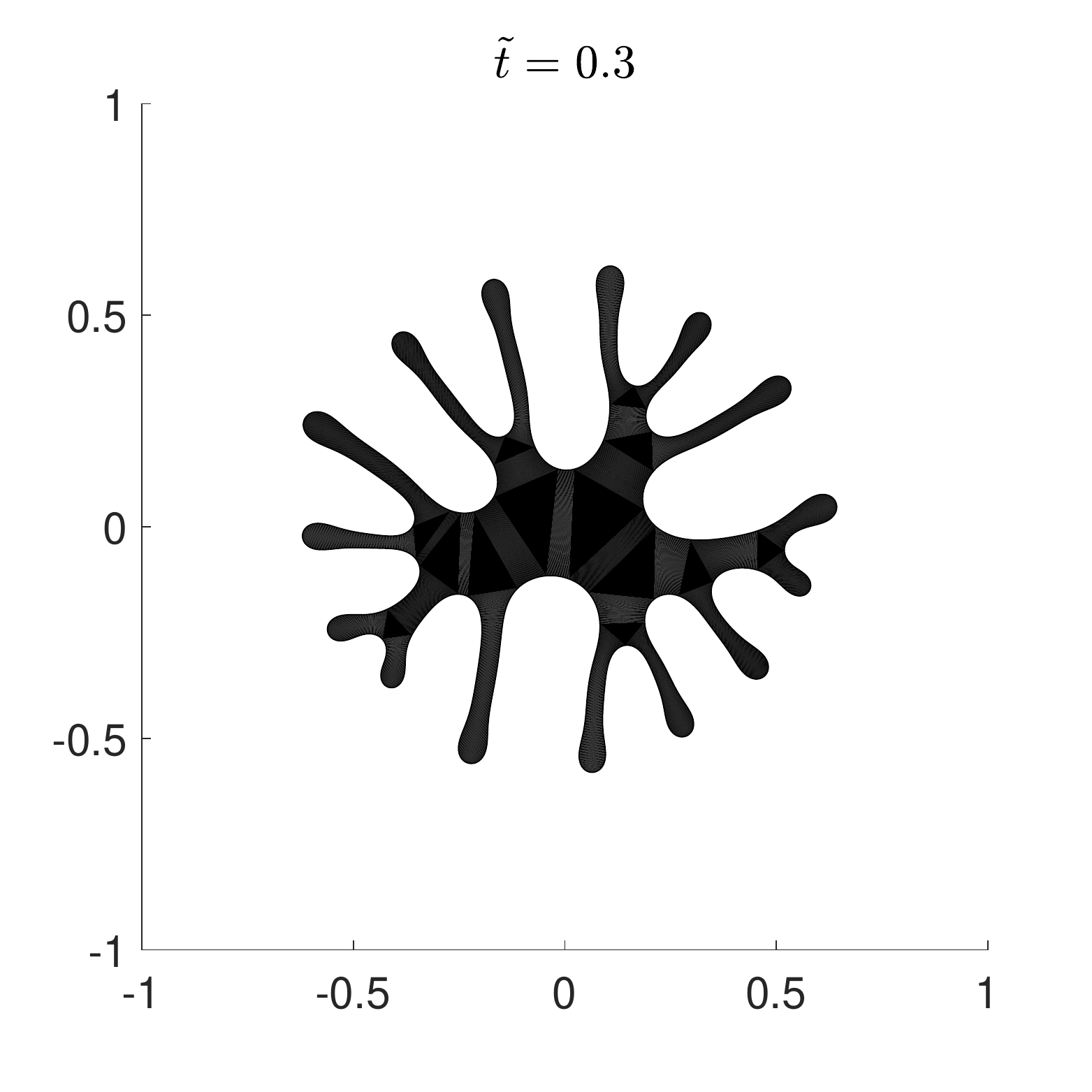}
\includegraphics[scale=0.25]{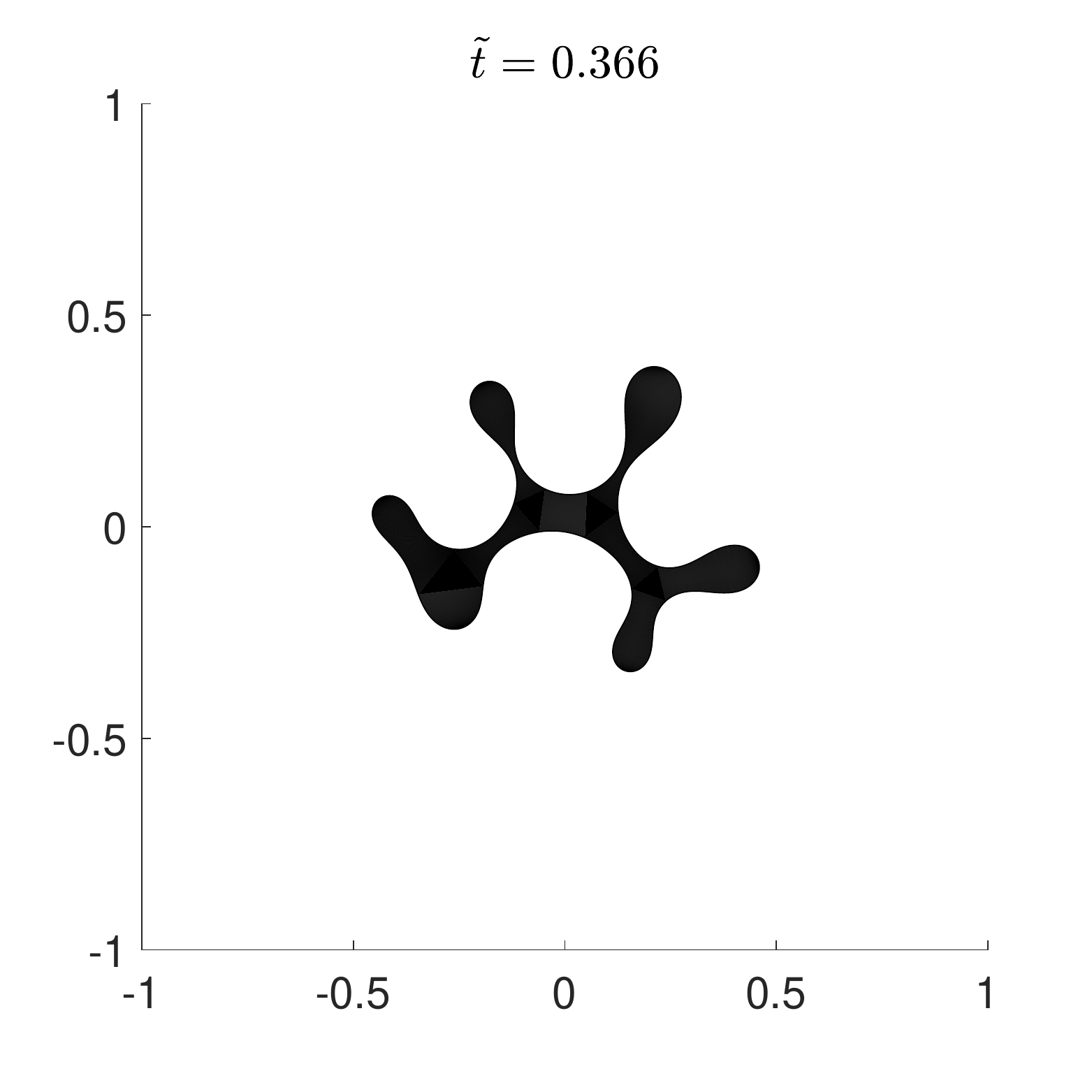}\\
\includegraphics[scale=0.25]{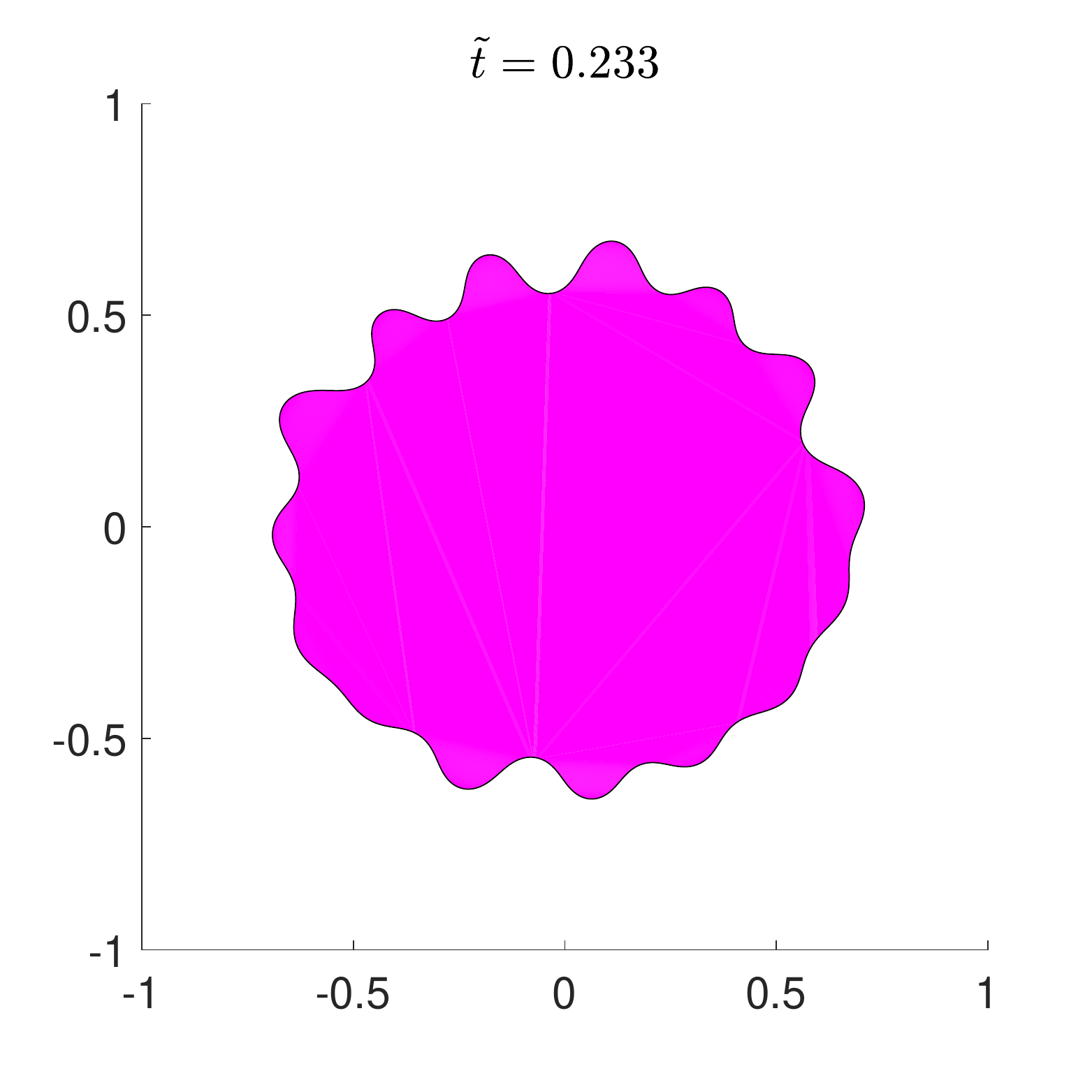}
\includegraphics[scale=0.25]{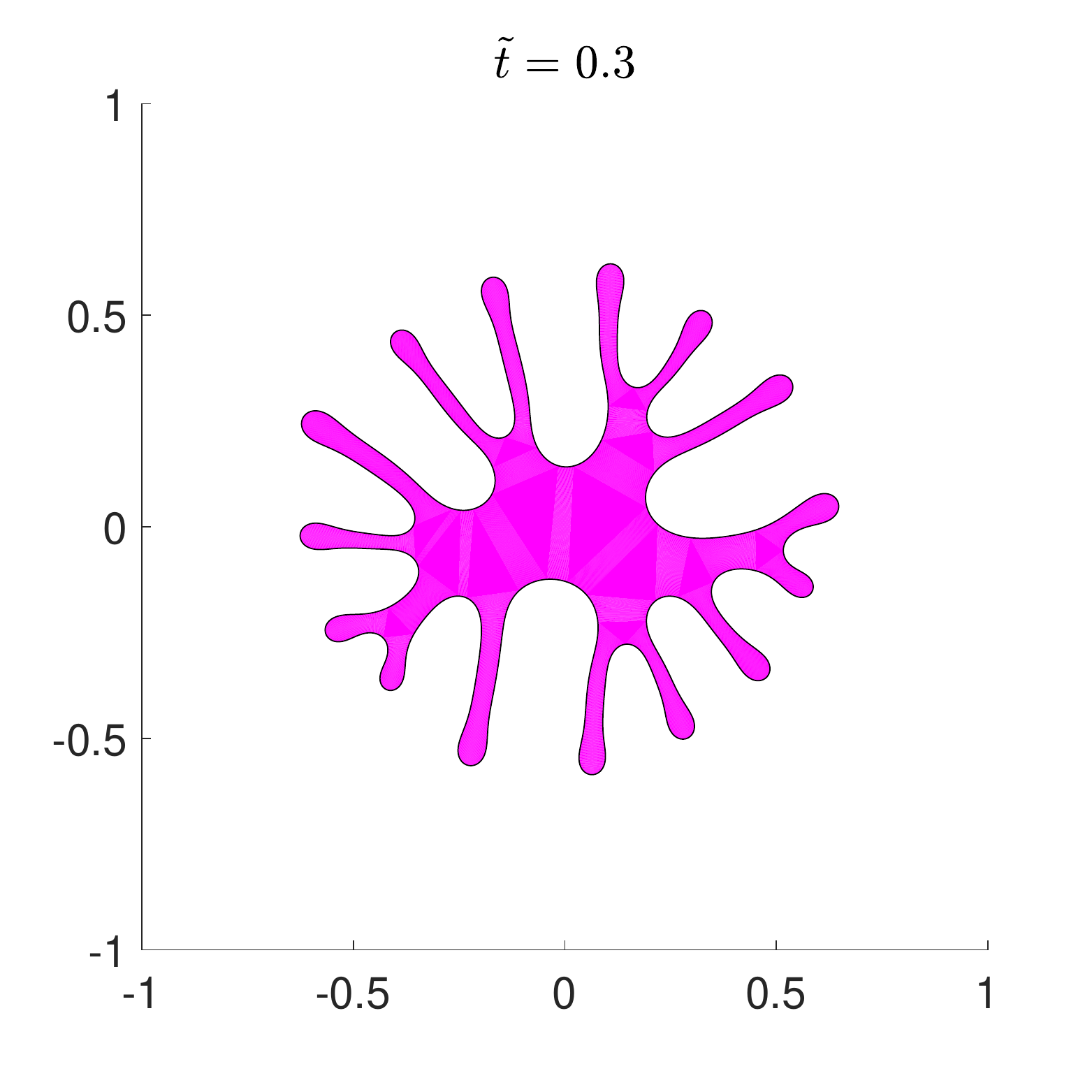}
\includegraphics[scale=0.25]{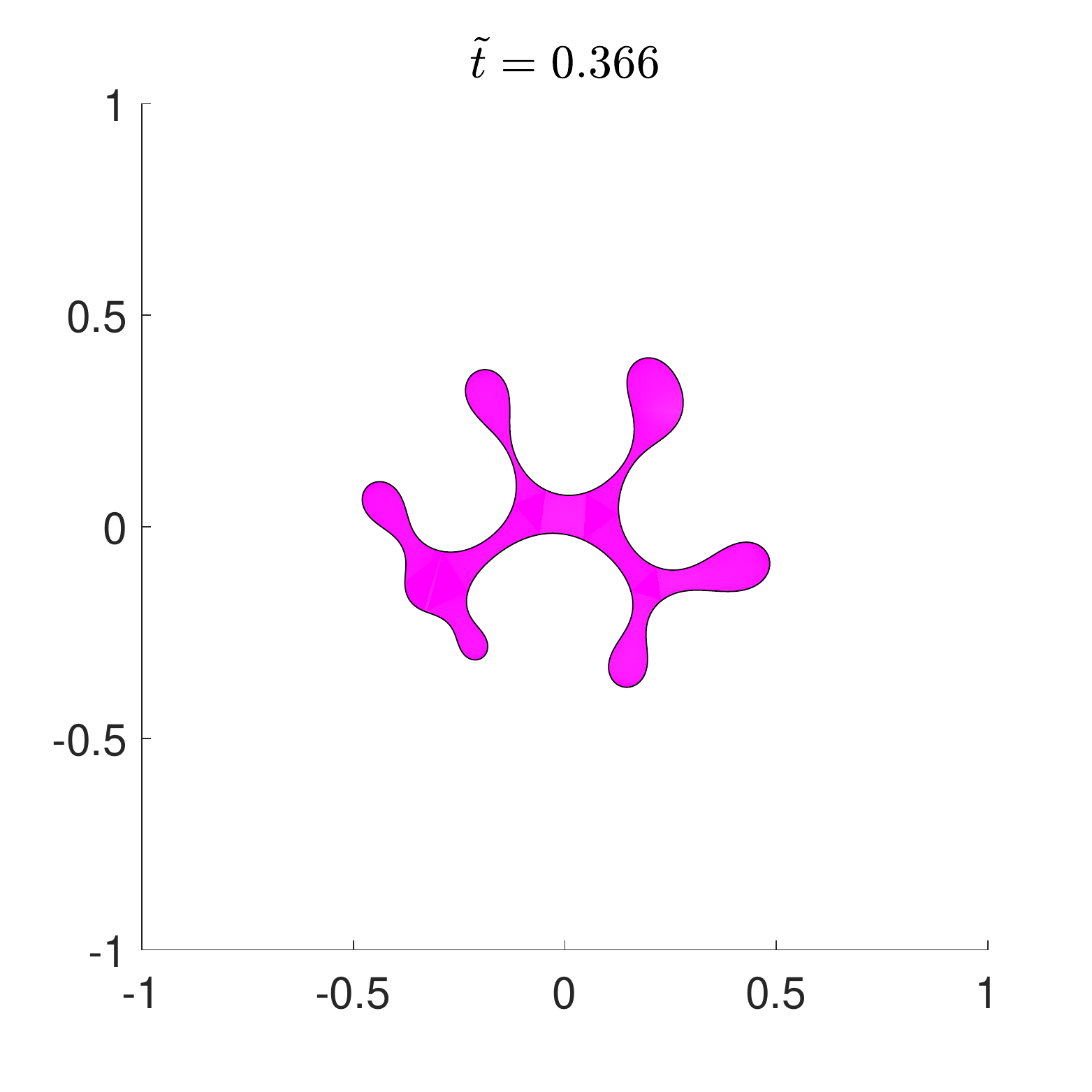}\\
\includegraphics[scale=0.25]{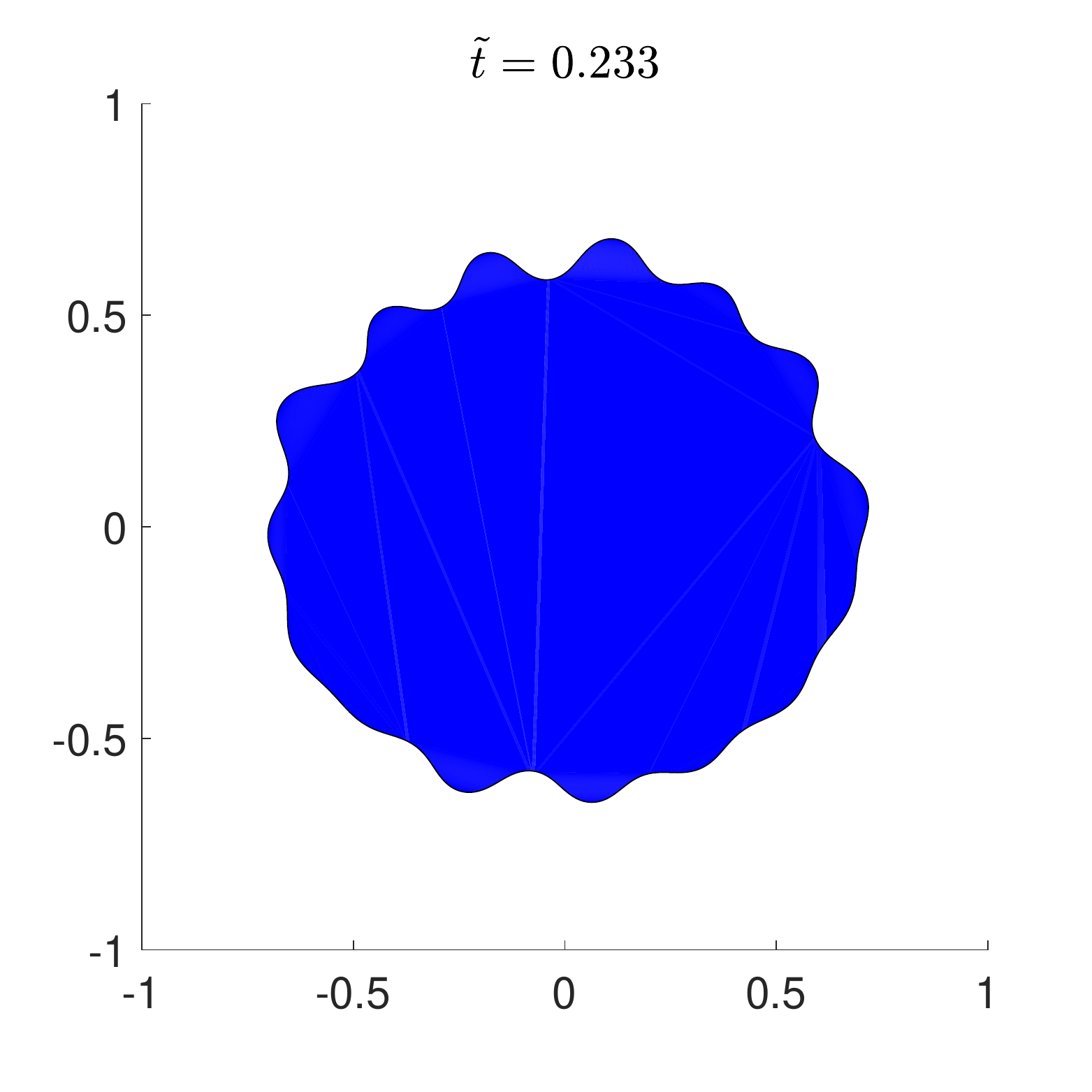}
\includegraphics[scale=0.25]{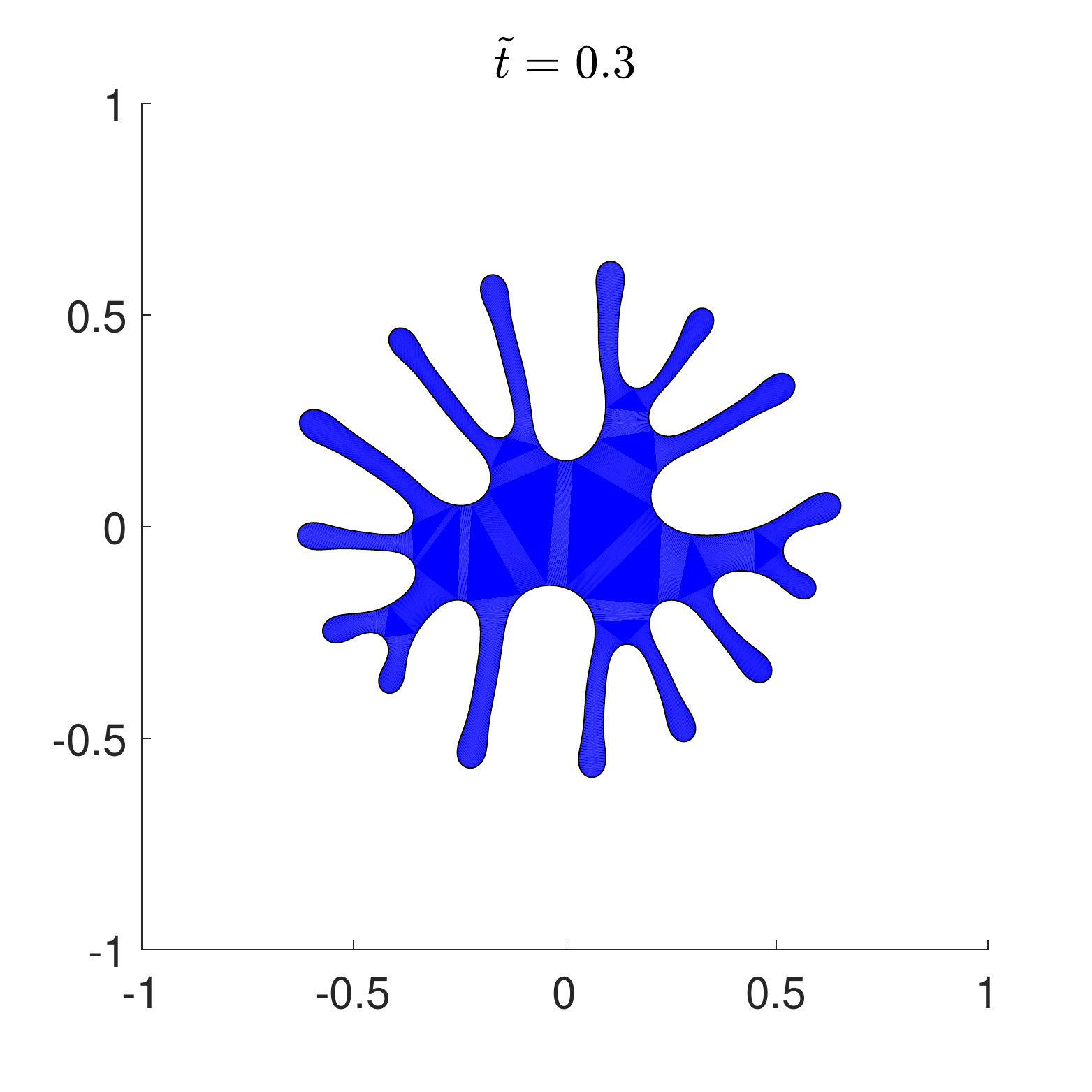}
\includegraphics[scale=0.25]{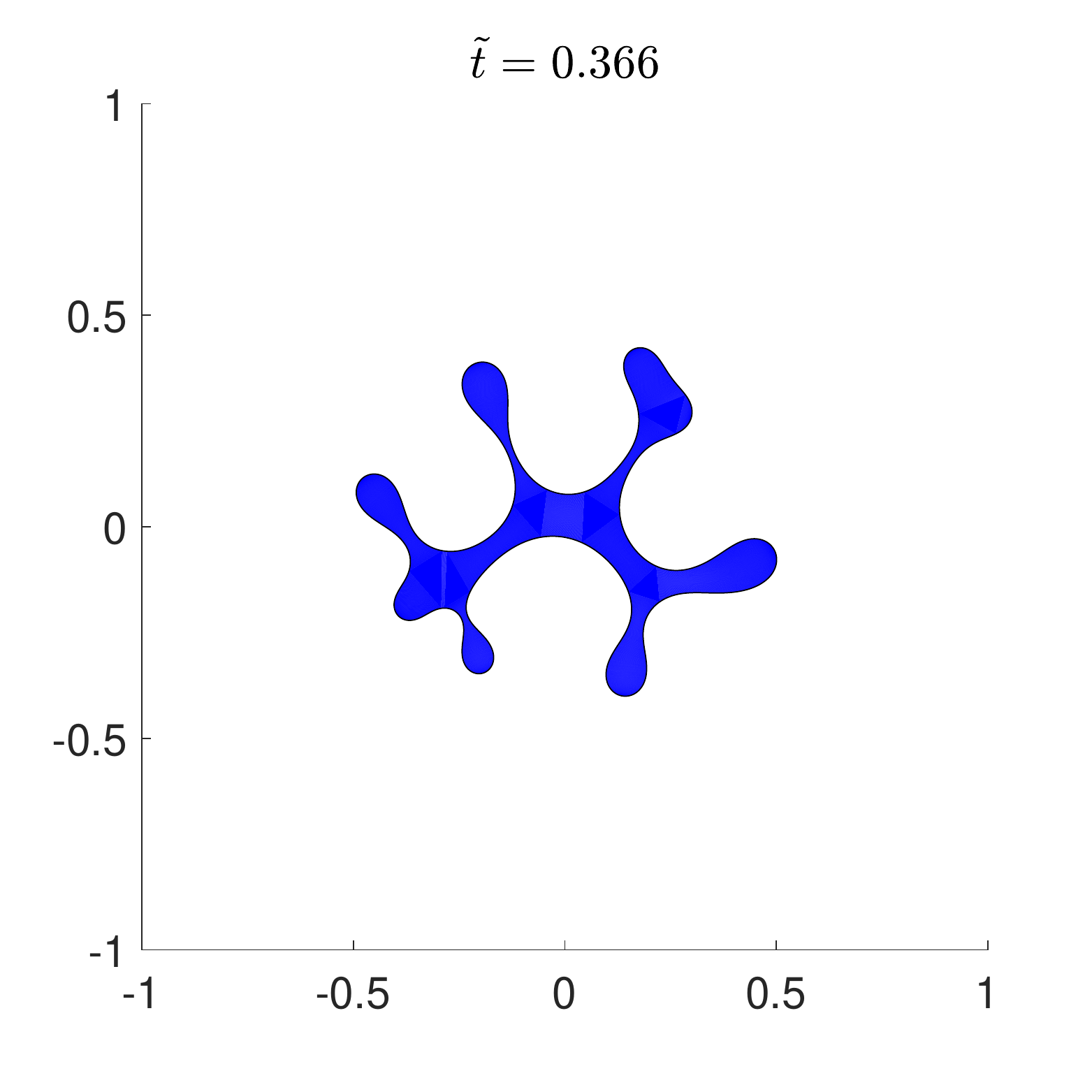}
\caption{Morphologies of the drops with nonlinear gap dynamics from experiments (first row) and simulations with different compensation parameters $e_1=0$ (black; row 2), $e_1=0.012$ (magenta; row 3), and $e_1=0.019$ (blue; rwo 4) at the dimensional times $\tilde{t}=0.233$ (first column), $0.3$ (second column), and $0.366$ (third column). }\label{comparison}
\end{suppfigure}
\bibliographystyle{jfm}
\bibliography{liftingdiagram}